\documentstyle[amssymb,amscd,12pt]{amsart}
\normalbaselines
\numberwithin{equation}{subsection}

\begin{document}

\newcommand{\thmref}[1]{Theorem~\ref{#1}}
\newcommand{\secref}[1]{\S~\ref{#1}}
\newcommand{\lemref}[1]{Lemma~\ref{#1}}
\newcommand{\propref}[1]{Proposition~\ref{#1}}
\newcommand{\corref}[1]{Corollary~\ref{#1}}
\newcommand{\remref}[1]{Remark~\ref{#1}}
\newcommand{\nc}{\newcommand}
\nc{\on}{\operatorname}
\nc{\ch}{\mbox{ch}}
\nc{\Z}{{\Bbb Z}}
\nc{\C}{{\Bbb C}}
\nc{\cond}{|\,}
\nc{\bib}{\bibitem}
\nc{\pone}{\Pro^1}
\nc{\pa}{\partial}
\nc{\F}{{\cal F}}
\nc{\arr}{\rightarrow}
\nc{\larr}{\longrightarrow}
\nc{\al}{\alpha}
\nc{\ri}{\rangle}
\nc{\lef}{\langle}
\nc{\W}{{\cal W}}
\nc{\gam}{\bar{\gamma}}
\nc{\Q}{\bar{Q}}
\nc{\q}{\widetilde{Q}}
\nc{\la}{\lambda}
\nc{\ep}{\epsilon}
\nc{\su}{\widehat{\goth{sl}}_2}
\nc{\sw}{\goth{sl}}
\nc{\g}{\goth{g}}
\nc{\h}{\goth{h}}
\nc{\n}{\goth{n}}
\nc{\ab}{\goth{a}}
\nc{\f}{\widehat{{\cal F}}}
\nc{\is}{{\bold i}}
\nc{\V}{\widetilde{V}}
\nc{\M}{\widetilde{M}}
\nc{\js}{{\bold j}}
\nc{\bi}{\bibitem}

\title{Integrals of motion and quantum groups}\thanks{Lectures given by
the second author at the C.I.M.E. Summer School ``Integrable Systems and
Quantum Groups'', Montecatini Terme, Italy, June 14-22, 1993; to appear in
Proceedings of the School, Lect. Notes in Math, {\bf 1620}, Springer Verlag
1995.}

\author{Boris Feigin}
\address{Landau Institute for Theoretical Physics, 2 Kosygina St, Moscow
117940, Russia and R.I.M.S., Kyoto University, Kyoto 606, Japan}

\author{Edward Frenkel}
\address{Department of Mathematics, Harvard University,
Cambridge, MA 02138, USA}

\date{September 1993 \\ Revised: March 1994, March 1995}

\maketitle

\bigskip
\bigskip
\begin{center}
{\sc Contents}
\end{center}
\contentsline {section}{\numberline {1.}Introduction}{}
\contentsline {section}{\numberline {2.}Classical Toda field theories
associated to finite-dimensional simple Lie algebras}{}
\contentsline {subsection}{\numberline {2.1.}The case of $\sw_2$ --
classical Liouville theory}{}
\contentsline {subsection}{\numberline {2.2.}General case}{}
\contentsline {subsection}{\numberline {2.3.}BGG resolution}{}
\contentsline {subsection}{\numberline {2.4.}Extended complex and its
cohomology}{}
\contentsline {section}{\numberline {3.}Classical affine Toda field
theories}{}
\contentsline {subsection}{\numberline {3.1.}The case of
$\su$ -- classical sine-Gordon theory}{}
\contentsline {subsection}{\numberline {3.2.}General case}{}
\contentsline {section}{\numberline {4.}Quantum Toda field theories}{}
\contentsline {subsection}{\numberline {4.1.}Vertex operator algebras}{}
\contentsline {subsection}{\numberline {4.2.}The VOA of the Heisenberg
algebra}{}
\contentsline {subsection}{\numberline {4.3.}Quantum integrals of motion}{}
\contentsline {subsection}{\numberline {4.4.}Liouville theory}{}
\contentsline {subsection}{\numberline {4.5.}Quantum groups and quantum BGG
resolutions}{}
\contentsline {subsection}{\numberline {4.6.}Toda field theories associated
to finite-dimensional simple Lie algebras}{}
\contentsline {subsection}{\numberline {4.7.}Affine Toda field theories}{}
\contentsline {subsection}{\numberline {4.8.}Concluding remarks}{}
\contentsline {section}{References}{}

\section{Introduction.}

\subsection{}
In these lectures we propose a new approach to the study of local integrals
of motion in the classical and quantum Toda field theories. Such a
theory is associated to a Lie algebra $\g$, which is either a
finite-dimensional simple Lie algebra or an affine Kac-Moody algebra.

The classical Toda field theory, associated to $\g$, revolves around
the system of equations
\begin{equation}   \label{toda}
\pa_\tau \pa_t \phi_i(t,\tau) = \frac{1}{2} \sum_{j \in S} (\al_i,\al_j)
\exp[\phi_j(t,\tau)], \quad \quad i \in S
\end{equation}
where each $\phi_i(t,\tau)$ is a family of functions in $t$, depending on
the time variable $\tau$, $S$ is the set of simple roots of $\g$, and
$(\al_i,\al_j)$ is the scalar product of the $i$th and $j$th simple roots
\cite{Kac}.

The simplest examples of the Toda equations are the Liouville equation
\begin{equation}    \label{liouville}
\pa_\tau \pa_t \phi(t,\tau) = e^{\phi(t,\tau)},
\end{equation}
corresponding to $\g=\sw_2$, and the sine-Gordon equation
\begin{equation}    \label{sine}
\pa_\tau \pa_t \phi(t,\tau) = e^{\phi(t,\tau)} - e^{-\phi(t,\tau)},
\end{equation}
corresponding to $\g=\widehat{\sw}_2$.

Many aspects of the classical Toda field theories have
been studied by both physicists and mathematicians (cf., e.g.,
\cite{akns,tf,ml,mop,ot1,ot2,DS,ds,kw,wilson,savlez,ft1,feher,bb1,bb2,
bobb,otu} and references therein): realization as a zero-curvature
equation, complete integrability, soliton solutions, dressing
transformations, connection with generalized KdV hierarchies, lattice
analogues, etc. There are also many interesting works devoted to the
quantum Toda field theory, cf., e.g.,
\cite{zz,stf,gernev,bilger,HM,EY,rs,bcds,volkov,fv,nie2} and
references therein.

\subsection{}
In this paper we will study a Hamiltonian formalism for the Toda field
theories. By that we mean constructing a Hamiltonian space $M$ and a
hamiltonian $H$, such that the system of equations
\eqref{toda} can be rewritten in the Hamiltonian form:
\begin{equation}   \label{hamform}
\pa_\tau U = \{U,H\}.
\end{equation}
Here $\{\cdot,\cdot\}$ stands for a Poisson bracket on the space
$F(M)$ of functions on $M$.  We will be primarily interested in the
integrals of motion for the equation \eqref{hamform}. An integral of
motion is an element $X$ of the space $F(M)$, which satisfies the
equation $$\{X,H\} = 0.$$ It is conserved with respect to the
evolution of the Hamiltonian system, defined by the equation
\eqref{hamform}.

Note that this definition does not require $H$ to be an element of $F(M)$,
it merely requires the Poisson bracket with $H$ to be a well-defined linear
operator, acting from $F(M)$ to some other vector space. Given $H$, we can
define the space of integrals of motion of the system
\eqref{hamform} as the kernel of this linear operator. If this
operator preserves the Poisson bracket, then the space of integrals of
motion is itself a Poisson algebra.

\subsection{}
For the Toda equation \eqref{toda} we choose as the Hamiltonian space, the
space $L\h$ of polynomial functions on the circle with values in the Cartan
subalgebra $\h$ of $\g$ and as the space of functions, the space ${\cal
F}_0$ of local functionals on $L\h$. Such a functional can be presented in
the form of the residue $$F[{\bold u}(t)] = \int P({\bold u},\pa_t{\bold
u},\ldots) dt,$$ where $P$ is a polynomial in the coordinates $u^i(t)$ of
${\bold u}(t)
\in L\h$ with respect to the basis of the simple roots, and their
derivatives (cf. \secref{classical} for the precise definition). The
Poisson structure on $L\h$ has an interpretation as a Kirillov-Kostant
structure, because $L\h$ can be viewed as a hyperplane in the dual space to
the Heisenberg Lie algebra $\widehat{\h}$ -- the central extension of $L\h$.
This defines a Poisson bracket on ${\cal F}_0$.

The space $\F_0$ was one of the first examples of Poisson algebras of
functions on infinite-dimensional hamiltonian spaces. It has been studied
since the discovery of integrability of the KdV equation and its
generalizations, and exhaustive literature is devoted to it, cf., e.g.
\cite{ggkm,gardner,zf,lax,gd1,gd2,adler,manin,ds,wilson,ft,dickey}.
We essentially follow the approach of Gelfand and Dickey and treat the
space $\F_0$ in a purely algebraic way. In addition, we also consider the
spaces $\F_{\al_i}$, consisting of functionals of the form $$\int P({\bold
u},\pa_t{\bold u},\ldots) e^{\phi_i(t)} dt,$$ where $\phi_i(t)$ is such that
$\pa_t \phi_i(t) = u^i(t)$. It is possible to extend the Poisson bracket
${\cal F}_0
\times {\cal F}_0 \arr {\cal F}_0$ to a bilinear map ${\cal F}_0
\times {\cal F}_{\al_i} \arr \F_{\al_i}$, cf. \cite{kw,wilson,ds}, and
this allows to write the Toda equation \eqref{toda} in the Hamiltonian
form $$\pa_\tau {\bold u}(t) = \{{\bold u}(t),H\}.$$

Here the hamiltonian $H$ is given by $$H = \frac{1}{2} \sum_i \int
e^{\phi_i(t)} dt.$$ It is an element of $\oplus_i
\F_{\al_i}$, and the Poisson bracket with $H$ is a well-defined linear
operator, acting from $\F_0$ to $\oplus_i \F_{\al_i}$.

So we can define the space of local integrals of motion of the Toda
equation \eqref{toda} as the kernel of the operator $\{\cdot,H\}$, or, in
other words, as the intersection of the kernels of the operators $\bar{Q}_i
= \{\cdot,\int e^{\phi_i(t)} dt\}: \F_0 \arr \F_{\al_i}$.  These operators
preserve the Poisson structure, and hence the space of integrals of motion
is a Poisson subalgebra of $\F_0$.

\subsection{}
The crucial observation, which will enable us to compute this space, is
that, roughly speaking, {\em the operators $\bar{Q}_i$ satisfy the Serre
relations of the Lie algebra $\g$, or, in other words, they generate the
nilpotent subalgebra $\n_+$ of $\g$}. Using this fact, we will be able to
interpret the space of local integrals of motion as a cohomology space of a
certain complex $F^*(\g)$. To construct this complex, we will use the
so-called Bernstein--Gelfand--Gelfand (BGG) resolution, which is the
resolution of the trivial representation of $\n_+$ by Verma modules. The
cohomologies of this complex coincide with the cohomologies of
$\n_+$ with coefficients in some $\n_+$--module.

More precisely, we can lift the operators $\bar{Q}_i$ to certain
linear operators $Q_i$, acting on the space $\pi_0$ of differential
polynomials in $u^i(t)$. These operators give us an action of the Lie
algebra $\n_+$ on $\pi_0$.

In the case, when $\g$ is a finite-dimensional simple Lie algebra, the
$0$th cohomology of $\n_+$ with coefficients in $\pi_0$ can be identified
with the space of differential polynomials in $W^{(1)},\ldots,W^{(l)} \in
\pi_0$ of degrees $d_1+1,\ldots,d_l+1$, respectively. Here the $d_i$'s are
the exponents of $\g$, and the grading is defined on $\pi_0$ in such a way
that the degree of $\pa_t^n u^i(t)$ is equal to $n+1$. We prove that the
integrals of motion of the Toda field theory, corresponding to $\g$,
coincide with all residues of differential polynomials in the
$W^{(i)}$'s. They form a Poisson subalgebra of $\F_0$, which is called the
Adler-Gelfand-Dickey algebra, or the classical ${\cal W}$--algebra,
associated with the Lie algebra $\g$.

The space of integrals of motion of the affine Toda field theory,
associated to an affine algebra $\g$, can be identified with the first
cohomology of the nilpotent subalgebra $\n_+$ of $\g$ with coefficients in
$\pi_0$. This space is naturally embedded into the space of integrals of
motion of the Toda theory, associated to the finite-dimensional Lie algebra
$\bar{\g}$, whose Dynkin diagram is obtained by deleting the $0$th nod of
the Dynkin diagram of $\g$ (or any other nod).

We can compute the latter cohomology and obtain the well-known result that
the integrals of motion of the affine Toda theory have degrees equal to the
exponents of $\g$ modulo the Coxeter number. These integrals of motion
commute with each other. This is especially easy to see in the case when
all the exponents of the corresponding affine algebra are odd, and the
Coxeter number is even (this excludes $A_n^{(1)}, n>1, D_{2n}^{(1)},
E_6^{(1)}$ and $E_7^{(1)}$). In such a case the degrees of all integrals of
motion are odd, so that the Poisson bracket of any two of them should be an
integral of motion of an even degree, and hence should vanish. The set of
local integrals of motion of the affine Toda field theory, associated to
$\g$, coincides with the set of hamiltonians of the corresponding
generalized KdV system. These integrals of motion generate a maximal
abelian subalgebra in the Poisson algebra of integrals of motion of the
corresponding finite-dimensional Toda field theory.

In our next paper \cite{kdv} we explain further the geometric meaning of
higher KdV hamiltonians. Namely, we will identify the vector space with the
coordinates $\pa^n u^i, i=1,\ldots,l, n\geq 0$, with a homogeneous space of
the nilpotent subgroup $N_+$ of the corresponding affine group. This
homogeneous space is the quotient of the group $N_+$ by its principal
commutative subgroup -- the Lie group of $\ab$. The vector fields on this
space, which, by Gelfand-Dickey formalism correspond to the KdV
hamiltonians, coincide with the vector fields of the infinitesimal action
of the opposite principal abelian subalgebra $\ab_- \subset \n_-$ on this
homogeneous space. In particular, this identification enables us to prove
the mutual commutativity of hamiltonians in general case.

\subsection{}
Thus we obtain an interpretation of the integrals of motion of the
classical Toda field theories as cohomologies of certain complexes.
This formulation not only allows us to describe the spaces of
classical integrals of motion, but also to prove the
existence of their quantum deformations.

The quantum integrals of motion are defined as elements of the quantum
Heisenberg algebra, which is a quantization of the Poisson algebra $\F_0$.
More precisely, we define a Lie algebra $\F_0^\beta$ of all Fourier
components of vertex operators from the vertex operator algebra of the
Heisenberg algebra, cf. \secref{voaheis}. For completeness, we include in
\secref{quantumvoa} a survey of vertex operator algebras, which closely
follows \S~3 of \cite{FKRW}.

The Lie bracket in $\F_0^\beta$ is polynomial in the deformation parameter
$\beta^2$ with zero constant term, and the linear term coincides with the
Poisson bracket in $\F_0$, so that $\F_0^\beta$ degenerates into $\F_0$
when $\beta \arr 0$. We can also interpret the operators $\Q_i$ as
classical limits of integrals of bosonic vertex operators,
$\Q_i^\beta$. Therefore it is natural to define the space of quantum
integrals of motion of the affine Toda field theory as the intersection of
kernels of the operators $\Q_i^\beta$.

Thus, if $x = x^{(0)} + \beta^2 x^{(1)} + \dots \in \F_0^\beta$ is a
quantum integral of motion, then $x^{(0)}$ is a classical integral of
motion.  It remains to be seen however, whether for each classical integral
of motion there exists its quantum deformation, and whether such deformed
integrals of motion commute with each other.

Such quantum deformations do not necessarily exist in general.
Indeed, if we have a family of linear operators acting between two
vector spaces, then the dimension of the kernel may increase for a
special value of parameter. Thus, the space of classical integrals of
motion, which is defined as the kernel of the operator $\sum_i
\Q_i^\beta$ for the {\em special} value $\beta=0$, may well be larger
than the space of quantum integrals of motion, which is defined as the
kernel of the operator $\sum_i \Q_i^\beta$ for {\em generic} values of
$\beta$.

\subsection{}
In order to prove the existence of quantum integrals of motion we will use
higher cohomologies. The usefulness of higher cohomologies can be
illustrated by the following toy example.

Suppose, we have two finite-dimensional vector spaces, $A$ and $B$, and a
family of linear operators $\phi_\beta$ depending on a parameter
$\beta$. Assume that for $\beta=0$ the $1$st cohomology of the complex
$A\larr B$ (= the cokernel of the operator $\phi_0$) is equal to $0$. One
can show then that the $0$th cohomology of this complex (= the kernel of
$\phi_0$) can be deformed.

Indeed, vanishing of the $1$st cohomology of the complex $A
\larr B$ for $\beta=0$ entails vanishing of the $1$st cohomology for
generic $\beta$, because the dimension of cohomology stays the same for
generic values of parameter, and it may only {\em increase} for special
values. But the Euler characteristics of our complex, i.e. the difference
between the dimension of the kernel and the dimension of the cokernel, is
also equal to $\dim A - \dim B$ and hence does not depend on $\beta$. Since
the $1$st cohomology vanishes for generic $\beta$ and $\beta=0$, we see
that the dimension of the $0$th cohomology for generic $\beta$ is the same
as for $\beta=0$.

We can apply this idea in our situation. Although our spaces are
infinite-dimensional, they are $\Z-$graded with finite-dimensional
homogeneous components, and our operator $H$ preserves the grading.
So, our infinite-dimensional linear problem splits into a set of
finite-dimensional linear problems. In the simplest case of $\g=\sw_2$,
these finite-dimensional problems can be solved in the same way as in
the example above. By proving vanishing of the cokernel of our
operator $\Q_1$, we can prove that all classical local integrals of
motion in the Liouville theory can be quantized. The quantum algebra
of integrals of motion in this case is the quantum Virasoro algebra,
which is a well-known fact.

In general, the cokernel of our operator is not equal to $0$ for $\beta=0$,
so that this simple trick does not work. However, a deformation of our
extended complex $F^*(\g)$ will do the job. Roughly speaking, it turns out
that the operators $Q_i^\beta$ generate the quantized universal enveloping
algebra $U_q(\n_+)$ of the nilpotent subalgebra $\n_+$ with $q = \exp(\pi i
\beta^2)$. A quantum analogue of the BGG resolution will allow us to
construct such a deformed complex, $F^*_\beta(\g)$, for generic
$\beta$ in the same way as for $\beta=0$.

\subsection{}
In the case when $\g$ is finite-dimensional, we prove that all higher
cohomologies of our complex vanish when $\beta=0$, so they also vanish for
generic $\beta$. Using Euler characteristics we then prove that all
classical integrals of motion of the corresponding Toda field theory can be
quantized. These quantum integrals of motion are Fourier components of
vertex operators from a certain vertex operator algebra -- the so-called
${\cal W}-$algebra. A comprehensive review of the theory of ${\cal
W}-$algebras and references can be found in \cite{bs}.

The $\W-$algebra, corresponding to $\g=\sw_2$, is the vertex operator
algebra of the Virasoro algebra. The $\W-$algebra, corresponding to
$\g=\sw_3$, was constructed by Zamolodchikov
\cite{zam}. For $A$ and $D$ series of simple Lie algebras the $\W-$algebras
were constructed by Fateev and Lukyanov
\cite{Fateev,Fateev1,fl}. We give a general proof of the
existence of ${\cal W}-$algebras, associated to arbitrary
finite-dimensional simple Lie algebras. A similar construction also
appeared in \cite{nie1} for $\g=\sw_n$.

The $\W-$algebra can also be defined by means of quantum Drinfeld-Sokolov
reduction \cite{ff1,critical,fkw} as the $0$th cohomology of a certain BRST
complex. The definition via quantum Drinfeld-Sokolov reduction is related
to the definition in this paper, because the first term of a spectral
sequence associated to the BRST complex coincides with our quantum complex
$F^*_\beta(\g)$ for generic values of $\beta$
\cite{critical,cargese}. Therefore the cohomologies of the BRST complex and
the complex $F^*_\beta(\g)$ coincide for generic $\beta$. Recently these
cohomologies were computed using the opposite spectral sequence of the BRST
complex \cite{tjin}. This gives an alternative proof of existence of
$\W-$algebras.

If $\g$ is affine, our extended complex has non-trivial higher cohomologies
when $\beta=0$. However, in the case when all exponents of $\g$ are odd and
the Coxeter number is even, we can still derive that all cohomology classes
can be deformed, from the Euler character argument.  We then immediately
see that the corresponding quantum integrals of motion commute with each
other. In the remaining cases we can prove these results by a more refined
argument. When $\g=\widehat{\sw}_2$, we obtain a proof of the existence of
quantum KdV hamiltonians. In this setting, it was conjectured in
\cite{gervais1} and some partial results were obtained in
\cite{Zam,EY,km,dm1,dm2,SY}.

\subsection{}
The ${\cal W}-$algebra is the chiral algebra of a certain two-dimensional
conformal field theory; this explains its importance for quantum field
theory. Our construction gives a realization of this chiral algebra in
terms of the simplest chiral algebra, the chiral algebra of the free
fields. Such a realization, which is usually referred to as a free field
realization, is very important for computation of the correlation functions
in the corresponding models of quantum field theory. The free field
realization is an attempt to immerse a complicated structure into a simpler
one and then give the precise description of the image of the complicated
structure by finding the constraints, to which it satisfies inside the
simple one. Our construction resembles the construction of the
Harish-Chandra homomorphism, which identifies the center of the universal
enveloping algebra of a simple Lie algebra $\g$ (complicated object) with
the polynomials on the Cartan subalgebra $\h$ -- an abelian Lie algebra
(simple object), which are invariant with respect to the action of the Weyl
group. In other words, the center can be described as the subalgebra in the
algebra of polynomials on $\h$, which are invariant with respect to the
simple reflections $s_i$.  Likewise, we have been able to embed the ${\cal
W}-$algebra of $\g$ into the vertex operator algebra of the Heisenberg
algebra $\widehat{\h}$, which is something like the algebra of polynomials on
$L\h$. The image of this embedding coincides with the kernel of the
operators $\Q_i^\beta$, which play the role of simple reflections from the
Weyl group.

The quantum integrals of motion of the affine Toda field theory, associated
to an affine algebra $\g$, constitute an infinite-dimensional abelian
subalgebra in the ${\cal W}-$algebra, associated to the finite-dimensional
Lie algebra $\bar{\g}$. This abelian subalgebra consists of the local
integrals of motion of a deformation of this conformal field theory. The
knowledge of the existence of infinitely many integrals of motion and of
their degrees (or spins) is very important for understanding this
non-conformal field theory, and in many cases it allows to construct the
S-matrix of this theory explicitly \cite{Zam}.

Some of the results of this paper have previously appeared in our papers
\cite{ff1,critical,cargese,fftoda}. An earlier version of this paper
appeared in September of 1993 as a preprint YITP/K-1036 of Yukawa Institute
of Kyoto University, and also as hep-th/9310022 on hep-th computer net.

\section{Classical Toda field theories associated to finite-dimensional
simple Lie algebras.}    \label{classical}

\subsection{The case of $\sw_2$ -- classical Liouville theory.}
    \label{liouv}

\subsubsection{Hamiltonian space.}    \label{lhamspace}
Denote by $\h$ the Cartan subalgebra of $\sw_2$ -- the one-dimensional
abelian Lie algebra, and by $L\h$ the abelian Lie algebra of polynomial
functions on the circle with values in $\h$. This will be our hamiltonian
space. It is isomorphic to the space of Laurent polynomials
$\C[t,t^{-1}]$. We would like to introduce a suitable space of functions on
$L\h$, which we will denote by $\F_0$, together with a Poisson bracket.

First let us introduce the space $\pi_0$ of differential polynomials,
i.e.  the space of polynomials in variables $u, \pa u, \pa^2 u,
\dots$. It is equipped with an action of derivative $\pa$, which sends
$\pa^n u$ to $\pa^{n+1} u$ and satisfies the Leibnitz rule.

We define $\F_0$ as the space of local functionals on $L\h$. A local
functional $F$ is a functional, whose value at a point $u(t) \in L\h$
can be represented as the formal residue $$F[u(t)] = \int P(u(t),\pa_t
u(t),\dots) dt,$$ where $P \in
\pi_0$ is a differential polynomial, and $\pa_t = \pa/\pa t$. In
words: we insert $u(t), \pa_t u(t), \dots$ into $P$; this gives us a
Laurent polynomial, and we take its residue, i.e. the $(-1)$st Fourier
component.

We can represent local functionals as series of the form
\begin{equation}    \label{series}
\sum_{i_1+\dots+i_m=-m+1} c_{i_1\dots i_m} \cdot u_{i_1}\dots u_{i_m},
\end{equation}
where the coefficients $c_{i_1\dots i_m}$ are polynomials in
$i_1,\dots,i_m$. Here $u_i$'s are the Fourier components of $u(t):
u(t) = \sum_{i\in \Z} u_i t^{-i-1}$. For example, $\int u(t) dt = u_0,
\int u(t)^2 dt = \sum_{i+j=-1} u_i u_j$. Note that since we deal with
Laurent {\em polynomials}, only finitely many summands of the series
\eqref{series} can be non-zero for a given $u(t)$.

We have a map $\int: \pi_0 \arr \F_0$, which sends $P \in \pi_0$ to $\int P
dt \in \F_0$. The following Lemma, for the proof of which
cf. \cite{lax,gd1}, shows that the kernel of the residue map consists of
total derivatives and constants.

\subsubsection{Lemma.}
{\em The sequence $$0 \larr \pi_0/\C \stackrel{\pa}{\larr} \pi_0/\C
\stackrel{\int}{\larr} \F_0 \larr 0$$ is exact.}

\subsubsection{Poisson bracket.}
We can now define the Poisson bracket of two local functionals $F$ and
$G$, corresponding to two differential polynomials, $P$ and $R$, as
follows:
\begin{equation}    \label{skobka}
\{F,G\}[u(t)] = - \int \frac{\delta P}{\delta u} \, \pa_t \, \frac{\delta
R}{\delta u} dt,
\end{equation}
where $$\frac{\delta P}{\delta u} = \frac{\pa P}{\pa u} -
\pa \frac{\pa P}{\pa(\pa u)} + \pa^2 \frac{\pa P}{\pa(\pa^2 u)} -
\dots$$ denotes the variational derivative. Note that the variational
derivative of a differential polynomial, which is a total derivative, is
equal to $0$, and so formula \eqref{skobka} defines a well-defined bracket
map $\F_0 \times \F_0 \arr \F_0$.

\subsubsection{}    \label{leibnitz}
This bracket satisfies all axioms of the Lie bracket and so it defines a
structure of Lie algebra on $\F_0$. One can prove this in terms of
differential polynomials \cite{lax,gd1}, or in terms of Fourier components
\cite{gardner}. We will recall the latter proof following
\cite{manin}, \S\S 7.21-7.23, since we will use it later in
\secref{quantum}.

Note that we can extend our Poisson algebra of local functionals $\F_0$ by
adjoining all Fourier components of differential polynomials, not only the
$(-1)$st ones. Let $\f_0$ be the space of functionals on $L\h$, which can
be represented as residues of differential polynomials with explicit
dependence on $t$: $$F[u(t)] = \int P(\pa^n u(t);t) dt.$$ We can define a
Poisson bracket on $\f_0$ by the same formula \eqref{skobka}. We will prove
now that this bracket makes $\f_0$ into a Lie algebra. This will imply that
$\F_0$ is a Lie algebra as well, because the bracket of two elements of
$\F_0$ is again an element of $\F_0$.

Any element of $\f_0$ can be presented as a {\em finite} linear combination
of the infinite series of the form
\begin{equation}    \label{genseries}
\sum_{i_1+\dots+i_m=N} c_{i_1\dots i_m} \cdot u_{i_1}\dots u_{i_m},
\end{equation}
where the $u_i$'s are the Fourier components of $u(t) = \sum_{i\in\Z} u_i
t^{-i-1}$, $c_{i_1\dots i_m}$ is a polynomial in $i_1,\ldots,i_m$, and $N$
is an arbitrary integer.

We can consider these elements as lying in a certain completion $\bar{A}$
of the polynomial algebra $A=\C[u_n]_{n\in\Z}$. In order to define this
completion, introduce a $\Z$-grading on $A$ by putting $\deg u_n = -n$. We
have $$A = \oplus_{N\in\Z,m\geq 0} A_{N,m},$$ where $A_{N,m}$ is the linear
span of monomials of degree $N$ and power $m$. Denote by $I^M, M>0$ the
ideal of $A$ generated by $u_n, n\geq M$ and by $I^M_{N,m}$ the
intersection $I^M
\cap A_{N,m}$. Let $\bar{A}_{N,m}$ be the completion of $A_{N,m}$ with
respect to the topology, generated by the open sets $I^M_{N,m},
M>0$. Clearly, $$\bar{A} = \oplus_{N\in\Z,m\geq 0} \bar{A}_{N,m}$$ is a
commutative algebra. It consists of finite linear combinations of infinite
series of the form
\eqref{genseries}, where $c_{i_1,\ldots,i_m}$ is an arbitrary function of
$i_1,\ldots,i_m$. In particular, we have an embedding $\f_0 \arr \bar{A}$.

Now, $u_i, i \in \Z$, are elements of $\f_0$ and $\bar{A}$. We find from
formula \eqref{skobka}:
\begin{equation}    \label{jacobi}
\{u_n,u_m\} = n \delta_{n,-m}.
\end{equation}

By the Leibnitz rule
\begin{equation}    \label{leib}
\{ xy,z \} = \{ x,z \} + \{ y,z \},
\end{equation}
we can extend this bracket to a bracket, defined for any pair of monomials
in $u_n$. We can then formally apply this formula by linearity to any pair
of elements of $\bar{A}$, using their presentation in the form
\eqref{genseries}. This will give us a well-defined bracket
$[,\cdot,\cdot]$ on $\bar{A}$. One can check directly that the restriction
\eqref{jacobi} of this formula to the subspace $\oplus_{n\in\Z} \C u_n$
is antisymmetric and satisfies the Jacobi identity. Therefore, by
construction, $[\cdot,\cdot]$ is antisymmetric and satisfies the Jacobi
identity on the whole $\bar{A}$. Thus $[\cdot,\cdot]$ defines a Lie algebra
structure on $\bar{A}$.

It is shown in \cite{manin}, \S 7.23 that the restriction of the Lie
bracket $[\cdot,\cdot]$ to $\f_0$ coincides with the bracket \eqref{skobka}
(for instance, this is clear for the subspace $\oplus_{n\in\Z} \C u_n$ of
$\f_0$). Therefore $\f_0$ is a Lie algebra, and $\F_0$ is its Lie
subalgebra.

\subsubsection{Remark.}
A Poisson algebra is usually defined as an object, which carries two
structures: associative commutative product and a Lie bracket, which are
compatible in the sense that the Leibnitz rule \eqref{leib} holds. It is
clear that the product of two local functionals is not a local functional,
so that $\F_0$ and $\f_0$ are not Poisson algebras in the usual sense, but
merely Lie algebras (however, $\bar{A}$ is a Poisson algebra in the usual
sense). We could, of course, take the algebra of all polynomials in local
functionals and extend our Poisson bracket to it by the Leibnitz rule. This
would give us a Poisson algebra in the usual sense. But this would not make
any difference for us, because we will never use the product structure,
only the Lie bracket. For this reason we will work with $\F_0$ and $\f_0$,
but we will still refer to them as Poisson algebras and will call the
bracket \eqref{skobka} the Poisson bracket, because of another meaning of
the term ``Poisson structure'' -- as the classical limit of some quantum
structure. In \secref{quantum} we will define this quantum structure.

\subsubsection{Kirillov-Kostant structure.}    \label{kirkos}
The Poisson structure on $\F_0$, defined above, has a nice
interpretation as a Kirillov-Kostant structure.

Let us introduce an anti-symmetric scalar product $\lef ,\ri$ on
$L\h$: $$\lef u(t),v(t) \ri =
\int u(t) d v(t).$$ Note that this scalar product does not depend on the
choice of coordinate $t$ on the circle and that its kernel consists of
constants. Using this scalar product, we can define a Heisenberg Lie
algebra $\widehat{\h}$ as the central extension of $L\h$ by the one-dimensional
center with generator $I$. The commutation relations in $\widehat{\h}$ are
$$[u(t),v(t)] = \lef u(t),v(t) \ri I,\quad \quad [u(t),I]=0.$$ In the natural
basis $b_j = t^{-j}, j\in\Z,$ of $L\h$ they can be rewritten as
\begin{equation}    \label{com}
[b_n,b_m] = n \delta_{n,-m} I,\quad \quad [b_n,I]=0.
\end{equation}

The (restricted) dual space $\widehat{\h}^*$ of $\widehat{\h}$ consists of
pairs $(y(t)dt,\mu)$, which define linear functionals on $\widehat{\h}$ by
formula $$(y(t) dt,\mu)[u(t)+\nu I] = \mu \nu + \int u(t) y(t) dt.$$
One has the Kirillov-Kostant Poisson structure on $\widehat{\h}^*$. Since
$I$ generates the center of $\widehat{\h}$, we can restrict this structure
to a hyperplane $\widehat{\h}^*_\mu$, which consists of the linear
functionals, taking value $\mu$ on $I$.

If we choose a coordinate $t$ on the circle, then we can identify
$\widehat{\h}^*_1$ with the space $L\h$. This gives us a Poisson bracket on
various spaces of functionals on $L\h$, e.g., on the space $A$ of
polynomial functionals or its completion $\bar{A}$. Formulas \eqref{jacobi}
and \eqref{com} show that this bracket coincides with the bracket
$[\cdot,\cdot]$, defined in \secref{leibnitz}. The restrictions of this
Poisson bracket to $\f_0$ and $\F_0$ coincide with the ones, defined by
formula \eqref{skobka}.

\subsubsection{The action of $\F_0$ on $\pi_0$.}    \label{action}
The Poisson bracket $\F_0 \times \F_0 \arr \F_0$, defined by formula
\eqref{skobka}, can be thought of as the adjoint action of the Lie algebra
$\F_0$ on itself. We can lift this action to an action of $\F_0$ on the
space $\pi_0$ of differential polynomials \cite{gd2}. Namely, we can
rewrite formula \eqref{skobka} as follows: $$\{F,G\} = -
\int \frac{\delta P}{\delta u} \, \pa_t \,
\frac{\delta R}{\delta u} dt = - \int \sum_{n\geq 0} (-\pa_t)^n
\frac{\pa P}{\pa(\pa^n u)} \, \pa_t \, \frac{\delta R}{\delta u} dt =$$
$$- \int \sum_{n\geq 0} \frac{\pa P}{\pa(\pa^n u)} \cdot \pa_t^{n+1} \,
\frac{\delta R}{\delta u} dt,$$ using the fact that the integral of a total
derivative is $0$. This suggests to define an action of the functional $G
= \int R dt$ on the space $\pi_0 \simeq \C[\pa^n u]_{n\geq 0}$ by the
vector field
\begin{equation}    \label{vectfield}
- \sum_{n\geq 0}  \left( \pa_t^{n+1}
\, \frac{\delta R}{\delta u} \right) \frac{\pa}{\pa(\pa^n u)}.
\end{equation}

The map $\pi_0 \arr \pi_0$, given by this formula, manifestly commutes with
the action of the derivative $\pa$, and the projection of this map to a map
from $\F_0 = \pi_0/(\pa \pi_0 \oplus \C)$ to itself coincides with the
adjoint action.

We will use the same notation $\{\cdot,G\}$ for the action of $G \in
\F_0$ on $\pi_0$. Note that the action of $\pa$ on $\pi_0$ coincides
with the action of $\frac{1}{2} \int u^2(t) dt$ and that it commutes with
the action of any other $G \in \F_0$.

\subsubsection{}    \label{poisext}
Now we define another space, $\F_1$, of functionals on $L\h$ and extend
our Poisson bracket. We follow \cite{kw}.

Let $\pi_1$ be the tensor product of the space of differential
polynomials $\pi_0$ with a one dimensional space $\C v_1$. Let
us define an action of $\pa$ on $\pi_1$ as $(\pa + u) \otimes 1$,
where $u$ stands for the operator of multiplication by $u$ on $\pi_0$.
Define the space $\F_1$ as the cokernel of the homomorphism $\pa:
\pi_1 \arr \pi_1$. We have the exact sequence: $$0 \larr \pi_1
\stackrel{\pa}{\larr} \pi_1 \larr \F_1 \larr 0.$$

To motivate this definition, introduce formally $\phi(t) = \int^t u(s)
ds$, so that $\pa_t \phi(t) = u(t)$. Consider the space of functionals
on $L\h$, which have the form
\begin{equation}    \label{fone}
\int P(u(t),\pa u(t),\dots) e^{\phi(t)} dt.
\end{equation}
There is a map, which sends $P \otimes v_1 \in \pi_1$ to the functional
\eqref{fone}, so that the action of derivative $\pa_t$ on $P e^{\phi(t)}$
coincides with the action of $\pa$ on $P \in \pi_1$. The kernel of
this map consists of the elements of $\pi_1$, which are total
derivatives; therefore $\F_1$ can be interpreted as the space of
functionals of the form \eqref{fone}.

Note that if $P$ is a differential polynomial in $u$, then $$-\pa \,
\frac{\delta P}{\delta u} = \frac{\delta P}{\delta \phi}.$$ This
formula allows us to extend the Poisson bracket \eqref{skobka}, which was
defined for two differential polynomials in $u$, to the case, when $P$ is a
differential polynomial in $u$, and $R$ depends explicitly on $\phi$. In
particular, we obtain a well-defined map $\F_0 \times \F_1
\arr \F_1$:
\begin{equation}    \label{extension}
\{\int P dt, \int R
e^\phi dt\} = \int \frac{\delta P}{\delta u} \, \frac{\delta [R
e^\phi]}{\delta \phi} dt = \int \frac{\delta P}{\delta u} \left[ R e^\phi -
\pa_t \left( \frac{\delta R}{\delta u} e^\phi \right) \right] dt.
\end{equation}
This bracket satisfies the Jacobi identity for any triple $F, G \in
\F_0, H \in \F_1$. In other words, $\F_1$ is a module over the Lie algebra
$\F_0$.

This statement can be proved in the same way as in
\secref{leibnitz}. Namely, we can extend the space $\F_1$ to the space
$\f_1$ by adjoining elements of the form $\int R e^\phi dt$, where $R$ is a
polynomial in $\pa^m u, m\geq 0$, and $t$. Consider the elements $w_n =
\int t^n e^\phi dt \in \f_1$. By formula \eqref{extension}, the action of
$u_m = \int t^m u(t) dt \in \f_0$ (cf. \secref{leibnitz}) on $w_n$ is given
by
\begin{equation}    \label{components}
\{ u_m,w_n \} = w_{n+m}.
\end{equation}
This formula defines a map $U \times W \arr W$, where $U =
\oplus_{m\in\Z} \C u_m$ and $W = \oplus_{n\in\Z} w_n$. Recall that $U$ is a
Lie algebra, with the commutation relations given by formula
\eqref{jacobi}. One can check directly that this map defines a structure of a
$U$-module on $W$.

In the same way as in \secref{leibnitz} we can define a completion
$\bar{B}$ of the space $B = \C[u_m]_{m\in\Z} \otimes W$, which contains
$\f_1$ and $\F_1$. Using the Leibnitz rule, we can extend the map $U
\times W \arr W$ given by \eqref{components} to a map $\bar{A} \times
\bar{B} \arr
\bar{B}$. By construction, this map defines a structure of an
$\bar{A}$--module on $\bar{B}$. The restriction $\f_0
\times \f_1 \arr \f_1$ of this map coincides with the map
$\{\cdot,\cdot\}$ given by \eqref{extension}. Therefore it defines on
$\f_1$ a structure of a module over the Lie algebra $\f_0$. Hence it makes
$\F_1$ into a module over $\F_0$.

\subsubsection{The Liouville hamiltonian.}    \label{delta}
We now introduce the hamiltonian $H$ of the Liouville model by the formula
$H = \int e^{\phi(t)} dt \in \F_1$. We can rewrite the Liouville equation
\eqref{liouville} in the hamiltonian form as $$\pa_\tau U(t) =
\{U(t),H\}.$$ Here $U(t)$ stands for the delta-like
functional on $L\h$, whose value on a function from $L\h$ is equal to
the value of this function at the point $t$. We can rewrite it as
$U(t) = \int \delta(t-s) u(s) ds$. The formula \eqref{extension} can
be extended to such functionals as well. Applying this formula, we
obtain $$\{\int \delta(t-s) u(s) ds,\int e^{\phi(s)} ds\} = \int
\delta(t-s) e^{\phi(s)} ds = e^{\phi(t)}.$$

\subsubsection{Definition.}
{\em The kernel of the linear operator
\begin{equation}    \label{oper}
\bar{Q} = \{\cdot,\int e^\phi dt\}: \F_0 \arr \F_1
\end{equation}
will be called the space of local integrals of motion of the
classical Liouville theory and will be denoted by $I_0(\sw_2)$.}

\subsubsection{}    \label{actiongen}
Our goal is to compute the space $I_0(\sw_2)$. Note that by the Jacobi
identity, it is closed with respect to the Poisson bracket.

It is more convenient to work with the spaces $\pi_0$ and $\pi_1$, than
with $\F_0$ and $\F_1$. We want to define a linear operator $\q: \pi_0
\arr \pi_1$, which commutes with the action of $\pa$ on these spaces,
and descends down to the operator $\bar{Q}$.

To define such an operator, we will use the same approach as in
\secref{action}. According to
formula \eqref{extension}, we have: $$\{ \int P dt,\int e^\phi dt\} =
\int \sum_{n\geq 0} (-\pa_t)^n
\frac{\pa P}{\pa(\pa^n u)} \cdot e^\phi dt = \int \sum_{n\geq 0}
\frac{\pa P}{\pa(\pa^n u)} \cdot \pa_t^n e^\phi dt,$$
where we used the fact that the integral of a total
derivative is $0$. We find: $\pa_t
e^\phi = u e^\phi, \pa_t^2 e^\phi = (u^2 + \pa_t u)
e^\phi,$ etc. In general, $\pa_t^n e^\phi = B_n
e^\phi$, where the $B_n$'s
are certain differential polynomials in $u(t)$, which are connected
by the recurrence relation
\begin{equation}    \label{recurrence}
B_{n+1} = u B_n + \pa_t B_n.
\end{equation}
Then we obtain
\begin{equation}    \label{obtain}
\{ \int P dt,\int e^\phi dt\} = \sum_{n\geq 0} \int
\frac{\pa P}{\pa(\pa^n u)} \cdot B_n e^\phi dt.
\end{equation}

Therefore we can define a map $\q: \pi_0 \arr \pi_1$ as follows: $$\q
\cdot P = \sum_{n\geq 0}  B_n \frac{\pa P}{\pa(\pa^n u)} \otimes v_1.$$
It commutes with the action of $\pa$ and descends down to the operator
$\Q: \F_0 \arr \F_1.$

In fact, in the same way we can define for any $G \in \F_1$ a map
$\{\cdot,G\}: \pi_0 \arr \pi_1$, which commutes with $\pa$ and
descends down to the map $\F_0 \arr \F_1$, given by formula
\eqref{extension}: $$\{P,\int R e^{\phi(t)} dt\} = \sum_{n\geq 0}
\frac{\pa P}{\pa(\pa^n u)} \cdot \pa^n (R \otimes v_1) - \pa^{n+1}
\left( \frac{\delta}{\delta u} R \otimes v_1 \right).$$ We can also extend
the action of $G \in \F_0$ on $\pi_0$, given by \eqref{vectfield}, to an
action on $\pi_1$ by adding to the vector field \eqref{vectfield} the term
$\frac{\delta R}{\delta u} \, \frac{\pa}{\pa \phi}$, where $\frac{\pa}{\pa
\phi}$ acts on $\pi_n$ by multiplication by $n$.

It is convenient to pass to the new variables $x_n = \pa^{-n-1} u/(-n-1)!,
n<0$. Then $\pi_0 = \C[x_n]_{n<0}, \pi_1 = \C[x_n]_{n<0} \otimes \C
v_1$. In these variables the action of the derivative $\pa$ on these spaces
is given by $$\pa = -\sum_{n<0} n x_{n-1} \frac{\pa}{\pa x_n} + x_{-1}
\frac{\pa}{\pa \phi}.$$

Let $T: \pi_0 \arr \pi_1$ be the translation operator, which sends $P
\in \pi_0$ to $P \otimes v_1 \in \pi_1$. In the new variables the
operator $\q: \pi_0 \arr \pi_1$ is given by the formula
\begin{equation}    \label{operq}
\q = T \sum_{n<0} S_{n+1} \frac{\pa}{\pa x_n},
\end{equation}
where the polynomials $S_n$ are the Schur polynomials, defined via the
generating function:
\begin{equation}    \label{schur}
\sum_{n\leq 0} S_n z^n = \exp(\sum_{m<0} -\frac{x_m}{m} z^m).
\end{equation}

One can check that $\pa S_n = - x_{1} S_n - (n-1) S_{n-1},$ and therefore, by
formula \eqref{recurrence}, the differential polynomial $B_{-n}$ coincides
with $S_n$ in the new variables $x_m$. We summarize these results in the
following Lemma.

\subsubsection{Lemma.}    \label{commute}
{\em The operator $\q$, given by formula \eqref{operq}, commutes with
the action of the derivative $\pa$ and the corresponding operator
$\F_0 \arr \F_1$ coincides with the operator $\bar{Q}$, given by
formula \eqref{oper}.}

\subsubsection{} Let us put $\deg 1 = 0, \deg v_1 =1, \deg x_n =-n$.
Since the $x_n$'s generate the spaces $\pi_0$ and $\pi_1$ from $1$ and
$v_1$, respectively, this defines a $\Z-$grading on these spaces such that
the homogeneous components are finite-dimensional. The operator $\pa$ is
homogeneous of degree $1$, and we can define grading on the spaces $\F_0$
and $\F_1$ by subtracting $1$ from the grading on the spaces $\pi_0$ and
$\pi_1$, respectively. The operators $\q$ and $\Q$ are homogeneous of
degree $0$ and therefore their kernels and cokernels are $\Z-$graded with
finite-dimensional homogeneous components.

\subsubsection{}    \label{dc}
To find the space of local integrals of motion of the classical
Liouville equation we have to find the $0$th cohomology of the complex
$$\F_0 \stackrel{\bar{Q}}{\larr} \F_1.$$

Consider the following double complex.

\begin{equation}    \label{doublecom}
\begin{CD}
\C \\
@AAA \\
\pi_0 @>{\q}>> \pi_1 \\
@A{\pa}AA @A{-\pa}AA \\
\pi_0 @>{\q}>> \pi_1 \\
@AAA \\
\C
\end{CD}
\end{equation}

We can calculate its cohomology by means of two spectral sequences (as
general references on spectral sequences, cf., e.g., \cite{maclane,bt}). In
one of them the $0$th differential is vertical. Therefore, in this spectral
sequence the $1$st term coincides with our complex with the degrees shifted
by $1$. Hence the $1$st cohomology of the double complex coincides with the
space of integrals of motion.

In the other spectral sequence the first differential is horizontal.
We have two identical complexes
\begin{equation}    \label{picom}
\pi_0 \stackrel{\q}{\larr} \pi_1.
\end{equation}
Let us calculate the cohomology of this complex.

\subsubsection{Proposition.}    \label{vir}
{\em The operator $\q: \pi_0 \arr \pi_1$ is surjective, so that its
cokernel is equal to $0$. The kernel $\W_0(\sw_2)$ of the operator $\q$
contains an element $W_{-2}$ of degree $2$, such that $\W_0(\g)$ coincides
with the polynomial algebra $\C[W_n]_{n\leq -2}$, where $W_n =
\pa^{-n-2} W_{-2}/(-n-2)!$.}

\vspace{3mm}
\noindent {\em Proof.} The operator $Q=T^{-1}\q$ is a linear combination of
the vector fields $S_{n+1} \pa/\pa x_n, n<0$, where $S_0=1$ and $S_{n+1}$
is a polynomial in $x_{-1},\ldots,x_{n+1}, n<-1$. The operator
$$\sum_{-m\leq j\leq -1} S_{j+1} \frac{\pa}{\pa x_j}$$ is therefore a
well-defined linear operator $Q(m)$ from $\C[x_j]_{j=-1,\ldots,-m}$ to
itself.

The operator $Q(m)$ is surjective. To see that, consider the dual operator
$Q(m)^*$, acting on the space dual to $\C[x_j]_{j=-1,\ldots,-m}$. Since our
operator is homogeneous, it is sufficient to consider the restricted dual
space, which we can identify with itself, choosing the monomials
$x^{k_1}_{-j_1}\ldots x^{k_n}_{-j_n}/(k_1!\ldots k_n!)^{1/2}$ as the
orthonormal basis. The operator $Q(m)^*$ then has the form $$Q(m)^* =
\sum_{-m\leq j\leq -1} x_j S^*_{j+1},$$ where $S^*_j$ is obtained from the
polynomial $S_j$ by replacing $x_i$ with $\pa/\pa x_i$. We see that the
operator $Q(m)^*$ is the sum of multiplication by $x_{-1}$, which increases
the power of any polynomial by $1$, and other operators, which do not
change or decrease the power. Since the operator of multiplication by
$x_{-1}$ has no kernel, the operator $Q(m)^*$ is injective. Therefore, the
operator $Q(m)$ is surjective. Hence, the operators $Q$ and $\q$ are also
surjective.

For any $n<-1$, there exist polynomials $W_n = (n+1) x_n + W'_n$, where
$W'_n$ is a linear combination of terms of power greater then $1$ and of
degree $n$, such that $\q \cdot W_n = 0$. Indeed, $Q(n) \cdot x_n =
S_{n+1}$ is an element of $\C[x_j]_{j=-1,\ldots,n+1}$. The operator
$Q(n+1)$ is surjective on this space. Therefore, there exists such $W'_n \in
\C[x_j]_{j=-1,\ldots,n+1}$ that $Q(n+1) \cdot W'_n = - Q(n) \cdot (n+1)
x_n$. But then $Q(n) \cdot ((n+1) x_n + W'_n) = 0$, and hence $\q \cdot W_n
= T Q(n) \cdot W_n = 0$.

In the coordinates $x_{-1}, W_n, n<-1,$ the operator $\q$ is equal to
$$\q = T \left( \frac{\pa}{\pa x_{-1}} + \sum_{n<-1} (\q \cdot W_n)
\frac{\pa}{\pa W_n} \right) = T \frac{\pa}{\pa x_{-1}}.$$ The
$W_n$'s are algebraically independent by construction, therefore the kernel
of the operator $\q$ coincides with $\C[W_n]_{n<-1}$.

Finally, we can choose as $W_n$ the polynomial $\pa^{-n-2}
W_{-2}/(-n-2)!$.  Indeed, the latter lies in the kernel of $\q$, because
$[\pa,\q]=0$. Its linear term is equal to $(n+1) x_n$, because the
linear term of $W_{-2}$ is $-x_{-2}$, and the derivative preserves the power
of a polynomial. The Proposition is proved.

\subsubsection{Remark.}
For any $\Z-$graded vector space $$V =
\oplus_{m \in \Z} V(m)$$ with finite-dimensional homogeneous components,
introduce its character as $$\ch V = \sum_{m \in \Z} \dim V(m) q^m.$$ The
Euler character of the complex \eqref{picom} is equal to $$\ch
\pi_0 -\ch \pi_1 = \ch \mbox{Ker} \q - \ch \mbox {Coker} \q =$$
$$\prod_{n\geq 1} (1-q^n)^{-1} - q \prod_{n\geq 1} (1-q^n)^{-1} =
\prod_{n\geq 2} (1-q^n)^{-1} = 1 + q^2 + \ldots.$$

This formula shows that there exists an element of degree $2$ in the kernel
of $\q$. Denote it by $W_{-2}$. The operator of multiplication by $W_{-2}$,
acting on $\pi_0$ and $\pi_1$, commutes with the action of $\q$. Further,
since $[\pa,\q]=0$, the operators of multiplication by $W_n = \pa^{-n-2}
W_{-2}/(-n-2)!, n\leq -2,$ also commute with $\q$. Therefore any polynomial
in the $W_n$'s, constructed this way, lies in the kernel of the operator
$\q$. The algebraic independence of these $W_n$'s and the surjectivity of
the operator $\q$ allowed us to identify the polynomial algebra in the
$W_n$'s with the kernel of $\q$.

\subsubsection{Proposition}    \label{localint}
{\em The space $I_0(\sw_2)$ of local integrals of motion of the
classical Liouville theory coincides with the quotient of $\W_0(\sw_2)$
by the total derivatives and constants.}

\vspace{3mm}
\noindent {\em Proof.} We have to prove that the kernel of the
operator $\bar{Q}$ coincides with the quotient of $\W_0(\sw_2)$ by the
total derivatives and constants. As we explained before, this kernel is the
same as the $1$st cohomology of the double complex \eqref{doublecom}. By
\propref{vir}, the first term of the spectral sequence, associated to this
double complex, looks as follows: $$\C \larr \W_0(\sw_2) \larr \W_0(\sw_2)
\larr \C.$$ The $1$st cohomology of this complex is equal to the quotient
of $\W_0(\sw_2)$ by the total derivatives and constants.

\subsubsection{}    \label{miura}
We can write down explicit formulas for the $W_n$'s as follows:
$$W_{-2} = \frac{1}{2} x_{-1}^2 - x_{-2},\quad W_n =
\frac{1}{(-n-2)!}\pa^{-n-2} W_{-2}, \quad n<-2.$$

Thus, the space $I_0(\sw_2)$ consists of local functionals, which are
defined by differential polynomials, depending on $$W =
\frac{1}{2} u^2 - \pa u.$$ These local functionals constitute a
Poisson subalgebra in $\F_0$. It is known that this Poisson subalgebra
is isomorphic to the classical Virasoro algebra.

The Virasoro algebra is the central extension of the Lie algebra of vector
fields on the circle. Its dual space is equipped with the Kirillov-Kostant
Poisson structure. This structure can be restricted to the hyperlane, which
consists of the linear functionals, whose value on the central element is
$1$. If we choose a coordinate on the circle, then this
hyperplane can be identified with the space of Laurent polynomials
$W(t)$. The local functionals on this space form a Poisson algebra, which
is isomorphic to $I_0(\sw_2)$.

There is a map from a hyperplane in the dual space to the Heisenberg
algebra to a hyperplane in the dual space to the Virasoro algebra, which
sends $u(t)$ to $W(t) = \frac{1}{2} u^2(t) -
\pa_t u(t)$ and preserves the Poisson structure. This map is called the
Miura transformation.

In the case of the Liouville theory, which is the simplest Toda field
theory, we were able to find explicit formulas for the local integrals of
motion and identify their Poisson algebra with the classical Virasoro
algebra. However, in general explicit formulas are much more complicated,
and we will have to rely on homological algebra to obtain information
about the integrals of motion.

\subsection{General case.}    \label{fdtoda}

\subsubsection{Hamiltonian space.}    \label{hamspace}
Let $\h$ be the Cartan subalgebra of $\g$. It is equipped with the
scalar product $( , )$, which is the restriction of the invariant
scalar product on $\g$, normalized as in
\cite{Kac}. In what follows we will identify $\h$ with its dual by
means of this scalar product.

We choose as the Hamiltonian space the space $L\h$ of Laurent polynomials
on the circle with values in $\h$, that is the space $\h \otimes
\C[t,t^{-1}]$.

Each element ${\bold u}(t)$ of $L\h$ can be represented by its
coordinates $(u^1(t),\ldots,u^l(t))$ with respect to the basis of
simple roots $\al_1,\ldots,\al_l$, where $l$ is the rank of $\g$. Let
$\pi_0$ be the space of differential polynomials of ${\bold u}(t)$, i.e.
the space of polynomials in $\pa^n u^i, i = 1,\ldots,l, n \geq 0$.

We define the space $\F_0$ of local functionals as the space of functionals
on $L\h$, which can be represented as formal residues $$F[{\bold u}(t)] =
\int P(\pa^n u^i(t)) dt,$$ where $P \in \pi_0$.  Again, we have the exact
sequence $$0 \larr \pi_0/\C
\stackrel{\pa}{\larr} \pi_0/\C \stackrel{\int}{\larr} \F_0 \larr 0.$$

Introduce the Poisson structure on $\F_0$ by the formula
\begin{equation}    \label{genskobka}
\{F,G\}[{\bold u}(t)] = - \int \left(\frac{\delta P}{\delta {\bold u}} ,
\pa_t \frac{\delta R}{\delta {\bold u}}\right) dt.
\end{equation}
Here $\frac{\delta P}{\delta {\bold u}}$ and $\pa_t \frac{\delta R}{\delta
{\bold u}}$ are vectors in the dual space to $\h$, and we take their scalar
product.

In coordinates, we can rewrite it as $$\{F,G\}[{\bold u}(t)] = - \int
\sum_{i,j=1}^l (\al_i,\al_j) \frac{\delta P}{\delta u^i} \, \pa_t \,
\frac{\delta R}{\delta u^j} dt.$$

This Poisson structure has an interpretation as a Kirillov-Kostant
structure on a hyperplane in the dual space to the Heisenberg algebra
$\widehat{\h}$, which is the central extension of $L\h$ (cf.
\secref{kirkos}).

Note that our conventions in the case of $\g=\sw_2$ (cf. \secref{liouv})
differ from our conventions in general by a factor of $2$, since for
$\sw_2$ we have $(\al,\al)=2$.

\subsubsection{}    \label{functspace}
Let us define other spaces of functionals on $L\h$. For each element
$\gamma$ of the weight lattice $P \subset \h^* \simeq \h$ we define the
space $\pi_\gamma = \pi_0 \otimes \C v_\gamma$, equipped with the action of
the derivative by the formula $(\pa + \gamma) \otimes 1$, where $\gamma$
denotes the operator of multiplication by $\gamma$ (as an element of
$\pi_0$) on $\pi_0$.

Let $\F_\gamma$ be the quotient of $\pi_\gamma$ by the image of the
operator $\pa$, i.e. by the total derivatives. We have the exact
sequence
\begin{equation}    \label{exactseq}
0 \larr \pi_\gamma \stackrel{\pa}{\larr} \pi_\gamma \larr \F_\gamma
\larr 0.
\end{equation}

As in \secref{poisext}, we can interpret $\F_\gamma$ as the space of
functionals on $L\h$ of the form $$\int P(\pa^n u^i(t))
e^{\bar{\gamma}(t)} dt,$$ where $\bar{\gamma}(t)$ is such that $\pa_t
\bar{\gamma}(t) = \gamma(t)$.

In the same way as in \secref{poisext}, we can extend our Poisson
bracket \eqref{genskobka} to a map $$\F_0 \times \oplus_{\gamma \in P}
\F_\gamma \arr \oplus_{\gamma \in P} \F_\gamma$$ by the formula
\begin{equation}
\{\int P dt, \int R e^{\gam} dt\} = \int
\left[ \left( \frac{\delta P}{\delta {\bold u}} , \gamma\right) R e^{\gam}
 - \left( \frac{\delta P}{\delta {\bold u}} , \pa_t \cdot \frac{\delta
R}{\delta {\bold u}} e^{\gam} \right) \right] dt.
\end{equation}
One can check that this bracket satisfies the Jacobi identity for any
triple $F,G \in \F_0, H \in \F_\gamma$.

\subsubsection{The Toda hamiltonian.} We can now introduce the
hamiltonian of the Toda field theory, associated to $\g$, by the formula
$$H = \frac{1}{2} \sum_{i=1}^l \int e^{\phi_i(t)} dt \in \oplus_{i=1}^l
\F_{\al_i},$$ where $\phi_i(t) =
\bar{\al}_i(t)$.  Note that $\int e^{\phi_i(t)} dt$ stands for the
image of $v_{\al_i} \in \pi_{\al_i}$ under the projection $\pi_{\al_i}
\arr \F_{\al_i}$.  The equation $${\bold U}(t) = \{{\bold U}(t),H\},$$
where ${\bold U}(t)$ denotes the delta-like functional as in
\secref{delta}, coincides with the Toda equation \eqref{toda}.

The operator $\Q_i = \{\cdot,\int e^{\phi_i(t)} dt\}$ is a well-defined
linear operator, acting from $\F_0$ to $\F_{\al_i}$. We can therefore
give the following definition.

\subsubsection{Definition.} {\em The kernel of the linear operator
$$\frac{1}{2} \sum_{i=1}^l \Q_i: \F_0 \arr \oplus_{i=1}^l \F_{\al_i}$$ will
be called the space of local integrals of motion of the classical Toda
field theory associated to $\g$ and will be denoted by $I_0(\g)$.}

\subsubsection{} Clearly, $I_0(\g)$ is equal to the intersection of the
kernels of the operators $\Q_i: \F_0 \arr \F_{\al_i}$. By Jacobi identity,
$I_0(\g)$ is a Poisson subalgebra of $\F_0$.

Let us write down an explicit formula for the operator $\Q_i$.

It is convenient to pass to the new variables $x_n^i = \pa^{-n-1}
u^i/(-n-1)!, i=1,\ldots,l, n<0$. Then $\pi_\gamma = \C[x_n^i] \otimes
\C v_\gamma$ (here $v_0 = 1$).

In these new variables the action of the derivative $\pa$ can be written as
$$\sum_{i=1}^l \left( -\sum_{n<0} n x^i_{n-1} \frac{\pa}{\pa x^i_n} +
x^i_{-1} \frac{\pa}{\pa \phi_i} \right),$$ where the action of $\pa/\pa
\phi_i$ on $\pi_\gamma$ with $\gamma = \sum_{i=1}^l l_i \al_i$ is given by
multiplication by $l_i$.

Let $T_i: \pi_\gamma \arr \pi_{\gamma+\al_i}$ be the translation
operator, which maps $P \otimes v_\gamma \in \pi_\gamma$ to $P \otimes
v_{\gamma+\al_i} \in \pi_{\gamma+\al_i}$.

Introduce the operators $\q_i: \pi_\gamma \arr \pi_{\gamma+\al_i}$ by
the formula
\begin{equation}    \label{operqi}
\q_i = T_i \sum_{n<0} S^i_{n+1} \pa^{(i)}_n,
\end{equation}
where $$\pa^{(i)}_n = \sum_{j=1}^l (\al_i,\al_j) \frac{\pa}{\pa
x^j_n},$$ and the Schur polynomials $S_n^i$ are given by the generating
function
$$\sum_{n\leq 0} S^i_n z^n = \exp(\sum_{m<0} -\frac{x^i_m}{m} z^m).$$

\subsubsection{Lemma.}    \label{commutegeneral}
{\em The operator $\q_i: \pi_0 \arr
\pi_{\al_i}$ commutes with the action of derivative $\pa$ and the
corresponding operator $\F_0 \arr \F_{\al_i}$ coincides with the
operator $\Q_i$.}

\vspace{3mm}
\noindent {\em Proof.} The same as in \lemref{commute}.

\subsubsection{} Our task is to compute the kernel of the operator
$\sum_{i=1}^l \Q_i$, or, in other words, the $0$th cohomology of the
complex
\begin{equation}    \label{compl}
\F_0 \larr \oplus_{i=1}^l \F_{\al_i}.
\end{equation}

The cohomology of this complex is very difficult to compute. Indeed, it is
clear that the $1$st cohomology is very large, because the first group of
the complex is ``$l$ times larger'' than the $0$th group. So, we can not
use the argument we used in the proof of \propref{vir}. What is even worse
is that as was explained in the Introduction, for such a complex it is
virtually impossible to prove that the cohomology classes can be
quantized.

To fix this situation we will extend this complex further to the
right. Clearly, by doing so we will not change the $0$th cohomology,
but we will be able to kill all higher cohomologies. We will then use
the resulting complex to compute the $0$th cohomology, and to prove
that it can be quantized.

First of all, it is convenient to realize our complex as the double
complex
\begin{equation}    \label{doublecomplex}
\begin{CD}
\C \\
@AAA \\
\pi_0 @>{\sum \q_i}>> \oplus_{i=1}^l \pi_{\al_i} \\
@A{\pa}AA @A{-\pa}AA\\
\pi_0 @>{\sum \q_i}>> \oplus_{i=1}^l \pi_{\al_i} \\
@AAA \\
\C
\end{CD}
\end{equation}
in the same way as in \secref{dc}.

We can compute the cohomology of the double complex
\eqref{doublecomplex} by means of the spectral sequence,
whose first term consists of two identical complexes
\begin{equation}    \label{comple}
\pi_0 \larr \oplus_{i=1}^l \pi_{\al_i}.
\end{equation}
We will extend both complexes \eqref{comple} in such a way that the
higher differentials will commute with the derivative $\pa$. We will
then be able to form a double complex, which will give us an extension
of the complex \eqref{compl} that we are looking for.

The key observation, which will enable us to do that, is as follows.
Introduce the operators $Q_i: \pi_0 \arr \pi_0$ as $T^{-1} \q_i$. We
will use the notation $\mbox{ad} A \cdot B = [A,B]$.

\subsubsection{Proposition.}    \label{serre}
{\em The operators $Q_i$ satisfy the Serre relations of the Lie
algebra $\g$} $$(\mbox{ad} \, Q_i)^{-a_{ij}+1} \cdot
Q_j = 0,$$ {\em where $\|a_{ij}\|$ is the Cartan matrix of $\g$.}

\vspace{3mm}
\noindent {\em Proof.} The Proposition follows from the following
formula:
$$(\mbox{ad} Q_i)^m \cdot Q_j = C_m \cdot (-a_{ij}-m+1)
\sum_{n_1,\ldots,n_{m+1}<0} S^i_{n_1} \ldots S^i_{n_m} S^j_{n_{m+1}}
\frac{1}{n_1 \ldots n_m} \cdot$$ $$\cdot \left( \sum_{l=1}^m
\frac{n_l}{n_1 + \ldots \widehat{n_l} \ldots + n_{m+1}}
\pa^{(i)}_{n_1 + \ldots + n_{m+1}} - \pa^{(j)}_{n_1 + \ldots +
n_{m+1}} \right),$$ where $C_m$ is a constant. This formula can
be proved by induction, using the commutation relations
$$[\pa^{(i)}_n,S^j_m] = - (\al_i,\al_j) \frac{1}{n} S^j_{m-n}$$ (here
we put $S^j_m = 0$, if $m>0$) and the simple identity
$$\frac{1}{a(a+b)} + \frac{1}{b(a+b)} =
\frac{1}{ab}.$$

\subsubsection{Remark.}    \label{serrerem}
In the proof of \propref{serre} we never used the fact that $\|a_{ij}\|$ is
the Cartan matrix of a simple Lie algebra. In fact, we could associate the
main objects, defined in this section, such as $\pi_\gamma$, $\F_\gamma$,
the Poisson structure, and the operators $Q_i$ to any symmetrizable Cartan
matrix, so that the results of this section, such as \propref{serre},
remain valid.

\subsubsection{} \propref{serre} shows that the
operators $Q_i$ generate an action of the nilpotent subalgebra $\n_+$ of
$\g$ on $\pi_0$. In order to extend the complex \eqref{comple}, we will use
the Bernstein-Gelfand-Gelfand (BGG) resolution of the trivial
representation of $\g$ by Verma modules. We will recall the relevant facts
about this resolution in the next subsection.

\subsection{BGG resolution.}

\subsubsection{Verma modules.}    \label{verma}
Recall that the Lie algebra $\g$ has the Cartan decomposition $\g = \n_-
\oplus \h \oplus \n_+$. For $\la \in \h^*$ denote by $\C_\la$ the
corresponding one-dimensional representation of $\h$. We can extend it
trivially to a representation of $\goth{b}_- = \h \oplus \n_-$. The induced
module over $\g$, $$M_\la = U(\g) \otimes_{U(\goth{b}_-)}
\C_\la,$$ is called the Verma module with lowest weight $\la$. It is
freely generated from the lowest weight vector ${\bold 1}_\la = 1\otimes 1$
by the action of the nilpotent subalgebra $\n_+$ of $\g$.

A vector $w$ in $M_\la$ is called a singular vector of weight $\mu$,
if it satisfies the properties: $$\n_- \cdot w = 0, \quad y \cdot w =
\mu(y) w, y \in \h.$$ In particular, ${\bold 1}_\la$ is a singular vector
of weight $\la$. A singular vector of weight $\mu$ generates a
submodule of $M_\la$, which is isomorphic to the Verma module $M_\mu$.

Consider the Verma module $M_0$. It is known that the singular vectors of
$M_0$ are labeled by the elements of the Weyl group of $\g$
\cite{bgg}. Such a vector $w_s$, corresponding to an element $s$ of the
Weyl group, has the weight $\rho-s(\rho)$, where $\rho \in h^*$ is the
half-sum of the positive roots of $\g$. Let us fix these vectors once and
for all.

\subsubsection{The definition of the resolution.}    \label{defres}
The BGG resolution is a complex, i.e. a $\Z-$graded vector space
$$B_*(\g) = \oplus_{j\geq 0} B_j(\g),$$ together with differentials
$d_j: B_j(\g) \arr B_{j-1}(\g)$, which are nilpotent: the composition of
two consecutive differentials $d_{j-1} d_j$ is equal to $0$.

The vector space $B_j(\g)$ is the direct sum of the Verma modules
$M_{\rho-s(\rho)}$, where $s$ runs over the set of elements of the
Weyl group of length $j$ \cite{BGG}.

Once we fixed the vectors $w_s$, we have canonical embeddings
$M_{\rho-s(\rho)} \arr M_0$. Therefore the module $M_{\rho-s(\rho)}$ can be
thought of as a submodule of the module $M_0$, generated by the vector
$w_s$. It is known that the vector $w_{s'}$ belongs to the module
$M_{\rho-s(\rho)}$ if and only if $s \preceq s'$ with respect to the Bruhat
order on the Weyl group \cite{bgg}. It is clear that these vectors are
singular vectors of the module $M_{\rho - s(\rho)}$ and that there are no
other singular vectors in this module. In that case we have an embedding
$i_{s',s}: M_{\rho-s'(\rho)} \arr M_{\rho-s(\rho)}$, which sends the lowest
weight vector of $M_{\rho-s'(\rho)}$ to the singular vector of
$M_{\rho-s(\rho)}$ of weight $\rho-s'(\rho)$.

\subsubsection{Lemma.}\cite{BGG}    \label{signs}
\begin{enumerate}
\item[(a)] {\em Let $s$ and $s''$ be two elements of the Weyl group, such
that $s \prec s''$ and $l(s'') = l(s) + 2$. Then there are either two
or no elements $s'$, such that $s \prec s' \prec s''$.}
\item[(b)] {\em Let us call a \mbox{square} a set of four elements of
the Weyl group, satisfying the conditions of the part (a).

It is possible to attach a sign $\ep_{s',s}$, $+$ or $-$, to each pair
of elements of the Weyl group $s, s'$, such that $s\prec s', l(s') =
l(s) + 1$, so that the product of signs over any square is $-$.}
\end{enumerate}

\subsubsection{The differential.} We are
now ready to define the differential of the BGG resolution $d_j:
B_j(\g) \arr B_{j-1}(\g)$ as
\begin{equation}    \label{diff}
d_j = \sum_{l(s)=j-1,l(s')=j,s\prec s'} \ep_{s',s} \cdot i_{s',s}.
\end{equation}
In other words, we take the sum of all possible embeddings of the Verma
modules, which are direct summands of $B_j(\g)$, with the special choice of
signs from \lemref{signs}. By definition, these differentials commute with
the action of $\g$.

\subsubsection{Theorem.}\cite{BGG}    \label{acyclic}
\begin{enumerate}
\item[(a)] {\em The differentials $d_j, j>0,$ are nilpotent: $d_{j-1}
d_j = 0$, and so $B_*(\g)$ is a complex.}
\item[(b)] {\em The $0$th homology of the complex $B_*(\g)$ is the
trivial one-dimensional representation of $\g$, and all higher
homologies vanish.}
\end{enumerate}

\vspace{3mm}
\noindent {\em Proof.} In the notation of
\secref{defres}, we have $w_{s'} = P_{s',s} \cdot w_s$, for some
element $P_{s',s}$ of $U(\n_+)$. If we have a square $s, s'_1, s'_2, s''$
of elements of the Weyl group, such that $s \prec s'_1, s'_2
\prec s''$, then we can write: $w_{s''} =
P_{s'',s'_1} P_{s'_1,s} w_s$ and $w_{s''} = P_{s'',s'_2} P_{s'_2,s} w_s$.
Therefore we obtain the following identity
\begin{equation}    \label{sq}
P_{s'',s'_1} P_{s'_1,s} = P_{s'',s'_2} P_{s'_2,s}
\end{equation}
in $U(\n_+)$. By definition, $i_{s',s}(u\cdot{\bold 1}_{\rho-s'(\rho)}) =
(u P_{s',s}) \cdot{\bold 1}_{\rho-s(\rho)}$. So, we obtain from formula
\eqref{sq}: $i_{s'_1,s}
\circ i_{s'',s'_1} = i_{s'_2,s} \circ i_{s'',s'_2}$. Thus, because of our
sign convention (cf. \lemref{signs}, (b)), such terms in the composition of
two consecutive differentials $d_{j-1} d_j$ will cancel out. This proves
part (a) of the Theorem.

The proof of part (b) is rather complicated; it uses the so-called
{\em weak} BGG resolution, which is obtained from the de Rham complex
on the big cell of the flag manifold of $\g$ (cf. \cite{BGG}).

\subsubsection{Remarks.} (1) One can define analogous resolutions for
arbitrary finite-dimensional representations of $\g$.

\noindent (2) There are generalizations of the BGG resolutions to
arbitrary symmetrizable Kac-Moody algebras \cite{rocha}. We will use such
resolutions for affine algebras in \secref{classicalaff}.

\subsection{Extended complex and its cohomology.}    \label{extended}

One of the main applications of the BGG resolution is to computation
of the cohomologies of the nilpotent Lie algebra $\n_+$ of $\g$. In this
subsection we will use this resolution to extend our complex
\eqref{comple}, and to compute the cohomology of the resulting
complex.

\subsubsection{} The $j$th component $F^j(\g)$ of our extended complex
$$F^*(\g) = \oplus_{j\geq 0} F^j(\g)$$ will be the direct sum of the
spaces $\pi_{\rho-s(\rho)}$, where $s$ runs over the set of elements
of the Weyl group of length $j$.

Now let us define the differentials. The algebra $U(\n_+)$ is generated by
$e_i, i = 1,\ldots,l$, which satisfy the Serre relations. So any element of
$U(\n_+)$ can be expressed in terms of $e_i$. Let $P_{s',s}(Q):
\pi_{\rho-s(\rho)} \arr \pi_{\rho-s'(\rho)}$ be the map, obtained by
inserting into $P_{s',s} \in U(\n_+)$ the operators $\q_i$ instead of
$e_i$. We can then introduce the differential $\delta^j: F^{j-1}(\g) \arr
F^j(\g)$ of our complex by the formula
\begin{equation}    \label{differential}
\delta^j = \sum_{l(s)=j-1,l(s')=j,s\prec s'} \ep_{s',s} \cdot
P_{s',s}(Q).
\end{equation}

\subsubsection{Lemma.}    \label{nilpotent}
{\em The differentials $\delta^j, j>0$, are nilpotent: $\delta^{j+1}
\delta^j = 0$, and so $F^*(\g)$ is a complex.}

\vspace{3mm}
\noindent {\em Proof.} Just as in the proof of part (a) of
\thmref{acyclic}, we have to check that $P_{s'',s'_1}(Q)$ $P_{s'_1,s}(Q)
= P_{s'',s'_2}(Q) P_{s'_2,s}(Q)$. But this follows at once from \eqref{sq},
since, according to \propref{serre}, the operators $Q_i$, and therefore the
operators $\q_i$, satisfy the defining relations of the algebra $U(\n_+)$.

\subsubsection{}    \label{zgrading}
We introduce a $\Z-$grading on the complex $F^*(\g)$ by putting $\deg x^i_n
= -n$, and $\deg v_{\rho-s(\rho)} = (\rho^\vee,\rho-s(\rho))$, where
$\rho^\vee \in h^*$ is defined by the property $(\rho^\vee,\al_i)=1,
i=1,\ldots,l$. Clearly, all homogeneous subspaces of $\pi_\gamma$ have
finite dimensions. With respect to this grading, the operator $\pa$ is
homogeneous of degree $1$, and we can define a grading on the spaces
$\F_\gamma$ by subtracting $1$ from the grading on the space
$\pi_\gamma$. The differentials $\delta^j$ are homogeneous of degree $0$
with respect to this grading. Therefore our complex $F^*(\g)$ decomposes
into a direct sum of finite-dimensional subcomplexes, corresponding to its
different graded components.

\subsubsection{Example of $\sw_3$.}    \label{sltri}
In this case the Weyl group consists of six elements. It is generated
by two reflections: $s_1$ and $s_2$ with the relation $s_1 s_2 s_1 =
s_2 s_1 s_2.$ The complex $F^*(\sw_3)$ is shown on Fig. 1.

The vertices represent the spaces $\pi_{\rho-s(\rho)}$, and arrows
represent the maps of the differential. There are four squares. The
anti-commutativity of maps, associated to one of them, reads: $$\q_1^2 \,
\q_2 = - A \q_1.$$ To find a solution $A$ to this equation, let us
consider the Serre relation, which the operators $\q_1$ and $\q_2$
satisfy (cf. \propref{serre}): $$(\mbox{ad} \q_1)^2 \cdot \q_2 =
\q_1^2 \q_2 - 2 \q_1 \q_2 \q_1 + \q_2 \q_1^2 = 0.$$ Therefore, $A = -
2 \q_1 \q_2 + \q_2 \q_1$ is such a solution. Similarly, if we put $B =
- 2 \q_2 \q_1 + \q_1 \q_2$; then all squares will be anti-commutative.

\setlength{\unitlength}{0.0125in}%
\begin{picture}(445,200)(55,560)
\thicklines
\put(155,677){$\bullet$}
\put(235,717){$\bullet$}
\put(235,637){$\bullet$}
\put(330,717){$\bullet$}
\put(330,637){$\bullet$}
\put(410,677){$\bullet$}
\put(165,675){\vector( 2,-1){ 65}}
\put(165,685){\vector( 2, 1){ 65}}
\put(245,650){\vector( 4, 3){ 80}}
\put(245,720){\vector( 1, 0){ 80}}
\put(245,710){\vector( 4,-3){ 80}}
\put(245,640){\vector( 1, 0){ 80}}
\put(340,640){\vector( 2, 1){ 65}}
\put(340,720){\vector( 2,-1){ 65}}
\put(270,585){\makebox(0,0)[lb]{\raisebox{0pt}[0pt][0pt]{Fig. 1}}}
\put(280,624){\makebox(0,0)[lb]{\raisebox{0pt}[0pt][0pt]{$\q_1^2$}}}
\put(265,655){\makebox(0,0)[lb]{\raisebox{0pt}[0pt][0pt]{$B$}}}
\put(265,700){\makebox(0,0)[lb]{\raisebox{0pt}[0pt][0pt]{$A$}}}
\put(280,725){\makebox(0,0)[lb]{\raisebox{0pt}[0pt][0pt]{$\q_2^2$}}}
\put(185,645){\makebox(0,0)[lb]{\raisebox{0pt}[0pt][0pt]{$\q_2$}}}
\put(185,710){\makebox(0,0)[lb]{\raisebox{0pt}[0pt][0pt]{$\q_1$}}}
\put(380,645){\makebox(0,0)[lb]{\raisebox{0pt}[0pt][0pt]{$\q_2$}}}
\put(380,710){\makebox(0,0)[lb]{\raisebox{0pt}[0pt][0pt]{$\q_1$}}}
\put(320,625){\makebox(0,0)[lb]{\raisebox{0pt}[0pt][0pt]{$\pi_{2\al_1
+\al_2}$}}}
\put(230,625){\makebox(0,0)[lb]{\raisebox{0pt}[0pt][0pt]{$\pi_{\al_2}$}}}
\put(320,730){\makebox(0,0)[lb]{\raisebox{0pt}[0pt][0pt]{$\pi_{\al_1
+2\al_2}$}}}
\put(230,730){\makebox(0,0)[lb]{\raisebox{0pt}[0pt][0pt]{$\pi_{\al_1}$}}}
\put(145,690){\makebox(0,0)[lb]{\raisebox{0pt}[0pt][0pt]{$\pi_0$}}}
\put(405,690){\makebox(0,0)[lb]{\raisebox{0pt}[0pt][0pt]{$\pi_{2\al_1
+2\al_2}$}}}

\end{picture}

Our complex $F^*(\sw_3)$ has four non-trivial groups: $F^0(\sw_3) =
\pi_0, F^1(\sw_3) = \pi_{\al_1} \oplus \pi_{\al_2}, F^2(\sw_3) =
\pi_{\al_1 + 2\al_2} \oplus \pi_{2\al_1 + \al_2}$, and $F^3(\sw_3) =
\pi_{2\al_1 + 2\al_2}$. The differentials have the form: $\delta_1 =
\q_1 + \q_2, \delta_2 = \q_1^2 + \q_2^2 + A + B$, and $\delta_3 = \q_1
+ \q_2$.

\subsubsection{Proposition.}    \label{cohom}
{\em The cohomologies of the complex $F^*(\g)$ are isomorphic to the
cohomologies of the Lie algebra $\n_+$ with coefficients in the module
$\pi_0$, $H^*(\n_+,\pi_0)$.}

\vspace{3mm}
\noindent {\em Proof.} The complex $F^*(\g)$ is isomorphic to
Hom$_{\n_+}(B_*(\g),\pi_0)$. Indeed, for any $\la$ the module $M_\la$ is
isomorphic to a free $\n_+-$module $M$ with one generator.  Therefore the
space of $\n_+-$homomorphisms Hom$_{\n_+}(M,\pi_0)$ is canonically
isomorphic to $\pi_0$. Indeed, any non-zero homomorphism $x \in
\mbox{Hom}_{\n_+}(M,\pi_0)$ defines a non-zero element in $\pi_0$: the image
of the lowest weight vector of $M$. The embedding $i_{s',s}$ of $M$ into
itself then induces the homomorphism from $\pi_0$ to $\pi_0$, which sends
$y \in \pi_0$ to $P_{s',s} \cdot y$. This is precisely the homomorphism
$P_{s',s}(Q)$. Hence the differentials $d_j$ of the BGG resolution $B_*(\g)$
map to the differentials $\delta^j$ of the complex $F^*(\g)$.

According to part (b) of \thmref{acyclic}, the BGG resolution $B_*(\g)$ is
the resolution of the trivial $\n_+-$module by free $\n_+-$modules.  The
cohomologies of the Hom$_{\n_+}$ of such a resolution to an $\n_+-$module,
are, by definition, the cohomologies of $\n_+$ with coefficients in this
module, cf., e.g., \cite{gui}. Therefore the cohomologies of the complex
$F^*(\g)$ coincide with $H^*(\n_+,\pi_0)$.

\subsubsection{Proposition.}    \label{vanish}
{\em All higher cohomologies of the complex $F^*(\g)$ vanish.}

\vspace{3mm}
\noindent {\em Proof.} Each of the root generators $e_\al$ of $\n_+$ acts on
$\pi_0$ by a certain vector field. This vector field has a shift term,
which is a linear combination of $\pa/\pa x^i_n$ (cf. the proof of
\propref{wedge}). It follows from the proof of \propref{wedgegen} that the
shift terms of the root generators of $\n_+$ are linearly
independent. Therefore the dual operators to $e_\al$ are equal to the sum
of some linear combination of $x^i_n$ and some differential operators which
do not change or decrease the power of a polynomial (cf. the proof of
\propref{vir}). Hence the dual module to the module $\pi_0$ is a free
$\n_+-$module. But then the module $\pi_0$ is injective, and so all higher
cohomologies of $\n_+$ with coefficients in $\pi_0$ must
vanish. \propref{cohom} then implies that all higher cohomologies of the
complex $F^*(\g)$ vanish.

\subsubsection{Proposition.}    \label{exist}
{\em There exist elements $W^{(1)}_{-d_1-1},\ldots, W^{(l)}_{-d_l-1}$ of
$\pi_0$ of degrees $d_1+1,\ldots,d_l$ $+1$, where the $d_i$'s are the
exponents of $\g$, such that the $0$th cohomology $\W_0(\g)$ of the complex
$F^*(\g)$ is isomorphic to the polynomial algebra
$$\C[W^{(i)}_{n_i}]_{1\leq i\leq l,n_i<-d_i},$$ where $W^{(i)}_{n_i} =
\pa^{-n_i-d_i-1} W^{(i)}_{-d_i-1}/(-n_i-d_i-1)!.$}

\vspace{3mm}
\noindent {\em Proof.}
The algebra $\pi_0$ is the inductive limit of the free commutative algebras
with the generators $x^i_n, -M\leq n\leq -1$. Hence the spectrum of $\pi_0$
is the inverse limit of the affine spaces $R_M, M>0$, with the coordinates
$x^i_n, -M\leq n\leq -1$. From the explicit formula for the action of the
generators of the Lie algebra $\n_+$ on $\pi_0$ we see that the algebras
$\C[R_M]$ are preserved under the action of $\n_+$ (cf. the proof of
\propref{vir}). The infinitesimal action of $\n_+$ on $R_M$ by vector
fields can be integrated to an action of the Lie group $N_+$ by means of
the exponential map $\n_+ \arr N_+$, which is an isomorphism. The action of
$N_+$ commutes with the projections $R_{M+K} \arr R_M$.

At each point of the spectrum of $\pi_0$ the vector fields of the
infinitesimal action of the root generators $e_\al$ of $\n_+$ are linearly
independent (cf. the proof of \propref{vanish}). Hence the action of $N_+$
on $R_M$ is free for $M$ large enough.  The orbits of this action are
isomorphic to the affine space $\C^{\dim N_+}$. Therefore the quotient
space $R_M/N_+$ is also an affine space and the algebra of functions on
this space is a free commutative algebra.

Since the projections $R_{M+K} \arr R_M$ are compatible with the action of
$N_+$, we can take the inverse limit of the quotient spaces $R_M/N_+$. The
algebra of functions on this inverse limit is the inductive limit of the
free polynomial algebras of functions on $R_M/N_+$ and therefore it is a
free polynomial algebra with infinitely many generators. It consists of all
$N_+$--invariant elements of $\pi_0$, which are the same as the
$\n_+$--invariant elements. Hence this algebra coincides with the $0$th
cohomology of our complex.

The algebra $\pi_0$ is $\Z$-graded and the action of $\n_+$ preserves this
grading, if we introduce the principal grading on $\n_+$ by putting $\deg
e_i = 1$. Therefore the $0$th cohomology of our complex is also
$\Z$-graded, and it is easy to compute the degrees of the generators of
this algebra by computing its character.

By \propref{vanish}, all higher cohomologies of the complex $F^*(\g)$
vanish. Therefore, the character of the $0$th cohomology is equal to the
Euler character of the complex. The latter is equal to $$\sum_{j\geq 0}
(-1)^j \sum_{l(s)=j} \ch
\pi_{\rho-s(\rho)} = \prod_{n>0} (1-q^n)^{-l} \sum_s (-1)^{l(s)}
q^{(\rho^\vee,\rho-s(\rho))}.$$ From the specialized Weyl character
formula we deduce $$\sum_s (-1)^{l(s)} q^{(\rho^\vee,\rho-s(\rho))} =
\prod_{1\leq i\leq l,1\leq n_i\leq d_i} (1-q^{n_i}).$$ This gives for
the character of the $0$th cohomology, $\W_0(\g)$, $$\ch \W_0(\g) =
\prod_{1\leq i\leq l,n_i>d_i} (1-q^{n_i})^{-1}.$$ This formula shows that
the $0$th cohomology is the free commutative algebra with generators
$W^{(i)}_{n_i}$ of degree $-n_i$, where $1\leq i\leq l, n_i<-d_i$. In the
same way as in the proof of \propref{vir} we can see that as the generators
$W^{(i)}_{n_i}$ we can take $\pa^{-n_i-d_i-1}
W^{(i)}_{-d_i-1}/(-n_i-d_i-1)!$. The Proposition follows.

\subsubsection{Lemma.}    \label{singular}
{\em Let $P$ be a homogeneous element of the algebra $U(\n_+)$ of
weight $\gamma$, such that $P \cdot {\bold 1}_\la$ is a singular vector of
the Verma module $M_\la$ of weight $\la+\gamma$. Then the operator
$P(Q): \pi_\la \arr \pi_{\la+\gamma}$ commutes with the action of the
derivative $\pa$.}

\vspace{3mm}
\noindent {\em Proof.} The action of the derivative $\pa$ on $\pi_\la$
differs from its action on $\pi_0$ by the operator of multiplication
by $\la_{-1}$. We have: $[\la_{-1},\q_i] = (\al_i,\la)T_i$. Therefore, by
\lemref{commutegeneral}, the commutator of the operator $\q_i: \pi_\la
\arr \pi_{\la+\al_i}$ with $\pa$ is equal to $(\al_i,\la)T_i$.
Hence, the commutator of the monomial $\q_{i_m} \ldots \q_{i_1}:
\pi_\la \arr \pi_{\la+\gamma}$, where $\gamma =
\sum_{j=1}^m \al_{i_j}$, with $\pa$ is equal to $$\sum_{j=1}^m
(\al_{i_j},\la+\al_{i_1}+\ldots+\al_{i_{j-1}}) \q_{i_m} \ldots
T_{i_j} \ldots \q_{i_1}.$$ This precisely coincides with
the action of $$\sum_{i=1}^l \frac{(\al_i,\al_i)}{2} f_i,$$ where the
$f_i, i=1,\ldots,l,$ are the generators of the Lie algebra $\n_-$, on
the vector $e_{i_m} \ldots e_{i_1} {\bold 1}_\la$ of the Verma module
$M_\la$. If $P {\bold 1}_\la$ is a singular vector in $M_\la$, then
$$\sum_{i=1}^l \frac{(\al_i,\al_i)}{2} f_i \cdot P {\bold 1}_\la
=0,$$ and so $[\pa,P(Q)]=0$.

\subsubsection{Corollary.}    \label{higher}
{\em The higher differentials $\delta^j, j>1$, of the complex $F^*(\g)$
commute with the action of the derivative $\pa$.}

\vspace{3mm}
\noindent {\em Proof.} Each $\delta^j$ is a linear combination of maps
$P_{s',s}(Q)$. Since by definition $P_{s',s}$ defines a singular vector,
such a map commutes with $\pa$ by \lemref{singular}.

\subsubsection{Theorem.}    \label{todaimcl}
{\em The space $I_0(\g)$ of local integrals of motion of the classical Toda
field theory, associated with $\g$, coincides with the quotient of $\W_0(\g)$
by the total derivatives and constants.}

\vspace{3mm}
\noindent {\em Proof.} Using \corref{higher}, we can construct the
double complex $\C \larr F^*(\g) \larr F^*(\g) \larr \C,$ which is shown
on Fig. 2.

By \corref{higher}, the total differential of this complex is nilpotent. If
we compute the cohomology of this complex by means of the spectral
sequence, in which the $0$th differential is vertical, then in the first
term we obtain the complex $\bar{F}^*(\g) = \oplus_{j\geq 0} \bar{F}^j(\g)$,
where $$\bar{F}^j(\g) = \oplus_{l(s)=j} \F_{\rho-s(\rho)}.$$ By definition,
the $0$th cohomology of the complex $\bar{F}^*(\g)$ is the space
$I_0(\g)$. Therefore it coincides with the $1$st cohomology of our double
complex.

But we can compute this cohomology by means of the other spectral sequence,
in which the $0$th differential is the horizontal one. Then the Theorem
follows from \propref{exist} in the same way as in the proof of
\propref{localint}.

\setlength{\unitlength}{0.0125in}%
\begin{picture}(432,240)(65,525)
\thicklines
\put(490,640){$\ldots$}
\put(210,640){$\ldots$}
\put( 77,547){$\bullet$}
\put( 77,727){$\bullet$}
\put(437,597){$\bullet$}
\put(437,677){$\bullet$}
\put(357,597){$\bullet$}
\put(277,597){$\bullet$}
\put(157,597){$\bullet$}
\put( 77,597){$\bullet$}
\put(357,677){$\bullet$}
\put(277,677){$\bullet$}
\put(157,677){$\bullet$}
\put( 77,677){$\bullet$}
\put(440,605){\vector( 0, 1){ 70}}
\put(365,600){\vector( 1, 0){ 70}}
\put(365,680){\vector( 1, 0){ 70}}
\put(360,605){\vector( 0, 1){ 70}}
\put(280,605){\vector( 0, 1){ 70}}
\put(285,600){\vector( 1, 0){ 70}}
\put(285,680){\vector( 1, 0){ 70}}
\put(160,605){\vector( 0, 1){ 70}}
\put( 85,680){\vector( 1, 0){ 70}}
\put( 85,600){\vector( 1, 0){ 70}}
\put( 80,605){\vector( 0, 1){ 70}}
\put( 80,555){\vector( 0, 1){ 40}}
\put( 80,685){\vector( 0, 1){ 40}}
\put(409,640){\makebox(0,0)[lb]{\raisebox{0pt}[0pt][0pt]{$\pm \pa$}}}
\put(339,640){\makebox(0,0)[lb]{\raisebox{0pt}[0pt][0pt]{$\mp \pa$}}}
\put(249,640){\makebox(0,0)[lb]{\raisebox{0pt}[0pt][0pt]{$\pm \pa$}}}
\put(270,520){\makebox(0,0)[lb]{\raisebox{0pt}[0pt][0pt]{Fig. 2}}}
\put(150,580){\makebox(0,0)[lb]{\raisebox{0pt}[0pt][0pt]{$F^{1}$}}}
\put(430,580){\makebox(0,0)[lb]{\raisebox{0pt}[0pt][0pt]{$F^{j+1}$}}}
\put(354,580){\makebox(0,0)[lb]{\raisebox{0pt}[0pt][0pt]{$F^{j}$}}}
\put(270,580){\makebox(0,0)[lb]{\raisebox{0pt}[0pt][0pt]{$F^{j-1}$}}}
\put(430,690){\makebox(0,0)[lb]{\raisebox{0pt}[0pt][0pt]{$F^{j+1}$}}}
\put(354,690){\makebox(0,0)[lb]{\raisebox{0pt}[0pt][0pt]{$F^{j}$}}}
\put(270,690){\makebox(0,0)[lb]{\raisebox{0pt}[0pt][0pt]{$F^{j-1}$}}}
\put(150,690){\makebox(0,0)[lb]{\raisebox{0pt}[0pt][0pt]{$F^{1}$}}}
\put( 60,580){\makebox(0,0)[lb]{\raisebox{0pt}[0pt][0pt]{$F^{0}$}}}
\put( 60,690){\makebox(0,0)[lb]{\raisebox{0pt}[0pt][0pt]{$F^{0}$}}}
\put( 65,540){\makebox(0,0)[lb]{\raisebox{0pt}[0pt][0pt]{$\C$}}}
\put( 65,730){\makebox(0,0)[lb]{\raisebox{0pt}[0pt][0pt]{$\C$}}}
\put(395,585){\makebox(0,0)[lb]{\raisebox{0pt}[0pt][0pt]{$\delta^{j+1}$}}}
\put(315,585){\makebox(0,0)[lb]{\raisebox{0pt}[0pt][0pt]{$\delta^{j}$}}}
\put(395,685){\makebox(0,0)[lb]{\raisebox{0pt}[0pt][0pt]{$\delta^{j+1}$}}}
\put(315,685){\makebox(0,0)[lb]{\raisebox{0pt}[0pt][0pt]{$\delta^{j}$}}}
\put(115,585){\makebox(0,0)[lb]{\raisebox{0pt}[0pt][0pt]{$\delta^{1}$}}}
\put(115,685){\makebox(0,0)[lb]{\raisebox{0pt}[0pt][0pt]{$\delta^{1}$}}}
\put(135,640){\makebox(0,0)[lb]{\raisebox{0pt}[0pt][0pt]{$-\pa$}}}
\put( 65,640){\makebox(0,0)[lb]{\raisebox{0pt}[0pt][0pt]{$\pa$}}}
\end{picture}

\vspace{10mm}

\subsubsection{The Adler-Gelfand-Dickey algebra.} According to
\thmref{todaimcl}, the Poisson algebra $I_0(\g)$ of local integrals of
motion of the Toda field theory associated
to $\g$ is the algebra of local functionals on a certain hamiltonian
space, $H(\g)$.

This Poisson algebra coincides with the Adler-Gelfand-Dickey (AGD) algebra,
or the classical $\W-$algebra \cite{gd,adler}.

The Drinfeld-Sokolov reduction \cite{DS,ds} produces the hamiltonian space
of the AGD algebra as the result of a hamiltonian reduction of a hyperplane
in the dual space to the affinization $\widehat{\g}$ of $\g$. Following the
standard technique of hamiltonian reduction \cite{ks}, one can obtain this
algebra as the $0$th cohomology of the corresponding (classical) BRST
complex.

It was explained in \cite{critical,thesis,cargese} that the complex
$\bar{F}^*(\g)$ appears as the first term of a spectral sequence,
associated to the BRST complex of the Drinfeld-Sokolov reduction.
Therefore $I_0(\g)$ is precisely the AGD algebra. We also see that higher
cohomologies of the BRST complex vanish.

Usually, one constructs a map, which is called the Miura transformation,
from the hamiltonian space $L\h$ to $H(\g)$, which preserves the Poisson
structures. The image of the inverse map of the spaces of functionals
embeds the AGD algebra into $\F_0$. As we have explained, the image of this
map coincides with the algebra of local integrals of motion of the
corresponding Toda field theory, and can be characterized in very simple
terms as the intersection of the kernels of certain linear operators,
acting from $\F_0$ to the spaces $\F_{\al_i}$.

For the classical simple Lie algebras explicit formulas for the Miura
transformation map are known \cite{ds}. They give explicit formulas for
the generators $W^{(i)}_n$ of the Poisson algebra $I_0(\g)$.

For example, the AGD hamiltonian space $H(\sw_n)$ is isomorphic to the
space of differential operators on the circle of the form $$\pa_t^n +
\sum_{i=1}^{n-1} W^{(i)}(t) \pa_t^{n-i-1}.$$ The Miura transformation
from the space $L\h$, which consists of functions on the circle with values
in the Cartan subalgebra $\h$ of $\sw_n$, ${\bold u}(t) =
(u^1(t),\ldots,u^{n-1}(t))$, to $H(\sw_n)$ can be constructed as
follows.

Introduce new variables $v^1(t),\ldots,v^n(t)$, such that $\sum_{i=1}^n
v^i(t) = 0$, and $u^i(t) = v^i(t) - v^{i+1}(t)$. Then put $$\pa_t^n +
\sum_{i=1}^{n-1} W^{(i)}(t) \pa_t^{n-i-1} = (\pa_t + v^1(t))\ldots(\pa_t +
v^n(t)).$$ These formulas allow to express $W^{(i)}(t)$ as a differential
polynomial in $u^j(t)$ (cf.
\secref{vir} for the case of $\sw_2$, when the AGD algebra is
isomorphic to the classical Virasoro algebra). One can find other
generators $\widetilde{W}^{(i)}(t)$ of $I_0(\sw_n)$, which transform as tensor
fields on the circle under changes of variables \cite{itzykson}.

\subsubsection{Integrals of motion in the extended space of local
functionals}    \label{extendedspace}
As in \secref{leibnitz}, we can extend our Poisson algebra of local
functionals $\F_0$ by adjoining all Fourier components of differential
polynomials, not only the $(-1)$st ones. Let $\f_0$ be the space of
functionals on $L\h$, which can be represented as residues of differential
polynomials with explicit dependence on $t$: $$F[{\bold u}(t)] = \int
P(\pa^n u^i(t);t) dt.$$ We can define the Poisson structure on $\f_0$ by
the same formula \eqref{genskobka}. Thus, $\f_0$ is a Poisson algebra, and
$\F_0 \subset \f_0$ is its Poisson subalgebra.

Analogously, one can define the spaces $\f_\gamma, \gamma \in P$ by
allowing differential polynomials to depend on $t$.

One has the analogue of the exact sequence \eqref{exactseq}: $$0 \larr
\pi_\gamma \otimes \C[t,t^{-1}] \stackrel{\pa}{\larr} \pi_\gamma
\otimes \C[t,t^{-1}] \larr \f_\gamma \larr 0,$$ where the action of
$\pa$ on $\pi_\gamma \otimes \C[t,t^{-1}]$ is given by $\pa \otimes 1 + 1
\otimes \pa_t.$ If $\gamma=0$, we have to replace the first $\pi_0
\otimes \C[t,t^{-1}]$ by $\pi_0 \otimes \C[t,t^{-1}]/\C \otimes \C$.

The Poisson bracket with $\int e^{\phi_i(t)} dt$ defines a linear
operator $\f_0 \arr \f_{\al_i}$, which we also denote by $\Q_i$.  We
can then define the space of integrals of motion of the Toda field
theory as the intersection of kernels of the operators $\Q_i: \f_0
\arr \f_{\al_i}, i=1,\ldots,l$.

We can use the methods of this section to compute this space. Indeed,
consider the tensor product of the complex $F^*(\g)$ with $\C[t,t^{-1}]$,
with the differentials acting on the first factor as $\delta^j$ and
identically on the second factor. Such differentials commute with the
action of the derivative, and therefore we can use the double complex
$$\C \otimes \C \larr F^*(\g) \otimes \C[t,t^{-1}] \larr F^*(\g) \otimes
\C[t,t^{-1}] \larr \C \otimes \C$$ to compute the space of integrals of
motion.

The cohomologies of the complex $F^*(\g) \otimes \C[t,t^{-1}]$ are equal to
the cohomologies of the complex $F^*(\g)$ tensored with $\C[t,t^{-1}]$. We
deduce from \propref{exist} that the space of integrals of motion is
isomorphic to the quotient of $\W_0(\g) \otimes \C[t,t^{-1}]$ by the total
derivatives and constants. This Poisson algebra is the algebra of local
functionals on the AGD hamiltonian space $H(\g)$, extended in the same way
-- by adjoining all Fourier components of differential polynomials. It
contains $I_0(\g)$ as a Poisson subalgebra. Sometimes it is this algebra,
which is called the classical $\W-$algebra.

\section{Classical affine Toda field theories.}    \label{classicalaff}

\subsection{The case of $\widehat{\sw}_2$ -- classical sine-Gordon theory.}
   \label{sinegor}

\subsubsection{Hamiltonian structure.} The hamiltonian space of the
classical sine-Gordon theory is the same as the hamiltonian space of the
classical Liouville theory, namely, the space $L\h$ of polynomial functions
on the circle with values in the one-dimensional Cartan subalgebra $\h$ of
$\sw_2$, cf. \secref{lhamspace}. As the space of functions on this space,
we again take the space $\F_0$ of local functionals. The Poisson structure
on $\F_0$ is given by formula \eqref{skobka}.

We also define spaces $\pi_n, n \in \Z$, as the tensor products $\pi_0
\otimes \C v_n$, where $\pi_0$ is the space of differential
polynomials (cf. \secref{lhamspace}), and extend the action of
derivative $\pa$ from $\pi_0$ to $\pi_n$ by the formula $(\pa + n\cdot
u) \otimes 1.$ We then put: $\F_n = \pi_n/\pa \pi_n$. This space has an
interpretation as the space of functionals of the form $$\int
P(u(t),\pa u(t),\ldots) e^{n\phi(t)} dt.$$ As in \secref{poisext}, we
can extend the Poisson structure on $\F_0$ to well-defined maps $\F_0
\times \F_n \arr \F_n$ given by the formula
\begin{equation}    \label{nextension}
\{\int P dt, \int R
e^{n\phi} dt\} = \int \frac{\delta P}{\delta u} \, \frac{\delta [R
e^{n\phi}]}{\delta \phi} dt = \int \frac{\delta P}{\delta u} \left[ n
R e^{n\phi} - \pa_t \left( \frac{\delta R}{\delta u} e^{n\phi} \right)
\right] dt.
\end{equation}
These brackets satisfy the Jacobi identity for any triple $F, G \in
\F_0, H \in \F_n$.

\subsubsection{The sine-Gordon hamiltonian and local integrals of
motion.}
The hamiltonian $H$ of the sine-Gordon model is given by $$H = \int
e^{\phi(t)} dt + \int e^{-\phi(t)} dt \in \F_1 \oplus \F_{-1}.$$ In
other words, it is equal to the sum of the projections of the vectors
$v_{\pm 1} \in \pi_{\pm 1}$ to $\F_{\pm 1}$. One can check that the
corresponding hamiltonian equation $$\pa_\tau U(t) = \{U(t),H\}$$
coincides with the sine-Gordon equation \eqref{sine}.

We then define the space of local integrals of motion
$I_0(\widehat{\sw}_2)$ of the sine-Gordon theory as the kernel of the
linear operator $$\Q_1 + \Q_0: \F_0 \arr \F_1 \oplus \F_{-1},$$ where
$\Q_1 = \{\cdot,\int e^{\phi(t)} dt\}: \F_0 \arr \F_1$, and $\Q_0 =
\{\cdot,\int e^{-\phi(t)} dt\}: \F_0 \arr \F_{-1}$. In other words,
$I_0(\widehat{\sw}_2)$ is the intersection of the kernels of the operators
$\Q_1$ and $\Q_0$. By the Jacobi identity, $I_0(\widehat{\sw}_2)$ is closed
with respect to the Poisson bracket. Note that the operator $\Q_1$
coincides with the operator $\Q$, defined by \eqref{oper}, and
therefore $I_0(\su)$ is a Poisson subalgebra in the Poisson algebra
$I_0(\sw_2)$ of local integrals of motion of the Liouville theory, that
is in the classical Virasoro algebra.

\subsubsection{}
Now introduce the operators $\q_1: \pi_m \arr \pi_{m+1}$ and $\q_0:
\pi_m \arr \pi_{m-1}$ by the formulas: $$\q_1 = T \sum_{n<0} S^+_{n+1}
\frac{\pa}{\pa x_n}, \quad \quad \q_0 = - T^{-1} \sum_{n<0}
S^-_{n+1} \frac{\pa}{\pa x_n},$$ where $T^{\pm 1}: \pi_m \arr
\pi_{m\pm 1}$ are the translation operators, and $S^\pm_n$ are the
Schur polynomials: $$\sum_{n\leq 0} S^\pm_n z^n = \exp(\sum_{m<0} \mp
\frac{x_m}{m} z^m).$$

We will also need the operators $Q_1 = T^{-1} \q_1$ and $Q_0 = T
\q_0$, which are linear endomorphisms of $\pi_0$ of degree $1$.
Note that $S^+_n$ coincides with $S_n$, given by formula
\eqref{schur}, and so $\q_1$ coincides with $\q$, given by formula
\eqref{operq}.

In the same way as in \lemref{commute}, one shows that the operators $\q_1$
and $\q_0$, acting from $\pi_0$ to $\pi_{\pm1}$ commute with the action of
derivative $\pa$, and that the corresponding operators $\F_0 \arr
\F_{\pm 1}$ coincide with $\Q_1$ and $\Q_0$.

The following Proposition will enable us to compute the space of local
integrals of motion $I_0(\widehat{\sw}_2)$.

\subsubsection{Proposition.}    \label{serresl}
{\em The operators $Q_1$ and $Q_0$
satisfy the Serre relations of the Lie algebra $\widehat{\sw}_2$}:
$$(\mbox{ad} \, Q_1)^3 \cdot Q_0 = 0, \quad \quad (\mbox{ad} \, Q_0)^3
\cdot Q_1 = 0.$$

\vspace{3mm}
\noindent {\em Proof} follows from \propref{serre} and
\remref{serrerem}.

\subsubsection{} Thus, operators $Q_0$ and $Q_1$ generate
an action of the nilpotent subalgebra $\n_+$ of $\widehat{\sw}_2$ on the
space $\pi_0$. We will use this fact and the BGG resolution for
$\widehat{\sw}_2$ to extend the complex $$\F_0 \larr \F_1 \oplus
\F_{-1},$$ whose $0$th cohomology is, by definition, the space
$I_0(\widehat{\sw}_2)$, to a larger complex with nicer cohomologies.

Again, as in the previous section, we will be using the double complex,
which consists of the spaces $\pi_n$. So, our task is to extend the complex
$$\pi_0 \larr \pi_1 \oplus \pi_{-1}$$ in such a way that all higher
differentials commute with the action of the derivative.

\subsubsection{BGG resolution of $\su$.} As in the case of
finite-dimensional Lie algebras, there exists a BGG resolution $B_*(\su)$
of $\su$ \cite{rocha}. It is a complex, consisting of Verma modules over
$\su$, whose higher homologies vanish, and the $0$th homology is
one-dimensional.

The $j$th group $B_j(\su)$ of this complex is equal to $M_0$, if
$j=0$, and $M_{2j} \oplus M_{-2j}$, if $j>0$. Here $M_\la$ stands for
the Verma module over $\su$ of level $0$ and lowest weight $\la$ (that
is the weight of the Cartan subalgebra of $\sw_2$, embedded into $\su$
as the constant subalgebra). Such a module is defined in exactly the
same way as a Verma module over a finite-dimensional Lie algebra (cf.
\secref{verma}).

The module $M_{0}$ contains the singular vectors, labeled by the elements
of the Weyl group of $\su$. Let us denote them by $w_{0}$, and
$w_{j}, w'_{j}, j>0$. The weight of $w_{j}$ (respectively, $w'_{j}$) is
equal to $2j$ (respectively, $-2j$), and it generates the submodule of
$M_{0}$, which is isomorphic to $M_{2j}$ (respectively, $M_{-2j}$). We have
$w_1 = Y'_1 w_0, w'_1 = Y_1 w_0$, and $w_{j}=X_{j}w_{j-1},
w'_{j}=Y_{j}w_{j-1}, w'_{j}=X'_{j}w'_{j-1}, w_{j}=Y'_{j}w'_{j-1}$, for
$j>1$, where $X_{j}, Y_{j}, X'_{j}$ and $Y'_{j}$ are certain elements from
$U(\n_{+})$.

The differential $d_{j}: B^{q}_{j} \rightarrow B^{q}_{j-1}, j>0,$ is given
by the alternating sum of the embeddings of $M_{2j}$ and $M_{-2j}$ into
$M_{2(j-1)}$ and $M_{-2(j-1)}$, which map the lowest weight vectors to the
corresponding singular vectors. The nilpotency of the differential,
$d_{j-1} d_j = 0$, is ensured by the commutativity of the embeddings,
corresponding to the ``squares'' in the Weyl group and a special sign
convention, analogous to the one from \lemref{signs}.

\setlength{\unitlength}{0.0125in}%
\begin{picture}(445,220)(145,555)
\thicklines
\put(155,677){$\bullet$}
\put(235,715){$\bullet$}
\put(235,638){$\bullet$}
\put(330,715){$\bullet$}
\put(330,638){$\bullet$}
\put(365,680){$\cdot$}
\put(375,680){$\cdot$}
\put(385,680){$\cdot$}
\put(585,680){$\cdot$}
\put(575,680){$\cdot$}
\put(565,680){$\cdot$}
\put(530,638){$\bullet$}
\put(530,715){$\bullet$}
\put(435,638){$\bullet$}
\put(435,715){$\bullet$}
\put(445,650){\vector( 4, 3){ 80}}
\put(445,710){\vector( 4,-3){ 80}}
\put(245,710){\vector( 4,-3){ 80}}
\put(245,650){\vector( 4, 3){ 80}}
\put(165,675){\vector( 2,-1){ 65}}
\put(165,685){\vector( 2, 1){ 65}}
\put(245,720){\vector( 1, 0){ 80}}
\put(245,640){\vector( 1, 0){ 80}}
\put(445,640){\vector( 1, 0){ 80}}
\put(445,720){\vector( 1, 0){ 80}}
\put(350,590){\makebox(0,0)[lb]{\raisebox{0pt}[0pt][0pt]{Fig. 3}}}
\put(485,628){\makebox(0,0)[lb]{\raisebox{0pt}[0pt][0pt]{$X'_j$}}}
\put(470,660){\makebox(0,0)[lb]{\raisebox{0pt}[0pt][0pt]{$Y'_j$}}}
\put(470,698){\makebox(0,0)[lb]{\raisebox{0pt}[0pt][0pt]{$Y_j$}}}
\put(485,725){\makebox(0,0)[lb]{\raisebox{0pt}[0pt][0pt]{$X_j$}}}
\put(280,628){\makebox(0,0)[lb]{\raisebox{0pt}[0pt][0pt]{$X'_2$}}}
\put(265,655){\makebox(0,0)[lb]{\raisebox{0pt}[0pt][0pt]{$Y'_2$}}}
\put(265,700){\makebox(0,0)[lb]{\raisebox{0pt}[0pt][0pt]{$Y_2$}}}
\put(280,725){\makebox(0,0)[lb]{\raisebox{0pt}[0pt][0pt]{$X_2$}}}
\put(185,645){\makebox(0,0)[lb]{\raisebox{0pt}[0pt][0pt]{$Y_1$}}}
\put(185,710){\makebox(0,0)[lb]{\raisebox{0pt}[0pt][0pt]{$Y'_1$}}}
\put(520,625){\makebox(0,0)[lb]{\raisebox{0pt}[0pt][0pt]{$\pi_{-j}$}}}
\put(430,625){\makebox(0,0)[lb]{\raisebox{0pt}[0pt][0pt]{$\pi_{-j+1}$}}}
\put(520,730){\makebox(0,0)[lb]{\raisebox{0pt}[0pt][0pt]{$\pi_j$}}}
\put(430,730){\makebox(0,0)[lb]{\raisebox{0pt}[0pt][0pt]{$\pi_{j-1}$}}}
\put(320,625){\makebox(0,0)[lb]{\raisebox{0pt}[0pt][0pt]{$\pi_{-2}$}}}
\put(230,625){\makebox(0,0)[lb]{\raisebox{0pt}[0pt][0pt]{$\pi_{-1}$}}}
\put(320,730){\makebox(0,0)[lb]{\raisebox{0pt}[0pt][0pt]{$\pi_2$}}}
\put(230,730){\makebox(0,0)[lb]{\raisebox{0pt}[0pt][0pt]{$\pi_1$}}}
\put(145,690){\makebox(0,0)[lb]{\raisebox{0pt}[0pt][0pt]{$\pi_0$}}}
\end{picture}

\subsubsection{Extended complex.}    \label{extendedsine}
We are ready now to define the extended complex $F^*(\su)$
(cf. Fig. 3). The $j$th group $F^*(\su)$ of this complex is equal to
$\pi_0$, if $j=0$, and $\pi_j \oplus \pi_{-j}$, if $j>0$. The differential
$\delta^j: F^{j-1}(\g) \arr F^j(\g)$ is given by the formula $\delta^1 =
Y_0(Q) + Y'_0(Q), \delta^j = X_j(Q) - (-1)^j Y_j(Q) - (-1)^j Y'_j(Q) +
X'_j(Q), j>1,$ where we insert into these elements of $U(\n_+)$ the
operators $\q_1$ and $\q_0$ instead of the generators $e_1$ and $e_0$.

The nilpotency of the differential, $\delta^j \delta^{j-1} = 0$, follows
from the relations in $U(\n_+)$: $X'_{j} Y_{j-1} = Y_{j} X_{j-1}, X_{j}
Y'_{j-1} = Y'_{j} X'_{j-1}, X_{j} X_{j-1} = Y'_{j} Y_{j-1}$, and $X'_{j}
X'_{j-1} = Y_{j} Y'_{j-1}$, and \propref{serresl}.

Note that $Y'_j(Q) = \q_1^{2j-1}, Y_j(Q) = \q_0^{2j-1}$. Other operators
are more complicated, but explicit formulas for them can be obtained in
principle from the commutativity relations above, using the Serre
relations, in the same way as in \secref{sltri}.

Let us introduce a $\Z-$grading on our complex, by putting $\deg v_n =
n^2$, and $\deg x_m = -m$. One can check that with respect to this
grading the differentials $\delta^j$ are homogeneous of degree $0$.
Therefore our complex is a direct sum of finite-dimensional
subcomplexes, corresponding to various graded components.

\subsubsection{Proposition.}    \label{sgcohom}
{\em The operators $Q_1$ and $Q_0$ define an action of the nilpotent Lie
subalgebra $\n_+$ of $\su$ on $\pi_0$. The cohomologies of the complex
$F^*(\su)$ are isomorphic to the cohomologies of $\n_+$ with coefficients
in $\pi_0$, $H^*(\n_+,\pi_0)$.}

\vspace{3mm}
\noindent {\em Proof.} The same as in \propref{cohom}.

\subsubsection{Principal commutative subalgebra.}
Recall that in the realization of $\su$ as the central extension of the
loop algebra $\sw_2 \otimes \C[t,t^{-1}]$, the nilpotent subalgebra $\n_+$
of $\su$ is identified with the Lie algebra $(\bar{\n}_+
\otimes 1) \oplus (\sw_2 \otimes t \C[t])$, where $\bar{\n}_+$ is the
nilpotent subalgebra of $\sw_2$. Let $e, h$ and $f$ be the standard
generators of $\sw_2$. If $y$ is one of them, we will denote by $y(m)$
the element $y \otimes t^m$ of $\su$. The basis of $\n_+$ consists of
$e(m), m \geq 0, h(m), m>0$, and $f(m), m>0$.

Let $\ab$ be the commutative subalgebra of $\n_+$, which is linearly
generated by $e(m) + f(m+1), m\geq 0$. We call $\ab$ the principal
commutative subalgebra.

\subsubsection{Proposition.}    \label{wedge}
{\em The cohomologies of the complex $F^*(\su)$ are isomorphic to the
exterior algebra $\bigwedge^*(\ab^*)$ of the dual space to the principal
commutative subalgebra of $\n_+$.}

\vspace{3mm}
\noindent {\em Proof.} If $X$ is an operator on
$\pi_0$ of the form $\sum_i X_i \pa/\pa x_i$, where $X_i$ are
polynomials in $x_n$, then we can define its shift term as the
sum of terms $X_i \pa/\pa x_i$, for which $X_i$ is a constant.
According to the definition of the operators $Q_1$ and $Q_0$, their
shift terms are equal to $\pa/\pa x_{-1}$ and $-\pa/\pa x_{-1}$,
respectively. Thus the shift term of the operator $p = e(0) + f(1) =
Q_0 + Q_1$ is equal to $0$. This operator has the form $\sum_{n<0} A_n
\frac{\pa}{\pa x_{n-1}}$, where $A_n = S^+_n - S^-_n$ is a certain
polynomial in $x_m$; $A_n$ is equal to $-\frac{2}{n} x_n +$ higher power
terms.

It is known that the Lie algebra $\n_+$ splits into the direct sum
Ker(ad $p) \oplus$ Im(ad $p)$, and that Ker(ad $p) =
\ab$. In the principal grading of $\n_+$, in which $\deg e(0) = 1, \deg
f(1) = 1$, Im(ad $p)$ is linearly generated by vectors of all
positive degrees, and Ker(ad $p)$ is linearly generated by vectors
of all positive odd degrees.

The element $(\mbox{ad} p)^m \cdot e_1, m\geq 0,$ can be chosen as a
generator $y_{m+1}$ of $\mbox{Im(ad} p)$ of degree $m+1$. By induction one
can check that the shift term of the operator $(\mbox{ad} p)^m \cdot Q_1$
is equal to a non-zero multiple of $\pa/\pa x_{-m-1}$. Indeed, suppose that
we have shown this for $m=0,1,\ldots,n-1$. Since $p$ does not have a shift
term, the shift term of $y_{n+1} = [p,y_n]$ is equal to the commutator of
the shift term $\pa/\pa x_{-n}$ (times a constant) of $y_n$ and a linear
term of the form $x_{-n} \pa/\pa x_j$ from $p$. There is only one such
summand in $p$, namely, $$-\frac{2}{n} x_{-n} \frac{\pa}{\pa x_{-n-1}}.$$
Its commutator with $\pa/\pa x_{-n}$ is equal to $\pa/\pa x_{-n-1}$ up to a
non-zero constant. Therefore the shift term of $y_{n+1}$ equals $\pa/\pa
x_{-n-1}$ up to a non-zero constant.

In the same way we can show that the shift term of any element of
$\mbox{Ker(ad} p)$ has to be $0$, because otherwise the commutator of this
element with $p$ would be non-trivial.

Let us consider the module $\pi_0^*$ over $\n_+$, dual to $\pi_0$. One can
identify $\pi_0$ with $\pi_0^*$ as vector spaces. We can then obtain the
formulas for the action of $\n_+$ on $\pi_0^*$ from the formulas for its
action on $\pi_0$ by interchanging $x_m$'s and $\pa/\pa x_m$'s (cf. the
proof of \propref{vir}). Since the Lie algebra $\ab$ acts on $\pi_0$ by
vector fields, which have no shift terms, this Lie algebra acts by $0$ on
the vector $1^* \in \pi_0^*$, dual to $1 \in \pi_0$. Let $L$ be the
$\n_+-$module, induced from the trivial one-dimensional representation of
the Lie subalgebra $\ab$ of $\n_+$: $$L = U(\n_+) \otimes_{U(\ab)} \C.$$
Since the vector $1^* \in \pi_0^*$ is $\ab$-invariant, there is a unique
$\n_+-$homomorphism: $L \arr \pi_0$, which sends the generating vector of
$L$ to $1^*$.

Under this homomorphism, a monomial $y_{i_1} \ldots y_{i_m} \otimes 1 \in
L$ maps to a vector of $\pi_0^*$, which is equal to a non-zero multiple of
$x_{-i_1} \ldots x_{-i_m} +$ lower power terms. Therefore this map has no
kernel. On the other hand, the character of the module $L$ in the principal
gradation is equal to $$\prod_{n>0} (1-q^n)^{-1}.$$ This coincides with the
character of the module $\pi_0^*$. Therefore, $\pi_0^*$ is isomorphic to
$L$ as an $\n_+-$module.

Going back, we see that the module $\pi_0$ is isomorphic to the module
$L^*$, which is the $\n_+-$module coinduced from the trivial
representation of $\ab$.

By ``Shapiro's lemma'' (cf. \cite{fuchs}, \S 1.5.4, \cite{gui}, \S II.7),
$H^*(\n_+,\pi_0) \simeq H^*(\ab,\C)$. But $\ab$ is an abelian Lie algebra,
hence $H^*(\ab,\C) = \bigwedge^*(\ab^*)$. The Proposition now follows from
\propref{sgcohom}.

\subsubsection{Theorem.}    \label{intsine}
{\em The space $I_0(\su)$ of local integrals of motion of the
sine-Gordon model is linearly generated by mutually commuting local
functionals ${\cal H}_{2i+1}, i\geq 0$, of all positive odd degrees.}

\vspace{3mm}
\noindent {\em Proof.}
According to \propref{wedge}, the $1$st cohomology of the complex
$F^*(\su)$ is isomorphic to $\ab^*$. As a $\Z-$graded space it is a
direct sum of one-dimensional subspaces, generated by some elements
$h_j$ of all positive odd degrees.  In the same way as in
\propref{higher} one checks that the higher differentials of the
complex $F^*(\su)$ commute with the action of the derivative $\pa$.
Therefore we can form the double complex $$\C \larr F^*(\su)
\larr F^*(\su) \larr \C$$ (cf. Fig. 2). The $1$st cohomology of this double
complex is isomorphic to the space $I_0(\su)$.

We can calculate the cohomologies of this double complex $F^*(\su)$ by
means of the spectral sequence, in which the $0$th differential is
horizontal. The first term of this spectral sequence consists of two copies
of the cohomologies of the complex $F^*(\su)$, and the first differential
coincides with the action of the derivative on them.

The $0$th cohomology of the complex $F^*(\su)$ is generated by the
vector $1 \in \pi_0$. Clearly, the corresponding two cohomology classes in
the 1st term of the spectral sequence are canceled by the two spaces $\C$
in the double complex (cf. Fig. 2).

On the other hand, the derivative $\pa$ acts by $0$ on $h_{2j+1}
\in H^1(\su)$, because $\pa$ has degree 1 and so it should send
cohomology classes of odd degrees to cohomology classes of even
degrees, which we do not have.

Since we only have two rows in our double complex, the spectral sequence
collapses in the first term. Therefore, the $1$st cohomology of the double
complex is equal to the $1$st cohomology $H^1$ of the complex $F^*(\su)$,
which is equal to $\oplus_{j\geq 0} \C h_{2j+1}$, by \propref{wedge}.

Each $h_{2j+1}$ gives rise to a local integral of motion ${\cal
H}_{2j+1} \in \F_0$ as follows. It is clear that $\pa h_{2j+1}$ is
also a cocycle in $F^1(\su)$, of even degree. Therefore it must be a
coboundary: $\pa h_{2j+1} = \delta^1 H_{2j+1}$, for some element
$H_{2j+1}$ from $F^0(\su) = \pi_{0}$. This element is not a total
derivative, because otherwise $h_{2j+1}$ would also be a coboundary.
Hence ${\cal H}_{2j+1}=\int H_{2j+1}(t) dt$ is non-zero and it lies in
the kernel of $\Q_1 + \Q_0$, because $(\q_1+\q_0) H_{2j+1} = \delta^1
H_{2j+1}$ is a total derivative.

Recall that $I_0(\su)$ is closed with respect to the Poisson bracket. So,
it is a Poisson subalgebra in $\F_0$. One can easily check that the Lie
bracket in $\F_0$ is compatible with the $\Z$-grading, introduced in
\secref{zgrading}. Therefore the degree of $\{H_{2i+1},H_{2j+1}\}$ should
be even, and we obtain $\{H_{2i+1},H_{2j+1}\}=0$.

\subsubsection{Examples of integrals of motion.}
In this subsection we will analyze our complex $F^*(\su)$ in low degrees
and give explicit formulas for local integrals of motion of degrees $1$ and
$3$.

Let us consider the homogeneous components of our complex $F^*(\su)$
of degrees up to $4$.  Since $\deg v_n>4$ for $|n|>2$ (cf.
\secref{extendedsine}), only $F^0(\su) = \pi_0, F^1(\su) = \pi_1 \oplus
\pi_{-1}$ and $F^2(\su) = \pi_2 \oplus \pi_{-2}$ contain subspaces of
degrees less than or equal to $4$, cf. Fig. 4.

On Fig. 4 dots represent basis vectors in $F^0(\su), F^1(\su)$, and
$F^2(\su)$, and the dots, corresponding to vectors of the same degree, are
situated at the same horizontal level.

We can see that the vector $1 \in \pi_0$ is the only element of degree $0$,
therefore it is necessarily in the kernel of the differential $\delta^1:
F^0(\su) \arr F^1(\su)$. It is, as we know, the only cohomology class in
the $0$th cohomology of our complex.

The second group of the complex $F^2(\su)$ has degrees greater than or
equal to $4$. Therefore, all vectors of $\pi_1 \oplus \pi_{-1}$ of
degrees $1, 2$ and $3$ necessarily lie in the kernel of the
differential $\delta^2: F^1(\su) \arr F^2(\su)$. But some of them lie in
the image of the differential $\delta^1$.

The component of degree $1$ of the space $\pi_0$ is one-dimensional, and of
$\pi_1 \oplus \pi_{-1}$ is two-dimensional, therefore, the $1$st cohomology
in degree $1$ is one-dimensional. As a representative of this cohomology
class we can take, for instance, vector $2 v_{-1}$. In degree $2$ both
spaces have two-dimensional components, so there are no cohomologies of
degree $2$. In degree $3$, the component of the space $\pi_0$ is
three-dimensional, and the component of the space $\pi_1 \oplus \pi_{-1}$
is four-dimensional, so again we have a cohomology class. As a
representative we can choose vector $x_{-1}^2 v_{-1}$.

In degree $4$ the vectors $v_2$ and $v_{-2}$ from $F^2(\su)$ are both in
the kernel of the differential $\delta^3: F^2(\su) \arr F^3(\su)$.  But
some linear combination of them is in the image of the differential
$\delta^2$. Indeed, we know that the differential $\delta^1$ has no kernel
in degree $4$. Therefore its image in the subspace of $F^2(\su)$ of degree
$4$ is $5$-dimensional. But the $1$st cohomology of degree $4$ is
trivial. Hence the kernel of the differential $\delta^2$ is also
$5$-dimensional. But the space $F^1(\su)$ has dimension $6$ in degree
$4$. Therefore the image of $\delta^3$ is one-dimensional. Thus the second
cohomology is one-dimensional in degree $4$. It can be represented by the
vector $v_{-2}$.

In view of \propref{wedge}, the cohomology classes that we constructed in
$F^1(\su)$, correspond to the generators of degrees $1$ and $3$ of the
space $\ab^*$, and the class that we constructed in $F^2$ corresponds to
their exterior product. The encircled dots represent these classes on the
picture.

Now let us assign to the $1$st cohomology classes of degrees $1$ and
$3$ local integrals of motion ${\cal H}_1$ and ${\cal H}_3$ (cf.
\thmref{intsine}).

The cohomology class $h_1$ of degree $1$ was represented by the vector
$2 v_1$. According to the general procedure, described in
\thmref{intsine}, we have to take $\pa v_1 = -2 x_{-1} v_1$ and find a
vector $H_1$ from $\pi_0$, such that $\delta^1 H_1 =
-2 x_{-1} v_1$. One checks that $\frac{1}{2} x_{-1}^2 - x_{-2} \in \pi_0$
is such a vector. Therefore, we can take as the integral of motion
\begin{equation}    \label{h1}
{\cal H}_1 = \int (\frac{1}{2} u(t)^2 - \pa_t u(t)) dt.
\end{equation}
According to \secref{action}, we can define an action of ${\cal H}_1$
on $\oplus_{n\in\Z} \pi_n$, and this action will coincide with the
action of $\pa$.

The cohomology class $h_3$ of degree $3$ was represented by vector
$x_{-1}^2 v_{-1}$. We have: $\pa h_3 = (-x_{-1}^3
+ 2 x_{-1} x_{-2}) v_{-1}$, and one can check that $\delta^1 \cdot
\frac{1}{2}(\frac{1}{2} x_{-1}^2 - x_{-2})^2 = \pa h_3$. Therefore,
\begin{equation}    \label{h3}
{\cal H}_3 = \frac{1}{2}\int (\frac{1}{2} u(t)^2 - \pa_t u(t))^2 dt
\end{equation}
is an integral of motion.

\setlength{\unitlength}{0.0125in}%
\begin{picture}(432,217)(118,525)

\put(155,700){$\bullet$}
\put(158,704){\circle{11}}

\put(155,675){$\bullet$}
\put(140,650){$\bullet$}
\put(170,650){$\bullet$}
\put(125,625){$\bullet$}
\put(155,625){$\bullet$}
\put(185,625){$\bullet$}
\put( 95,600){$\bullet$}
\put(125,600){$\bullet$}
\put(155,600){$\bullet$}
\put(185,600){$\bullet$}
\put(215,600){$\bullet$}

\put(155,560){\makebox(0,0)[lb]{\raisebox{0pt}[0pt][0pt]{$\pi_0$}}}

\put(310,675){$\bullet$}
\put(310,650){$\bullet$}
\put(295,625){$\bullet$}
\put(325,625){$\bullet$}
\put(280,600){$\bullet$}
\put(310,600){$\bullet$}
\put(340,600){$\bullet$}

\put(230,560){\makebox(0,0)[lb]{\raisebox{0pt}[0pt][0pt]{$\larr$}}}

\put(310,560){\makebox(0,0)[lb]{\raisebox{0pt}[0pt][0pt]{$\pi_1$}}}

\put(430,675){$\bullet$}
\put(433,679){\circle{11}}

\put(430,650){$\bullet$}
\put(415,625){$\bullet$}
\put(418,629){\circle{11}}

\put(445,625){$\bullet$}
\put(400,600){$\bullet$}
\put(430,600){$\bullet$}
\put(460,600){$\bullet$}

\put(370,560){\makebox(0,0)[lb]{\raisebox{0pt}[0pt][0pt]{$\oplus$}}}

\put(430,560){\makebox(0,0)[lb]{\raisebox{0pt}[0pt][0pt]{$\pi_{-1}$}}}

\put(535,600){$\bullet$}

\put(480,560){\makebox(0,0)[lb]{\raisebox{0pt}[0pt][0pt]{$\larr$}}}

\put(535,560){\makebox(0,0)[lb]{\raisebox{0pt}[0pt][0pt]{$\pi_2$}}}

\put(595,600){$\bullet$}
\put(598,604){\circle{11}}

\put(565,560){\makebox(0,0)[lb]{\raisebox{0pt}[0pt][0pt]{$\oplus$}}}

\put(595,560){\makebox(0,0)[lb]{\raisebox{0pt}[0pt][0pt]{$\pi_{-2}$}}}

\put(340,530){\makebox(0,0)[lb]{\raisebox{0pt}[0pt][0pt]{Fig. 4}}}

\end{picture}

\vspace{10mm}

\subsubsection{Connection with the KdV and mKdV systems.}
According to \thmref{intsine}, the integrals of motion ${\cal H}_{2j+1} \in
\F_0, j\geq 0$, mutually commute with respect to the Poisson bracket. In
particular, we have $$\{ {\cal H}_1,{\cal H}_{2j+1} \} = 0, \quad \quad \{
{\cal H}_3,{\cal H}_{2j+1} \} = 0,$$ where ${\cal H}_1$ and ${\cal H}_3$
are given by formulas \eqref{h1} and
\eqref{h3}, respectively. Since ${\cal H}_1 = \pa$, the first equation is
satisfied automatically for any element of ${\cal F}_0$
(cf. \secref{action}).  However, the second equation is non-trivial, and it
shows that the higher integrals of motion ${\cal H}_{2j+1}, j>1$, coincide
with the higher hamiltonians of the mKdV hierarchy. In fact, this equation
can be taken as a definition of these integrals of motion.

This is because ${\cal H}_3$ is the hamiltonian of the modified
Korteweg--de Vries (mKdV) equation. In other words, the mKdV equation
\begin{equation}    \label{mkdv}
\pa_\tau U(t) = \pa_t^3 U(t) - \frac{3}{2} U(t)^2 \pa_t U(t)
\end{equation}
can be written in the hamiltonian form as $$\pa_{\tau} U(t) = \{ U(t),{\cal
H}_3 \}.$$ The elements of $\F_0$, which commute with ${\cal H}_3$ are
called the higher mKdV hamiltonians. It is known that they exist precisely
for all odd degrees, therefore they coincide with our ${\cal
H}_{2j+1}$. Thus, we see that the local integrals of motion of the
sine-Gordon model coincide with the hamiltonians of the mKdV hierarchy, and
that the property of commutativity with ${\cal H}_3$ defines these local
integrals of motion uniquely. The mKdV hamiltonians define hamiltonian
flows on $L\h$, which commute with the flow, defined by ${\cal
H}_3$. Altogether they define the mKdV hierarchy of partial differential
equations.

The mKdV hierarchy is closely connected with the KdV hierarchy.  Namely, as
was explained in \secref{miura}, the kernel of the operator $\Q_1$ is
isomorphic to the classical Virasoro algebra, which is the Poisson algebra
of local functionals, depending on $W(t) = \frac{1}{2} u^2(t) - \pa_t
u(t).$ Since $I_0(\su)$ was defined as the intersection of the kernels of
the operators $\Q_1$ and $\Q_0$, it is a subspace in the classical Virasoro
algebra. For example, we can rewrite ${\cal H}_1$ and ${\cal H}_3$ via
$W(t)$ by the formulas $${\cal H}_1 = \int W(t) dt, \quad \quad {\cal H}_3
= \frac{1}{2}\int W^2(t) dt.$$ The functional ${\cal H}_3$ is the KdV
hamiltonian, therefore the commutativity condition $$\{ {\cal H}_3 , {\cal
H}_{2j+1} \} = 0$$ implies that the ${\cal H}_{2j+1}$'s, rewritten in terms
of $W(t)$, coincide with the higher KdV hamiltonians.

These hamiltonians define the KdV hierarchy of mutually commuting flows on
a hyperplane in the dual space to the Virasoro algebra, cf., e.g.,
\cite{segal}. The Miura transformation (cf. \secref{miura}) maps the
hierarchy of these flows to the mKdV hierarchy. In particular, the Miura
transformation maps the mKdV equation \eqref{mkdv} to the KdV equation
$$\pa_\tau W(t) = \pa_t^3 W(t) + 3 W(t) \pa_t W(t).$$

\subsection{General case.}

Let $\g$ be an arbitrary affine algebra, twisted or untwisted. We will
denote by $\bar{\g}$ the finite-dimensional simple Lie algebra, whose Dynkin
diagram is obtained by deleting the $0$th node of the Dynkin diagram of
$\g$.

\subsubsection{Hamiltonian structure.} As the hamiltonian space, the
spaces of functionals and the Poisson brackets of the affine Toda field
theory, associated to $\g$, we will take the objects, corresponding to the
Toda field theory, associated to $\bar{\g}$. They were defined in
\secref{hamspace} and \secref{functspace}. Throughout this section we will
use notation, introduced in \secref{fdtoda}.

\subsubsection{The hamiltonian and integrals of motion.} The imaginary
root $\delta$ of the Lie algebra $\g$ has the decomposition $$\delta =
\sum_{i=0}^l a_i \al_i,$$ where $\al_i, i=0,\ldots,l$, are the simple
roots of $\g$, and the $a_i$'s are the labels of the Dynkin diagram of
$\g$ \cite{Kac}. In particular, $a_0=1$ for all affine algebras, except
$A^{(2)}_{2n}$, in which case $a_0=2$. Somewhat abusing notation, we
will introduce a vector $\al_0$ in the Cartan subalgebra $\h$ of $\g$
by the formula $$\al_0 = - \frac{1}{a_0} \sum_{i=1}^l a_i \al_i.$$

We can now define the hamiltonian of the affine Toda field theory,
associated to $\g$, by the formula $$H = \frac{1}{2} \sum_{i=0}^l \int
e^{\phi_i(t)} dt \in \oplus_{i=0}^l \F_{\al_i},$$ where $\phi_i(t) =
\bar{\al}_i(t)$ (cf. \secref{functspace}). The corresponding hamiltonian
equation coincides with the Toda equation \eqref{toda}, associated to $\g$.

We define the space $I_0(\g)$ of local integrals of motion as the
kernel of the operator $\{\cdot,H\}: \F_0 \arr \oplus_{i=0}^l
\F_{\al_i}$, or, in other words, as the intersection of the kernels
of the operators $\Q_i = \{\cdot,\int e^{\phi_i(t)} dt\}: \F_0 \arr
\F_{\al_i}, i = 0,\ldots,l.$

As before, we will use the operators $\q_i: \pi_0 \arr \pi_{\al_i},
i=0,\ldots,l$, given by formula \eqref{operqi}, where we put $$x^0_n = -
\frac{1}{a_0} \sum_{i=1}^l a_i x^i_n.$$ These operators commute with the
action of the derivative and descend down to the operators $\Q_i,
i=0,\ldots,l$. We will also need the operators $Q_i = T_i^{-1} \q_i$,
acting on $\pi_0$.

According to \propref{serre} and \remref{serrerem}, the operators
$\q_i, i=0,\ldots,l$, as well as $Q_i, i=0,\ldots,l$, satisfy the
Serre relations of the affine algebra $\g$. Thus, the operators $Q_i$
generate an action of the nilpotent subalgebra $\n_+$ of $\g$ on the
space $\pi_0$.

In the rest of this subsection we will go through the main steps of
\secref{classical} and \secref{sinegor} to describe the space $I_0(\g)$.
Most of the proofs are the same as in those sections.

\subsubsection{The complex $F^*(\g)$.}
For an affine algebra $\g$ there also exists a BGG resolution, which is
defined in the same way and has the same properties as in the case of
finite-dimensional simple Lie algebras \cite{rocha}.

Using this resolution, we can define the complex $$F^*(\g) = \oplus_{j\geq
0} F^j(\g)$$ in the same way as in \secref{extended}, by putting $$F^j(\g)
= \oplus_{l(s)=j} \pi_{\rho-s(\rho)},$$ where $s$ runs over the affine Weyl
group. The differential $\delta^j: F^{j-1}(\g) \arr F^j(\g)$ is given by
formula \eqref{differential}. In the same way as in the case of
finite-dimensional simple Lie algebras (cf. \lemref{nilpotent}), it follows
that this differential is nilpotent.

Introduce a $\Z-$grading on this complex similar to the one introduced
in \secref{zgrading}. Namely, we put $\deg v_{\rho-s(\rho)} =
(\rho^\vee,\rho-s(\rho))$, where $\rho^\vee$ is an element in the dual
space to the Cartan subalgebra of $\g$, such that $(\rho^\vee,\al_i)=1,
i=0,\ldots,l$, and $\deg x^i_n = -n$. With respect to this grading,
the differentials of our complex have degree $0$, and our complex
decomposes into a direct sum of finite-dimensional subcomplexes.

One has the analogue of \propref{cohom}: the cohomologies of the
complex $F^*(\g)$ are isomorphic to the cohomologies of $\n_+$ with
coefficients in $\pi_0$.

\subsubsection{Principal commutative subalgebra.}
Consider the element $p = \sum_{i=0}^l a_i e_i$ in the Lie algebra
$\n_+$. It is known that $\n_+ = \on{Ker(ad} p) \oplus \on{Im(ad} p)$,
where $\on{Ker(ad} p)$ is an abelian Lie subalgebra $\ab$ of $\n_+$,
which we will call the principal commutative subalgebra.

In the principal gradation of $\n_+$, which is obtained by assigning
degree $1$ to the generators $e_i$, the Lie algebra $\ab$ has a basis of
homogeneous elements of degrees equal to the exponents $d_i,
i=1,\ldots,l$, of $\g$ modulo the Coxeter number $h$.

The space $\on{Ker(ad} p)$ splits into the direct sum $\on{Ker
(ad} p) = \oplus_{j>0} \n_+^j$ of homogeneous components of degree $j$
with respect to the principal grading. Each of these components has
dimension $l$ and the operator $\on{ad} p: \n_+^j \arr \n_+^{j+1}$ is
an isomorphism.

For the proof of these facts, cf. \cite{kac}, Proposition 3.8 (b). We will
use them in the proof of the following statement, which is a generalization
of \propref{wedge}.

\subsubsection{Proposition.}    \label{wedgegen}
{\em The cohomologies of the complex $F^*(\g)$ are isomorphic to the
exterior algebra $\bigwedge^*(\ab^*)$ of the dual space to the pricipal
commutative subalgebra $a$.}

\vspace{3mm}
\noindent {\em Proof.} From the formulas, defining the operators $Q_i$,
one can see that the operator $p = \sum_{i=0}^l a_i Q_i$ does not have a
shift term (cf. the proof of \propref{wedge}). Therefore, other elements of
the Lie subalgebra $\ab$ of $\n_+$ also do not have shift terms.

Denote by $V_j$ the vector space of operators $\pa/\pa x^i_{-j},
i=1,\ldots,l, j>0$. One can check that the operator $\mbox{ad} p$
isomorphically maps $V_j$ to $V_{j+1}$. The space of shift terms of the
operators from $\n_+^1 \subset \n_+$ coincides with $V_1$. By induction, in
as in the proof of \propref{wedge}, one can show that the space of shift
terms of operators from $\n_+^j \subset \n_+$ coincides with $V_j$.

In the same way as in the proof of \propref{wedge}, we deduce from these
facts that the $\n_+-$module $\pi_0$ is isomorphic to the module, coinduced
from the trivial representation of $\ab$. Therefore the cohomology
$H^*(\n_+,\pi_0)$, are equal, by ``Shapiro's lemma'', to $H^*(\ab,\C)
\simeq \bigwedge^*(\ab^*)$, because $\ab$ is an abelian Lie algebra. But
$H^*(\n_+,\pi_0)$ coincides with the cohomology of the complex
$F^*(\g)$. The Proposition is proved.

\subsubsection{Theorem.}    \label{intaffine}
{\em The space $I_0(\g)$ of local integrals of motion of the Toda field
theory, associated to an affine algebra $\g$, is linearly generated by
mutually commuting local functionals of degrees equal to the exponents of
$\g$ modulo the Coxeter number.}

\vspace{3mm}
\noindent {\em Proof.} If the exponents of $\g$ are odd and the Coxeter
number is even, then the proof simply repeats the proof of
\thmref{intsine}. We can again identify the space $I_0(\g)$ with the
$1$st cohomology of the double complex $\C \larr F^*(\g) \larr F^*(\g)
\larr \C$. According to \propref{wedgegen}, the $1$st cohomology of the
complex $F^*(\g)$ is linearly spanned by elements of certain odd
degrees. We can then compute this cohomology using the spectral
sequence in the same way as in the proof of \thmref{intsine}.

The proof in other cases ($A_n^{(1)}, n>1, D_{2n}^{(1)}, E_6^{(1)}$
and $E_7^{(1)}$) will be published in \cite{kdv}.

\subsubsection{Integrals of motion in the extended space of local
functionals.} We can define the space of integrals of motion of an affine
Toda field theory as the intersection of the kernels of operators $\Q_i:
\f_0 \arr \f_{\al_0}$, cf. \secref{extendedspace}. We can use the
tensor product of the complex $F^*(\g)$ with $\C[t,t^{-1}]$ to compute
this space. By repeating the proof of \thmref{intaffine}, we conclude
that the space of integrals of motion in the larger space $\f_0$ of
local functionals coincides with the space $I_0(\g)$ of integrals of
motion in the small space $\F_0$.

\subsubsection{Remark.}    \label{conslaws}
One usually defines a conservation law of the Toda field theory (in
the light cone coordinates $x_+$ and $x_-$) as a pair $(P^-,P^+) \in
\pi_0 \oplus (\oplus_i \pi_{\al_i})$, such that $$\frac{\pa P^-}{\pa x_+}
 = \frac{\pa P^+}{\pa x_-}.$$ In our language, $\pa/\pa x_+ =
\sum_i \q_i = \delta^1$ is the Poisson bracket with the Toda hamiltonian, and
$\pa/\pa x_- = \pa$. Thus, the pair $(P^-,P^+)$, where $P^+$ is a
$1$-cocycle of the complex $F^*(\g)$ and $P^-$ is the density of our
integral of motion, i.e. the $0$-cocycle of the quotient complex
$F^*(\g)/\pa F^*(\g)$ such that $\delta^1 P^- = \pa P^+$, is a conservation
law. The important observation, which enabled us to find all such
conservation laws was that in order to be a component of a conservation
law, $P^+$ should satisfy a certain equation, namely, $\delta^2
\cdot P^+ = 0$.

The space $I_0(\g)$ constitutes a maximal abelian Poisson subalgebra in the
classical ${\cal W}-$algebra $I_0(\bar{\g})$. One can show that the first
generators of $I_0(\g)$ of degrees $d_i, i=1,\ldots,l$, are equal to $\int
W^{(i)}(t) dt$, where the $W^{(i)}$'s can be chosen as the generators of
$I_0(\bar{\g})$, defined in \secref{exist}.

Using this fact, one can identify the local integrals of motion of the
classical affine Toda field theory, associated with $\g$, with the
hamiltonians of the corresponding generalized KdV and mKdV
hierarchies.

\subsubsection{Remark.}
In the course of proving \propref{wedgegen}, we showed that the
$\n_+-$module $\pi_0$ is isomorphic to the module, coinduced from the
trivial representation of $\ab$. It means that $\pi_0 \simeq \C[\pa^n u^i]$
is isomorphic to the space of functions on the homogeneous space $N_+/A$,
where $N_+$ and $A$ are Lie groups of the Lie algebras $\n_+$ and $\ab$,
respectively. In other words, the space $N_+/A$ can be identified with an
infinite-dimensional vector space with coordinates $\pa^n u^i$. The local
integrals of motion of the corresponding Toda field theory therefore define
certain vector fields $\eta_j$'s on $N_+/A$ (cf.  formula
\eqref{vectfield}), which commute among themselves. The first of them,
$\eta_1$, is the vector field $\pa$.

In our paper \cite{kdv} we identify the vector field $\eta_j$ with the
infinitesimal right action of a generator of degree $-j$ of the opposite
abelian subalgebra $\ab_-$ of $\g$. Indeed, since $\g$ infinitesimally acts
on $N_+$ from the right as on the big cell of the flag manifold, and
$\ab_-$ commutes with $\ab$, the Lie algebra $\ab_-$ acts on the coset
space $N_+/A$ by mutually commuting vector fields. In particular, the
action of the vector field $\pa$ corresponds to the action of the element
$$\sum_{i=0}^l \frac{(\al_i,\al_i)}{2} f_i \in \ab_-,$$ in accordance with
\lemref{singular}.

In the next section we will show how to quantize the classical integrals of
motion.

\section{Quantum Toda field theories.}    \label{quantum}

\subsection{Vertex operator algebras.}    \label{quantumvoa}
The theory of vertex operator algebras was started by Borcherds
\cite{borcherds} and then further developed by I.Frenkel, Lepowsky and
Meurman in \cite{flm} (cf. also \cite{fhl,G,L}).

In this section we will give the definition of vertex operator algebras and
study some of their properties following closely \S~3 of \cite{FKRW}. Ater
that we will introduce the vertex operator algebra (VOA) of the Heisenberg
algebra (or the VOA of free fields), which we will need in the study of
quantum integrals of motion of Toda field theories.

\subsubsection{Fields.}
Let $V=\oplus_{n=0}^{\infty}V_n\/$ be a $\Bbb Z_{+}\/$-graded
vector space, where $\dim V_n < \infty\/$ for all $n\/$, called
the {\em space of states\/}. A {\em field\/} on $V\/$ of
conformal dimension $\Delta \in \Bbb Z\/$ is a power series $\phi
(z) = \sum_{j \in \Bbb Z} \phi_j z^{-j - \Delta}\/$, where
$\phi_j \in \mbox{End}\, V \/$ and $\phi_j V_n \subset
V_{n-j}\/$. Note that if $\phi(z)\/$ is a field of conformal
dimension $\Delta\/$, then the power series $\partial_z \phi (z)
= \sum_{j \in \Bbb Z} (-j - \Delta)\phi_j z^{-j - \Delta-1}\/$ is
a field of conformal dimension $\Delta +1\/$.

If $z \in \C^\times$ is a non-zero complex number, then $\phi(z)$ can be
considered as a linear operator $V \arr \bar{V}$, where
$\bar{V}=\prod_{n=0}^\infty V_n$.

We have a natural pairing $\langle,\rangle: V_n^{*} \times V_n
\rightarrow \C$. A linear operator $P: V \rightarrow
\bar{V}$ can be represented by a set of finite-dimensional linear
operators $P^{j}_{i}: V_i \arr V_j, i,j \in
\Z$, such that $\langle A, P \cdot
B\rangle = \langle A, P^{j}_{i} \cdot B \rangle$ for $B \in V_i, A
\in V_j^{*}$. Let $P, Q$ be two linear operators $V \rightarrow
\bar{V}$. We say that the {\em composition} $PQ$ {\em exists}, if for any
$i,k \in \Z, B \in V_i, A \in V_k^*$ the series $\sum_{j \in \Z}
\langle A, P^{k}_{j} Q^{j}_{i} \cdot B \rangle$ converges absolutely.

\subsubsection{Definition.} {\em Two fields $\phi (z)\/$ and $\psi(z)\/$
are called $\on{local}$ with respect to each other, if}

\begin{itemize}
\item {\em for any $z, w \in \C^\times$, such
that $|z|>|w|$, the composition $\phi(z) \psi(w)$ exists and can be
analytically continued to a rational operator-valued function on}
$(\C^{\times})^{2}\backslash diagonal$, $R(\phi(z) \psi(w))$;

\item {\em for any $z, w \in \C^\times$, such that $|w|>|z|$, the
composition $\psi(w) \phi(z)$ exists and can be analytically continued to a
rational operator-valued function on} $(\C^{\times})^{2}\backslash
diagonal$, $R(\psi(w) \phi(z))$;

\item $R(\phi(z) \psi(w)) = R(\psi(w) \phi(z))$.
\end{itemize}

In other words, fields $\phi (z)\/$ and $\psi(z)\/$ are local with respect
to each other, if for any $x \in V_n\/$ and $y \in V_m^*\/$ both matrix
coefficients $\langle y |\phi (z) \psi(w) | x
\rangle\/$ for $|z| > |w|\/$ and $\langle y |\psi(w)\phi (z)|x\rangle\/$
for $|z| < |w|\/$ converge to the same rational function in $z\/$ and $w\/$
which has no poles outside the lines $z=0\/$, $w=0\/$ and $z=w\/$.

\subsubsection{Definition.}  A VOA structure {\em on $V\/$ is a linear map
$Y(\cdot,z) : V \longrightarrow \on{End}\, V
\left[\left[z,z^{-1}\right]\right]\/$ which associates to each $A \in
V_n\/$ a field of conformal dimension $n\/$ (also called a \mbox{vertex
operator}) $Y(A,z) = \sum_{j \in \Bbb Z}A_j z^{-j-n}\/$, such that the
following axioms hold:

{\em {\bf (A1)} (vacuum axiom)} There exists an element $|0\rangle \in
V_0\/$ ({\em vacuum vector}) such that $Y(|0\rangle ,z) = \on{Id}\/$ and
$\lim_{z \rightarrow 0} Y(A,z)|0\rangle = A\/$.

{\em {\bf (A2)} (translation invariance)} There exists an operator $T
\in \on{End}\, V\/$ such that
$$
\partial_z Y(A,z) = [T,Y(A,z)] \quad \quad \text{and} \quad \quad
T|0\rangle = 0.
$$

{\em {\bf (A3)} (locality)} All fields $Y(A,z)\/$ are local
with respect to each other.}

{\em A VOA $V\/$ is called {\em conformal\/} of {\em central charge} $c \in
\Bbb C\/$ if there exists an element $\omega \in V_2\/$ (called the {\em
Virasoro element\/}), such that the corresponding vertex operator
$Y(\omega,z)=\sum_{n\in \Bbb Z} L_n z^{-n-2}\/$ satisfies the following
properties:}

{\bf (C)} $L_{-1} = T\/$, $L_0 |_{V_n} = n \cdot \on{Id}\/$, and
$L_2 \omega = \frac 12 c | 0 \rangle\/$.

\subsubsection{Proposition.}    \label{assocprop}
{\em A VOA $V$ automatically satisfies the associativity property: for any
$A, B \in V$,
\begin{equation}
  R(Y(A,z) Y(B,w)) = R(Y(Y(A,z-w)B,w)),
\label{eq:assoc}
\end{equation}
where the left-hand (resp.\ right-hand) side is the analytic
continuation from the domain $\left|z \right| > \left|w
\right|\/$ (resp.\ $ |w| > |z-w|\/$)}.

\vspace{3mm}
\noindent {\em Proof} which was communicated to us by V.~Kac follows from
the following two Lemmas.

\subsubsection{Lemma.}    \label{conj}
{\em For any $z \in \C^\times$ and $w \in \C$, such that $|w| < |z|$, the
composition $e^{wT} Y(A,z) e^{-Tw}$, where $$e^{wT} = \sum_{n=0}^\infty
\frac{(wT)^n}{n!},$$ exists and can be analytically continued to a rational
operator-valued function, which is equal to $Y(A,z+w)$, i.e.} $$R(e^{wT}
Y(A,z) e^{-Tw}) = Y(A,z+w).$$

\vspace{3mm}
\noindent {\em Proof.} By axiom {\bf (A2)}, $[T,Y(A,z)] = \pa_z Y(A,z)$.
Hence as a formal powers series, $$e^{wT} Y(A,z) e^{-Tw} =
\sum_{n=0}^\infty \frac{w^n}{n!} \pa_z^n Y(A,z).$$ The right hand side is
the Taylor expansion of $Y(A,z+w)$. Since $Y(A,z)$ is holomorphic
everywhere except the origin, $e^{wT} Y(A,z) e^{-Tw}$ converges to
$Y(A,z+w)$ if $|w| < |z|$.

\subsubsection{Lemma.}    \label{permutation}
{\em For any $z \in \C^\times$, $Y(A,z) B = e^{zT} Y(B,-z) A$.}

\vspace{3mm}
\noindent {\em Proof.} By \lemref{conj}, $$R(e^{(z+w)T} Y(B,-z) A) =
R(Y(B,w) e^{(z+w)T} A).$$ We can derive from axioms {\bf (A1)} and {\bf
(A2)} that $e^{(z+w)T} A = Y(A,z+w) |0\rangle$. This and the previous
formula give: $$R(e^{(z+w)T} Y(B,-z) A) = R(Y(B,w) Y(A,z+w) |0\rangle).$$
By locality, $$R(e^{(z+w)T} Y(B,-z) A) = R(Y(A,z+w) Y(B,w) |0\rangle).$$
Both sides of the last formula have well-defined limits when $w \arr 0$,
hence these limits coincide and we obtain using axiom {\bf (A1)}: $e^{Tz}
Y(B,-z) A = Y(A,z) B$.

\subsubsection{Proof of \propref{assocprop}.} For any vector $C \in
V$ we have: $$Y(A,z) Y(B,w) C = Y(A,z) e^{wT} Y(C,-w) B$$ for $|z| > |w|$,
where we applied \lemref{permutation} to $Y(B,w) C$. By applying $1 =
e^{Tw} e^{-Tw}$ to both sides of this formula (we can do this, because
$e^{\pm Tw}$ is a series infinite in only one direction) and using
\lemref{conj} in the right hand side of the previous formula we obtain
\begin{equation}    \label{lhs}
Y(A,z) Y(B,w) C = e^{wT} Y(A,z-w) Y(C,-w) B
\end{equation}
for $|z| > |w|$.

On the other hand, consider $$Y(Y(A,z-w)B,w) C = \sum_{n\in\Z}
(z-w)^{-n-\Delta_A} Y(A_n \cdot B,w) C$$ as a formal power series in
$(z-w)$. By \lemref{permutation}, $$Y(A_n \cdot B,w) C = e^{wT} Y(C,-w) A_n
B.$$ Hence
\begin{equation}    \label{rhs}
Y(Y(A,z-w)B,w) C = e^{wT} Y(C,-w) Y(A,z-w) B,
\end{equation}
as formal power series in $(z-w)$. But the right hand side of
\eqref{rhs} converges when $|w| > |z-w|$ and can be analytically
continued to a rational function in $z$ and $w$, by axiom {\bf (A3)}.
Therefore the left hand side of \eqref{rhs} has the same properties.

By applying axiom {\bf (A3)} to the analytic continuations of the right
hand sides of \eqref{lhs} and \eqref{rhs}, we obtain the equality of the
analytic continuations of the left hand sides: $$R(Y(A,z) Y(B,w)) C =
R(Y(Y(A,z-w)B,w)) C$$ for any $C \in V$, and \propref{assocprop} follows.

\subsubsection{Operator product expansion.}
We may rewrite formula \eqref{eq:assoc} as $$Y(A,z) Y(B,w) = \sum_{n \in
\Z} (z-w)^{-n-\deg A} Y(A_{n} \cdot B,w),$$ using the formula
$Y(A,z)=\sum_{n \in \Z} A_{n} z^{-n-\deg A}$. Here and further on to
simplify notation we omit $R(\cdot)$ in formulas for analytic continuation
of functions. Such an identity is called an operator product expansion
(OPE). There exists such $M \in \Z$ that $A_{n} \cdot B=0$ for any $n >
M$. Therefore, the right hand side of this formula has only finitely many
terms with negative powers of $(z-w)$. Combining axioms (3), (4), and the
Cauchy theorem we obtain the following identity \cite{flm}:

$$\int_{C^{\rho}_w} \int_{C^{R}_z} Y(A,z) Y(B,w) f(z,w)
\, dz dw - \int_{C^{\rho}_w}
\int_{C^{r}_z} Y(B,w) Y(A,z) f(z,w) \, dz dw=$$
$$\int_{C^{\rho}_w}
\int_{C^{\delta}_z(w)} \sum_{n \in \Z} (z-w)^{-n-\deg A} Y(A_{n}
\cdot B,w) f(z,w) \, dz dw,$$
where $C^{x}$ denotes a circle of radius $x$ around the origin,
$R>\rho>r$, $C^{\delta}(w)$ denotes a small circle of radius $\delta$
around $w$, and $f(z,w)$ is an arbitrary rational function on
$(\C^{\times})^{2}\backslash diagonal$.

This formula can be used in order to compute commutation relations between
Fourier components of vertex operators. Indeed, if we choose $f(z,w) =
z^{m+\deg A-1} w^{k+\deg B-1}$, then we obtain:
$$[A_m,B_k] = \sum_{n>-\deg A} \int_{C^{\rho}_w} \frac{dw}{w}
w^{k+\deg B} \int_{C^{\delta}_z(w)} dz \frac{z^{m+\deg
A-1}}{(z-w)^{n+\deg A}} Y(A_n \cdot B,w) =$$
\begin{equation}    \label{commutator}
\sum_{-\deg A<n\leq m} \left( \begin{array}{c}
                        m+\deg A-1 \\ n+\deg A-1
                        \end{array} \right) (A_n \cdot B)_{m+k}.
\end{equation}
In particular, we see that only the terms in the OPE, which are singular at
the diagonal $z=w$, contribute to the commutator. This formula also shows
that the commutator of Fourier components of two fields is a linear
combination of Fourier components of other fields (namely, the ones,
corresponding to the vectors $A_n \cdot B$). Therefore we obtain the
following result.

\subsubsection{Theorem.}    \label{liealgebra}
{\em The space of all Fourier components of vertex operators defined by a
VOA is a Lie algebra.}

In particular we derive from formula \eqref{eq:assoc} and axiom {\bf (C)}
that $$Y(\omega,z) Y(\omega,w) = \frac{c/2}{(z-w)^4} + \frac{2
Y(\omega,w)}{(z-w)^2} + \frac{\pa_w Y(\omega,w)}{z-w} + O(1),$$ which
implies using \eqref{commutator} the following commutation relations
between the Fourier coefficients $L_n, n \in \Z$, of $Y(\omega,z)$:
$$[L_n,L_m] = (n-m) L_{n+m} + \frac{1}{12} (n^3-n) \delta_{n,-m} c.$$ These
are the defining relations of the Virasoro algebra with central charge $c$.

\subsubsection{Remark.} The axioms of VOA may look rather complicated,
but in fact they are quite natural generalizations of the axioms of a
$\Z-$graded associative commutative algebra with a unit. Indeed, such an
algebra is defined as a $\Z-$graded vector space $V$ along with a linear
operator $Y: V \arr \mbox{End} (V)$ of degree $0$ and an element ${\bold 1}
\in V$, such that $Y({\bold 1}) = \on{Id}$. The linear operator $Y$
defines a product structure by the formula $A \cdot B = Y(A)B$. The axioms
of commutativity and associativity of this product then read as follows:
$Y(A) Y(B) = Y(B) Y(A)$, and $Y(Y(A)B) = Y(A) Y(B)$.

On a VOA $V$ the operator $Y(\cdot,z)$ defines a family of ``products'' --
linear operators $V \arr V$, depending (in the formal sense) on a complex
parameter $z$. The axiom {\bf (A3)} and formula \eqref{eq:assoc} can be
viewed as analogues of the axioms of commutativity and associativity (with
a proper regularization of the compositions of operators), and the first
half of the axiom {\bf (A1)} is the analogue of the axiom of unit. A rather
surprising aspect of the theory of VOA is that the ``associativity''
property \eqref{eq:assoc} follows from the ``commutativity'' (locality)
{\bf (A3)} together with {\bf (A1)} and {\bf (A2)}.

Another novelty is vector $\omega$. The meaning of this vector is the
following. The Fourier components of $Y(\omega,z)$ define on $V$ an action
of the Virasoro algebra, which is the central extension of the Lie algebra
of vector fields on a punctured disc. The existence of such an element
inside $V$ means that all infinitesimal changes of the coordinate $z$ can
be regarded as ``interior automorphisms'' of $V$.

\subsubsection{Ultralocal fields.}
Let us call two fields $\phi(z)$ and $\psi(z)$ {\em ultralocal} with
respect to each other if there exists an integer $N$, such that for any $v
\in V_n$ and $v^* \in V_m^*$, both series $\langle v^* |\phi (z)
\psi(w) | v \rangle\/ (z-w)^N$ and $\langle v^* |\psi (w)
\phi(z) | v \rangle\/ (z-w)^N$ are equal to the same finite polynomial in
$z^{\pm 1}$ and $w^{\pm 1}$. Clearly, ultralocality implies
locality. Moreover, in a vertex operator algebra any two vertex operators
are automatically ultralocal with respect to each other according to
formula (\ref{eq:assoc}) and the fact that the $\Bbb Z$--gradation on $V$
is bounded from below.

Given two fields $\phi (z)\/$ and $\psi(z)\/$ of conformal dimensions
$\Delta_\phi$ and $\Delta_\psi$ one defines their {\em normally ordered
product\/} as the field
\begin{equation}
  :\phi(z)\psi(z): = \sum_{n\in\Z} \left( \sum_{m<-\Delta_\phi} \phi_m
  \psi_{n-m} + \sum_{m\geq -\Delta_\phi} \psi_{n-m} \phi_m \right)
  z^{-n-\Delta_\phi-\Delta_\psi}
\label{eq:normal}
\end{equation}
of conformal dimension $\Delta_\phi+\Delta_\psi$.
The Leibniz rule holds for the normally ordered product:
\begin{equation}
  \partial_z :\phi (z)\psi(z): =
  :\partial_z \phi (z)\psi(z): +:\phi (z)\partial_z \psi(z):.
\label{eq:rule}
\end{equation}

The following proposition proved in \cite{FKRW}, Proposition 3.1, allows
one to check easily the axioms of a VOA.

\subsubsection{Proposition.}    \label{prop:generator}
  {\em Let $V$ be a $\Bbb Z_{+}$-graded vector space.  Suppose that to some
  vectors $a^{(0)} = |0\rangle \in V_0, a^{(1)} \in V_{\Delta_1}, \ldots$,
  one associates fields $Y(|0\rangle ,z) = \on{Id}, Y(a^{(1)}, z) =
  \sum_j a_j^{(1)} z^{-j-\Delta_1}$, $\ldots$, of conformal dimensions $0,
  \Delta_1, \ldots$, such that the following properties hold:

  {\em {\bf (1)}} all fields $Y(a^{(i)}, z)$ are ultralocal with respect to
  each other;

  {\em {\bf (2)}} $ \lim_{z\rightarrow 0} Y(a^{(i)}, z)|0\rangle
  = a^{(i)}$;

  {\em {\bf (3)}} the space $V$ has a linear basis of vectors
  \begin{equation}
    a^{(k_s)}_{-j_s - \Delta_{k_s}} \ldots a^{(k_1)}_{-j_1 -
      \Delta_{k_1}}|0\rangle ,
      \,\,\, j_1, \ldots, j_s \in \Bbb Z_{+};
    \label{eq:span}
  \end{equation}

  {\em {\bf(4)}} there exists an endomorphism $T\/$ of $V$ such that
  \begin{equation}
    \left[T, a^{(k)}_{-j-\Delta_k} \right] =
      (j+1)a^{(k)}_{-j-\Delta_k - 1},
    \quad
    T(|0\rangle ) =0.
    \label{eq:eq}
  \end{equation}
  Then letting
    \begin{multline}    \label{eq:norm}
      Y(a^{(k_s)}_{-j_s - \Delta_{k_s}} \ldots
        a^{(k_1)}_{-j_1 - \Delta_{k_1}} |0\rangle , z) \\
      =  (j_1! \cdot \ldots \cdot j_s!)^{-1} \cdot
      :\partial_z^{j_s}Y(a^{(k_s)},z) \ldots \partial_z^{j_2}
      Y(a^{(k_2)},z) \partial_z^{j_1} Y(a^{(k_1)},z):
    \end{multline}
  (where the normal ordering of more than two fields is nested from right
  to left), gives a well-defined VOA structure on $V$.}

\vspace{3mm}
\noindent {\em Proof}. We define the map $Y (\cdot, z)\/$ by formula
(\ref{eq:norm}). It is clear that axiom {\bf (A1)} holds. Given two fields
$\phi (z)\/$ and $\psi(z)\/$, if $[T, \phi (z) ] = \partial_z \phi (z)\/$
and $[T, \psi (z) ] = \partial_z \psi (z)\/$, then from (\ref{eq:normal})
and (\ref{eq:rule}) it follows that
\begin{equation}
  [T, :\phi (z) \psi (z) :] = \partial_z : \phi (z)\psi (z) :.
\label{eq:ind}
\end{equation}
Hence the axiom {\bf (A2)} follows inductively from (\ref{eq:norm}) and
(\ref{eq:ind}).

Using an argument of Dong (cf. \cite{L}, Proposition 3.2.7), one can
show that if three fields $\chi (z)\/$, $\phi (z)\/$ and $\psi(z)\/$ are
ultralocal with respect to each other, then $:\phi (z)\psi(z)\/$: and $\chi
(z)\/$ are ultralocal.

Indeed, by assumption, there exists such $r \in \Z_+$ that
\begin{equation}    \label{ultra}
(w-z)^s \phi(z) \psi(w) = (w-z)^s \psi(w) \phi(z),
\end{equation}
\begin{equation}    \label{ultra1}
(u-z)^s \phi(z) \chi(u) = (u-z)^s \chi(u) \phi(z),
\end{equation}
\begin{equation}    \label{ultra2}
(u-w)^s \psi(w) \chi(u) = (u-w)^s \chi(u) \psi(w),
\end{equation}
for any $s\geq r$.

Consider the formal power series in $z,w$ and $u$:
\begin{equation}    \label{three}
(w-u)^{3r} \left[ (z-w)^{-1} \phi(z) \psi(w) - (z-w)^{-1} \psi(w)
\phi(z) \right] \chi(u),
\end{equation}
where $(z-w)^{-1}$ is considered as a power series in $w/z$ in the first
summand and as a power series in $z/w$ in the second summand.

This series is equal to $$(w-u)^r \sum_{s=0}^{2r} \left( \begin{array}{c}
2r \\ s \end{array} \right) (w-z)^s (z-u)^{2r-s} \left[ (z-w)^{-1} \phi(z)
\psi(w) - (z-w)^{-1} \psi(w) \phi(z) \right] \chi(u).$$ The terms with $r <
s \leq 2r$ in the last formula vanish by
\eqref{ultra}. Hence we can rewrite it as $$(w-u)^r \sum_{s=0}^{r} \left(
\begin{array}{c} 2r \\ s \end{array} \right) (w-z)^s
(z-u)^{2r-s} \left[ (z-w)^{-1} \phi(z) \psi(w) - (z-w)^{-1} \psi(w)
\phi(z) \right] \chi(u),$$ and further as $$(w-u)^r \sum_{s=0}^{r} \left(
\begin{array}{c} 2r \\ s \end{array} \right) (w-z)^s
(z-u)^{2r-s} \chi(u) \left[ (z-w)^{-1} \phi(z) \psi(w) - (z-w)^{-1}
\psi(w) \phi(z) \right],$$ using \eqref{ultra1} and \eqref{ultra2}. By
the same trick as above, we see that \eqref{three} is equal to
\begin{equation}    \label{threeprime}
(w-u)^{3r} \chi(u) \left[ (z-w)^{-1} \phi(z) \psi(w) - (z-w)^{-1}
\psi(w) \phi(z) \right].
\end{equation}

Now it follows from the definition of normal ordering \eqref{eq:normal}
that the series $:\phi(w)\psi(w):$ is the coefficient of $z^{-1}$ in the
series $$(z-w)^{-1} \phi(z) \psi(w) - (z-w)^{-1} \psi(w) \phi(z),$$ where
again $(z-w)^{-1}$ is considered as a power series in $w/z$ in the first
summand and as a power series in $z/w$ in the second summand.

Hence if we take the coefficients in front of $z^{-1}$ in formulas
\eqref{three} and \eqref{threeprime}, we obtain the following equality of
formal power series in $w^{\pm 1}$ and $u^{\pm 1}$: $$(w-u)^{3r}
:\phi(w)\psi(w):\chi(u) = (w-u)^{3r} \chi(u):\phi(w)\psi(w):.$$

But since the gradation on $V$ is bounded from below, the matrix
coefficients of the left hand side are finite in $u^{-1}$ and $w$, whereas
the matrix coefficients of the right hand side are finite in $w^{-1}$ and
in $u$. Therefore both are polynomials in $w^{\pm 1}$ and $u^{\pm 1}$, and
so $:\phi(z)\psi(z):$ and $\chi(z)$ are ultralocal.

It is also clear that if $\phi (z)\/$ and $\psi (z)\/$ are ultralocal, then
$\partial_z \phi (z)\/$ and $\psi (z)\/$ are ultralocal. This implies axiom
{\bf (A3)} and completes the proof.

\subsection{The VOA of the Heisenberg algebra}    \label{voaheis}

\subsubsection{}
Let $\h$ be the Cartan subalgebra of a simple finite-dimensional Lie
algebra $\g$. In \secref{kirkos} we defined the Heisenberg algebra
$\widehat{\h}$. It has generators $b^i_n, i=1,\ldots,l, n \in \Z$, and
relations $$[b^i_n,b^j_m] = n (\al_i,\al_j) \delta_{n,-m}.$$ In the case
$\g=\sw_2$ we will have generators $b_n, n\in\Z$, and relations $$[b_n,b_m]
= n \delta_{n,-m}.$$

For $\la \in \h^*$, let $\pi_\la$ be the Fock representation of
$\widehat{\h}$, which is freely generated by the operators $b^i_n,
i=1,\ldots,l, n \in \Z$, from a vector $v_\la$, such that $$b^i_n v_\la =
0, n>0, \quad \quad b^i_0 v_\la = (\la,\al_i) v_\la.$$ The fact that we
used the same notation $\pi_\la$ for these representations as for the
spaces of differential polynomials in \secref{classical} will be justified
later on.

We want to introduce a structure of VOA on $\pi_0$.

First we introduce a $\Z-$grading on $\pi_0$ by putting $\deg v_0 = 0, \deg
b(n) = -n,$ so that $\deg b(n_1) \ldots b(n_m) v_0 = - \sum_{i=1}^m
n_i$.

Next we introduce a linear map $Y(\cdot,z)$ from $\pi_0$ to
End$\pi_0[[z,z^{-1}]]$.

Defining the operator $Y(\cdot,z)$ amounts to assigning to each homogeneous
vector $A \in \pi_0$ a formal power series $$Y(A,z) = \sum_{n \in \Z} A_n
z^{-n-\deg A}.$$ In this formula, the Fourier component $A_n$ stands
for some linear operator of degree $-n$, acting on $\pi_0$.

Let us define these operators by explicit formulas. The degree $0$ subspace
of the module $\pi_0$ is spanned by one vector: $v_0$. The corresponding
operator $Y(v_0,z)$ is equal to the identity operator (times 1, and all
other Fourier components are equal to $0$).

In degree one we have $l$ linearly independent vectors: $b^i_{-1} v_0,
i=1,\ldots,l$. We assign to them the fields $$Y(b^i_{-1} v_0,z) = b^i(z)
= \sum_{n \in Z} b^i_n z^{-n-1}.$$ In degree $2$ the module $\pi_0$ has two
types of vectors: $b^i_{-2} v_0$, to which we assign $$Y(b^i_{-2} v_0,z) =
\pa_z b^i(z),$$ where $\pa_z = \pa/\pa z$; and $b^i_{-1} b^j_{-1} v_0$, to
which we assign
$$Y(b^i_{-1} b^j_{-1} v_0,z) = :b^i(z) b^j(z): = \sum_{n\in\Z} \left(
\sum_{m\in\Z} :b^i(m) b^j(n-m): \right) z^{-n-2}.$$ Here the columns denote
the {\em normal ordering}. If we have a monomial in $b^i_n$'s, its
normal ordering is the ordering, in which all ``creation'' operators
$b^i_n, n<0$, are to the left of all ``annihilation'' operators $b^i_n,
n\geq 0$.

The general formula for a monomial basis element $b^{i_1}_{n_1} \ldots
b^{i_m}_{n_m} v_0$ of $\pi_0$ is the following:
\begin{multline}    \label{current}
Y(b^{i_1}_{n_1} \ldots b^{i_m}_{n_m} v_0,z) = \\ \frac{1}{(-n_1-1)!}
\ldots \frac{1}{(-n_m-1)!} :\pa_z^{-n_1-1} b^{i_1}(z) \ldots
\pa_z^{-n_m-1} b^{i_m}(z):.
\end{multline}

By linearity, we can extend the map $Y(\cdot,z)$ to any vector of
$\pi_0$.

Finally, we put $|0\rangle=v_0$ and
\begin{equation}    \label{TT}
T = \frac{1}{2} \sum_{i=1}^l \sum_{n \in \Z} :b^i_n b^{i*}_{-n-1}:,
\end{equation}
where $b^{i*}_n$ are the dual generators of the Heisenberg algebra,
i.e. the following commutation relations hold: $$[b^i_n,b^{j*}_m] = n
\delta_{i,j} \delta_{n,-m}.$$ We also have: $$[\pa,b^i_n] = -n b^i_{n-1}.$$

\subsubsection{Theorem.}    \label{voa}
{\em The map $Y(\cdot,z)$ defined by formula \eqref{current} together with
the vector $|0\rangle$ and the operator $T$ define the structure of VOA on
$\pi_0$.}

\vspace{3mm}
\noindent {\em Proof.} Consider the space $\pi_0$ and put $a^{(0)} = v_0,
a^{(1)} = b(-1) v_0$.  Define also $T$ by formula \eqref{T}. Then all
conditions of \propref{prop:generator} will be satisfied. In particular,
the fact that $Y(a^{(1)},z)=b(z)$ is ultralocal with itself follows from
the computation in \secref{commutationrel}. Therefore, by
\propref{prop:generator}, the map $Y(\cdot,z)$ defined in \secref{voaheis}
satisfies the axioms of vertex operator algebra.

The VOA $\pi_0$ can be given a structure of conformal VOA with an arbitrary
central charge. Define for any $\gamma \in \h$ the elements $\omega_\gamma$
by the formula $$\omega_\gamma = \left( \frac{1}{2} \sum_{i=1}^l b^i_{-1}
b^{i*}_{-1} + \gamma_{-2} \right) v_0.$$ These elements satisfy the axiom
{\bf (C)} with $c=1-12 \| \gamma \|^2$.

\subsubsection{Commutation relations.}    \label{commutationrel}
Now we are going to compute the commutation relations in the Heisenberg
algebra $\widehat{\h}$ using formula \eqref{commutator}. First let us
compute $b^i(z) b(w)$ for $|z|>|w|$. We have to rewrite this composition as
a linear combination of well-defined, i.e. normally ordered, linear
operators on $\pi_0$. We have $$b^i(z) b^j(w) = \sum_{n,m \in\Z} b^i(n)
b^j(m) z^{-n-1} w^{-m-1} = \sum_{n<0,m\in\Z} b^i(n) b^j(m) z^{-n-1}
w^{-m-1} +$$
$$+ \sum_{n\geq 0,m\in\Z} b^i(m) b^j(n) z^{-n-1} w^{-m-1} + \sum_{n\geq
0,m\in\Z} [b^i(n),b^j(m)] z^{-n-1} w^{-m-1} =$$ $$= :b^i(z) b^j(w): +
\sum_{n>0} n (\al_i,\al_j) \frac{1}{z^2} \left ( \frac{w}{z} \right)^{n-1}
= :b^i(z) b^j(w): + \frac{(\al_i,\al_j)}{(z-w)^2}.$$ We can further rewrite
this as
\begin{equation}    \label{form}
b^i(z) b^j(w) = \frac{(\al_i,\al_j)}{(z-w)^2} + \sum_{n\geq 0} \frac{1}{n!}
(z-w)^n:\pa_w^n b^i(w) b^j(w):,
\end{equation}
by Taylor's formula.

On the other hand, we have: $b^i(n) \cdot b^j(-1) v_0 = 0$, for $n>1$ or
$n=0$, $b^i(1) \cdot b^j(-1) v_0 = (\al_i,\al_j) v_0$, and $b^i(n) \cdot
b^j(-1) v_0 = b^i(n) b^j(-1) v_0, n<0$. Hence we obtain
\begin{equation}    \label{ope}
Y(Y(b^i(-1) v_0,z-w) \cdot b^j(-1) v_0,w) = \sum_{n\in\Z} (z-w)^{-n-1}
Y(b^i(n) \cdot b^j(-1) v_0,w)
\end{equation}
$$= \frac{(\al_i,\al_j)}{(z-w)^2} +
\sum_{n\geq 0} \frac{1}{n!} (z-w)^n:\pa_w^n b^i(w) b^j(w):,$$
which coincides with formula \eqref{form}, in accordance with formula
\eqref{eq:assoc}. Note that the second computation is simpler.

Now we obtain using formula \eqref{commutator}: $$[b^i(n),b^j(m)] =
\int_{C^{\rho}_w} dw w^m \int_{C^{\delta}_z(w)} dz z^n
\frac{(\al_i,\al_j)}{(z-w)^2} = n (\al_i,\al_j) \int dw w^{n+m-1} = n
(\al_i,\al_j) \delta_{n,-m},$$ in accordance with the commutation relations
of $\widehat{\goth{h}}$.

\subsubsection{Bosonic vertex operators.}
Other Fock representations $\pi_\la$ carry the structure of a module
over the VOA $\pi_0$ (for general definition, cf. \cite{fhl}). The
Fourier components of the vertex operators $Y(A,z)$, given by formula
\eqref{current}, define linear operators on any of the modules
$\pi_\la$, and so we obtain maps $\pi_0 \arr \mbox{End} (\pi_\la)
[[z,z^{-1}]]$, which satisfy axioms similar to the axioms of VOA.

There is also another structure: a map $\pi_\la \arr \mbox{Hom}
(\pi_0,\pi_\la) [[z,z^{-1}]]$, in other words, to each vector of
$\pi_\la$ we can assign a field, whose Fourier components are linear
operators, acting from $\pi_0$ to $\pi_\la$.

Let us define this map by explicit formulas. To the highest weight
vector $v_\la \in \pi_\la$ we associate the following field, which
is called the bosonic vertex operator:
\begin{equation}    \label{screening}
\widetilde{V}_\la (z) = \sum_{n
\in \Z} V_\la(n) z^{-n} = T_\la \exp \left( -\sum_{n<0} \frac{\la_n
z^{-n}}{n} \right)
\exp \left( -\sum_{n>0} \frac{\la_n z^{-n}}{n} \right) ,
\end{equation}
where $T_\la: \pi_0 \arr \pi_\la$ is the shift operator, which maps
$v_0$ to $v_\la$ and commutes with the operators $b^i_n, n<0$, and
$\la_n = \la \otimes t^n$ is the element of the Heisenberg algebra $\h
= h \otimes \C[t,t^{-1}]$, corresponding to the element $\lambda \in
h$ ($\h$ is identified with $\h^*$ by means of the scalar product).

The field $\widetilde{V}_\la(z)$ can be viewed as the normally ordered
exponential of $$\bar{\la}(z) = \int^z \la(w) dw = - \sum_{n\neq 0}
\frac{\la_n z^{-n}}{n} + \la_0 \log z.$$

We can then define the fields for other elements of $\pi_\la$ as
follows: $$Y(P v_\la,z) = :Y(P v_0,z) \widetilde{V}_\la(z):,$$ where $P$ is
a polynomial in $b^i_n, n<0$.

These fields satisfy the axioms {\bf (A1)} and {\bf (A2)}. The axiom {\bf
(A3)} and formula \eqref{eq:assoc} are also satisfied for any pair $A \in
\pi_0, B \in \pi_\la$, or $B \in \pi_0, A \in \pi_\la$.

Using formula \eqref{eq:assoc}, it is easy to compute the commutation
relations between $b^i_n$ and $V_\la(m)$. Indeed, from the fact that $b^i_0
v_\la = (\la,\al_i) v_\la$ we derive the following OPE: $$b^i(z)
\widetilde{V}_\la(w) = \frac{(\la,\al_i) \widetilde{V}_\la(w)}{z-w} +
\mbox{regular terms},$$ which gives (using the technique of
\secref{commutationrel}) the commutation relations
\begin{equation}    \label{vertexcom}
[b^i_n,V_\la(m)] = (\la,\al_i) V_\la(n+m).
\end{equation}

In the next subsection we will show that the operators $\int
\widetilde{V}_{\beta\al_i}(z) dz$, where $\beta$ is a deformation
parameter, can be viewed as quantizations of the classical operators
$\q_i$ from \secref{classical}, corresponding to the value $\beta=0$.
We will use these operators to define the space of quantum integrals
of motion of the Toda field theories and related structures.

\subsection{Quantum integrals of motion.}

\subsubsection{Quantization of the Poisson algebra.}    \label{quant}
Let us introduce a deformation parameter $\beta$ and modify the
commutation relations in our Heisenberg algebra by putting
\begin{equation}    \label{rescale}
[b^i_n,b^j_m] = n(\al_i,\al_j) \delta_{n,-m} \beta^2.
\end{equation}
If $\g=\sw_2$, we put $$[b_n,b_m] = n \delta_{n,-m} \beta^2.$$

We also modify the definition of the Fock module $\pi_\la$ by putting
$b^i_0 v_\la = \beta (\al_i,\la) v_\la$. Of course, if $\beta\neq 0$,
this algebra and these modules are equivalent to the original
Heisenberg algebra and Fock modules if we multiply the old generators
by $\beta$.

According to \thmref{liealgebra}, the vector space of all Fourier
components of fields from a VOA is a Lie algebra. It contains a Lie
subalgebra, which consists of residues of fields, i.e. their $(-1)$st
Fourier components. Indeed, according to formula \eqref{commutator},
\begin{equation}    \label{closed}
[\int Y(A,z) dz,\int Y(B,z) dz] = \int Y(\int Y(A,w) dw \cdot B,z) dz.
\end{equation}

Let us denote the Lie algebra of all Fourier components of fields from the
VOA $\pi_0$, by $\f_0^\beta$, and its Lie subalgebra of the residues of the
fields -- by $\F_0^\beta$. They depend on $\beta$, because $\beta$ enters
the defining commutation relations \eqref{rescale}. Since the commutation
relations in these Lie algebras are polynomial in $\beta^2$, we can
consider $\f_0^\beta$ and $\F_0^\beta$ as Lie algebras over the ring
$\C[\beta^2]$, which are free as $\C[\beta^2]$--modules.

The constant term in the commutation relations in $\f_0^\beta$ is always
equal to $0$. We can therefore define the structure of Lie algebra on the
space $\f_0^\beta/(\beta^2 \cdot \f_0^\beta)$, by taking the
$\beta^2-$linear term in the commutator. That is for any pair $A, B \in
\f_0^\beta/(\beta^2 \cdot \f_0^\beta)$ consider their arbitrary liftings
$\widetilde{A}, \widetilde{B} \in \f_0$ and define $\{ A,B \}$ as the
$\beta^2$--linear term in the commutator of $\widetilde{A}$ and
$\widetilde{B}$: $$[\widetilde{A},\widetilde{B}] = \beta^2 \{ A,B \} +
\beta^4 (\ldots).$$ Clearly this bracket does not depend on the liftings of
$A$ and $B$ and satisfies the axioms of Lie bracket.

As vector spaces $\f_0^\beta/(\beta^2 \cdot \f_0^\beta)$ and
$\F_0^\beta/(\beta^2 \cdot \F_0^\beta)$ are isomorphic to $\f_0$ and
$\F_0$, respectively: $\int:P(\pa_z^n b^i(z)): z^n dz$ is identified with
$\int P(\pa_t^n u^i(t)) t^n dt.$

\subsubsection{Lemma.}    \label{cllimit}
{\em The Poisson structure on $\f_0^\beta/(\beta^2 \cdot \f_0^\beta)$,
induced by the Lie algebra structure on $\f_0^\beta$, coincides with
the Poisson structure on $\f_0$, given by formula \eqref{genskobka}.}

\vspace{3mm}
\noindent {\em Proof.} Recall that the Poisson structure on $\f_0$ is
uniquely determined by the Poisson bracket of the functionals $u^i_n = \int
u^i(t) t^n dt$, which is equal to $$\{u^i_n,u^j_m\} = n (\al_i,\al_j)
\delta_{n,-m}.$$ Indeed, any local functional can be represented by an
infinite sum $$\sum_{n_1+\dots+n_m=M} c_{n_1\dots n_m} \cdot
u^{i_1}_{n_1}\dots u^{i_m}_{n_m},$$ and we can compute the Poisson bracket
of two such sums term by term, using the Leibnitz rule
(cf. \secref{leibnitz}).

Likewise, the commutation relations in the Lie algebra $\f_0^\beta$
are uniquely determined by formula \eqref{rescale}, and we can again
compute the Lie bracket of two Fourier components term by term. But then
we immediately see that the $\beta^2-$linear term in the Lie bracket
in $\f_0^\beta$ coincides with the Poisson bracket in $\f_0$.

\subsubsection{} Thus, the Lie algebras $\f_0^\beta$ and $\F_0^\beta$
can be viewed as quantizations of the Poisson algebras $\f_0$ and
$\F_0$, respectively.

Let $\F_\la^\beta$ be the space of linear operators $\pi_0 \arr
\pi_\la$ of the form $\int Y(A,z) dz, A \in \pi_\la$. Formula
\eqref{closed} shows that the Lie algebra $\F_0^\beta$ acts on
$\F_\la^\beta$ by commutation. In the same way as in \lemref{cllimit} we
can check that this action is a quantization of the action of $\F_0$ on
$\F_\la$, defined in \secref{functspace}. In other words, we can check that
the $\beta^2-$linear term in the commutator between operators from
$\F_0^\beta$ and $\F_\la^\beta$ coincides with the Poisson bracket between
the corresponding elements of $\F_0$ and $\F_\la$.

Moreover, the actions of $\F_\la^\beta$ from $\pi_0$ to $\pi_\la$ and of
$\F_0^\beta$ from $\pi_\la$ to $\pi_\la$ are quantizations of the actions
of $\F_\la$ from $\pi_0$ to $\pi_\la$ and of $\F_0$ from $\pi_0$ to
$\pi_0$, defined in \secref{actiongen}. More precisely, let us consider the
Fock representation $\pi_\la$ as a free module $\pi_\la^\beta$ over
$\C[[\beta^2]]$. The operator $\int Y(A,z) dz$ can be considered over the
ring $\C[[\beta^2]]$. It has the form $\int Y(A,z) dz = \beta^2 \cdot
Y(A)^{(0)} + \beta^4 \cdot (\ldots)$. Hence it induces the operator
$Y(A)^{(0)}$ on the quotients $\pi_\la^\beta/\beta^2 \cdot
\pi_\la$. Such a quotient can be identified with the space $\pi_\la$ of
differential polynomials by identifying $b^i_n, n<0$, with $x^i_n$. Note
that the actions of derivative on both spaces coincide. We have
$$Y(A)^{(0)} dz = \{ \int \bar{A} dt,\cdot \},$$ where $\bar{A} = A
\on{mod} \beta^2$. A particular example of this formula, when
$A=v_{\al_i}$, will be considered in \lemref{linear}.

Consider the map $\pi_\la \arr \F^\beta_\la$, which sends $A \in
\pi_\la$ to $\int Y(A,z) dz$. From the classical result of
\secref{functspace} we derive that the kernel of this map consists of
total derivatives, and therefore we have the exact sequences
\begin{equation}    \label{pa}
0 \larr \pi_\la \stackrel{\pa}{\larr} \pi_\la \larr \F^\beta_\la \larr
0
\end{equation}
(if $\la=0$, then $\pi_0$ should be replaced by $\pi_0/\C v_0$).

Analogously, one obtains the exact sequence
\begin{equation}    \label{paan}
0 \larr \pi_\la \otimes \C[z,z^{-1}] \stackrel{\pa}{\larr} \pi_\la
\otimes \C[z,z^{-1}] \larr \f^\beta_\la \larr 0
\end{equation}
(if $\la=0$, then the first $\pi_0 \otimes \C[z,z^{-1}]$ should be replaced
by $\pi_0 \otimes \C[z,z^{-1}]/\C v_0 \otimes \C$).

Now we can quantize the Toda hamiltonian. Let us define the
map $\q_i^\beta: \pi_0 \arr \pi_\la^\beta$ as $\int
\widetilde{V}_{\al_i}(z) dz$. For any $A \in \pi_\la$ the
operator $\int Y(A,z) dz$ commutes with the derivative $\pa$. Hence it
defines a map $\F_0 \arr \F_\la$. Denote the map $\F_0^\beta
\arr \F_{\al_i}^\beta$, corresponding to $\q_i^\beta$, by $\Q_i^\beta$.

\subsubsection{Lemma.}    \label{linear}
{\em The $\beta^2-$linear terms of the operators $\q_i^\beta$ and
$\Q_i^\beta$ coincide with the operators $-\q_i$ and $-\Q_i$,
respectively.}

\vspace{3mm}
\noindent {\em Proof.} According to formula \eqref{rescale}, the
operator $b^i_n, n>0$, acts on $\pi_0$ as $$\beta^2 \sum_{j=1}^l
(\al_i,\al_j)\frac{\pa}{\pa b^j_{-n}}.$$
Thus, we obtain from formula \eqref{screening}: $$\q_i^\beta = \int \exp
\left( -\sum_{n<0} \frac{b^i_n z^{-n}}{n} \right) \exp
\left( -\sum_{n>0} \frac{b^i_n z^{-n}}{n} \right) dz =$$ $$\int \exp
\left( -\sum_{n<0} \frac{b^i_n z^{-n}}{n} \right) dz + \int \exp
\left( -\sum_{n<0} \frac{b^i_n z^{-n}}{n} \right) \left( - \sum_{n>0}
\frac{b^i_n z^{-n}}{n} \right) dz + \ldots =$$ $$- \beta^2 \sum_{n<0}
\widetilde{V}_i(n+1) \sum_{j=1}^l (\al_i,\al_j)\frac{\pa}{\pa b^j_n} +
\beta^4 (\ldots),$$ where $\widetilde{V}_i(n+1)$ are the Schur
polynomials in $b^i_n, n<0$.  If we replace $b^i_n, n<0$ by $x^i_n$,
then the linear term in this formula will coincide with formula
\eqref{operqi}, which defines the operator $\q_i$, with the sign minus.
Therefore, the $\beta^2-$linear term in $\Q_i^\beta$ coincides with the
operator $-\Q_i$.

\subsubsection{Definition.} {\em The intersection of kernels of the
linear operators $\Q_i^\beta: \F_0^\beta \arr
\F_{\al_i}^\beta, i \in S,$ will be called the space of local
integrals of motion of the quantum Toda field theory, associated to a
finite-dimensional simple Lie algebra $\g$ (in this case $S =
\{1,\ldots,l\}$), or an affine Lie algebra $\g$ (in this case $S =
\{0,\ldots,l\}$), and will be denoted by $I_\beta(\g)$.}

\subsubsection{Definition.}
{\em The intersection of kernels of the operators $\q_i^\beta: \pi_0 \arr
\pi_{\al_i}, i=1,\ldots,l$, will be called the $\W$-algebra of a
finite-dimensional simple Lie algebra $\g$ and will be denoted by
$\W_\beta(\g)$.}

\subsubsection{}
We will say that $U$ is a vertex operator subalgebra of a vertex operator
algebra $V$, if there is an embedding of vector spaces $i: U \arr V$, which
preserves the $\Z-$gradings, such that $i(|0\rangle_U) = |0\rangle_V$, and
that for any $A, B \in U$
\begin{equation}    \label{subvoa}
i \cdot \left[ Y(A,z) \cdot B \right] = Y(i \cdot A,z) \cdot (i \cdot B).
\end{equation}
We will say that $U$ is a conformal vertex operator subalgebra of $V$, if
in addition $i(\omega_U) = \omega_V$.

\subsubsection{Lemma.}    \label{kernel}
{\em Let $A^{(1)},\ldots,A^{(N)}$ be homogeneous vectors of
$\pi_{\la_1},\ldots,\pi_{\la_N},$ respectively. Then}
\begin{itemize}
\item[(a)] {\em The intersection of
kernels of the operators $\int Y(A^{(j)},z) dz: \pi_0 \arr \pi_{\la_j},
j=1,\ldots,N$, is a vertex operator subalgebra of $\pi_0$;}
\item[(b)] {\em The intersection of
kernels of the operators $\int Y(A^{(j)},z) dz: \F_0 \arr \F_{\la_j},
j=1,\ldots,N$, is a Lie subalgebra of $\F_0$.}
\end{itemize}

\vspace{3mm}
\noindent {\em Proof.} (a) Denote by $X$ the intersection of kernels of the
operators $\int Y(A^{(j)},z) dz, j=1,\ldots,N,$ and by $i$ the embedding of
$X$ into $\pi_0$. Since the elements $A^{(j)}$ are homogeneous, these
operators are homogeneous, and hence $X$ is $\Z$-graded. Clearly, $v_0
\in \pi_0$ lies in $X$, so we can put $|0\rangle_X = v_0$. Thus, we have
$i(|0\rangle_X) = |0\rangle_{\pi_0}$. We have to show that for any $B, C
\in X$ the vector $B_k \cdot C$, where $B_k$ is a Fourier component of the
field $Y(B,z)$, lies in $X$ for any $k \in \Z$. In other words, we have
to show that $$\int Y(A^{(j)},z) dz
\cdot \left( B_k \cdot C \right) = 0,\quad \quad j=1,\ldots,N.$$ But this
follows from vanishing of the commutators $$[\int Y(A^{(j)},z) dz,B_k] = 0,
\quad \quad j=1,\ldots,N,$$ and the fact that $\int Y(A^{(j)},z) dz \cdot C
= 0$. The commutator vanishes according to formula \eqref{commutator}:
$\int Y(A^{(j)},z) dz = A^{(j)}_{-\deg A^{(j)}+1}$, so there is only one
term in the commutator, which vanishes, because $A^{(j)}_{-\deg A^{(j)}+1}
\cdot B = \int Y(A^{(j)},z) dz \cdot B = 0.$

(b) follows at once from formula \eqref{commutator}.

Thus we see that $I_\beta(\g)$ is a Lie algebra and $\W_\beta(\g)$ is a
VOA.

\subsubsection{}
A quantum integral of motion can be represented as $x = x^{(0)} +
\beta^2 x^{(1)} + \dots \in \F_0^\beta$. By \lemref{linear}, the
constant term $x^{(0)}$ should be a classical integral of motion. We will
call $x$ a deformation or a quantization of $x^{(0)}$. Clearly, if exists,
it is uniquely defined up to adding $\beta^2$ times other quantum integrals
of motions. In other words, the dimension of the space of quantum integrals
of motion of a given degree is less than or equal to the dimension of the
space of classical integrals of motion of the same degree. In the rest of
this section we will show that the dimension of the space of integrals of
motion of a given degree of a Toda field theory for generic values of
$\beta$ is the same as for $\beta=0$. Since the quantum integrals of motion
depend algebraically on $\beta$, this will imply that all classical
integrals of motion of the Toda field theories can be deformed. The same is
true for the VOA $\W_\beta(\g)$.

\subsection{Liouville theory.}

\subsubsection{} In the quantum Liouville theory the space of
integrals of motion is defined as the kernel of the operator $\Q^\beta:
\F_0 \arr \F_1$. We will proceed in the same way as in \secref{liouv},
by computing the kernel of the operator $\q^\beta: \pi_0 \arr \pi_1$ and
then using the spectral sequence. Note that since the operator $\q^\beta$
is a residue of a field, its kernel is a VOA, by
\lemref{kernel}.

Thus, we consider the complex
\begin{equation}    \label{family}
\pi_0 \stackrel{\beta^{-2}\Q^\beta}{\larr} \pi_1,
\end{equation}
where we normalized $\Q^\beta = \int \widetilde{V}(z) dz$ by $\beta^{-2}$.
According to \lemref{linear}, this complex makes sense even when $\beta =
0$, when it coincides with the classical complex \eqref{picom}. We have
therefore a family of complexes, depending on a complex parameter
$\beta$. Moreover, as a vector space our complex does not change, and the
differential depends on $\beta$ polynomially in each homogeneous
component. The following simple observation will enable us to prove that
all classical integrals of motion can be quantized.

\subsubsection{Lemma.}    \label{generic}
{\em Suppose, one is given a family of finite-dimensional complexes,
depending on a complex parameter $\beta$, which are the same as vector
spaces, and the differentials depend analytically on $\beta$. Then for any
$n$ the dimension of the $n$th cohomology group of the complex is the same
for generic values of $\beta$, and it may only increase for special values
of $\beta$. In particular, if a certain cohomology group of the complex
vanishes, when $\beta=0$, then it also vanishes for generic values of
$\beta$.}

\vspace{3mm}
\noindent {\em Proof.} The $n$th cohomology is the quotient of the
kernel of the $(n+1)$st differential by the image of the $n$th
differential. The dimension of the kernel (respectively, the image) of a
finite-dimensional linear operator, depending analytically on a parameter
$\beta$, is the same for generic values of $\beta$, and it may only
increase (respectively, decrease) for special values of $\beta$.

\subsubsection{Corollary.}    \label{van}
{\em Under the conditions of \lemref{generic}, suppose also that all higher
cohomologies of the complex vanish, when $\beta=0$. Then the dimension of
the $0$th cohomology for generic $\beta$ is the same as for $\beta=0$.}

\vspace{3mm}
\noindent {\em Proof.} Vanishing of the higher cohomology groups for
generic values of $\beta$ follows from vanishing for $\beta=0$ and
\lemref{generic}. Since the complex does not depend on $\beta$ as a
vector space, its Euler characteristics, which is equal to the alternating
sum of the dimensions of the groups of the complex, also does not depend on
$\beta$. But the Euler characteristics is also equal to the alternating sum
of the dimensions of the cohomology groups. Since the dimensions of higher
cohomology groups are $0$ for generic $\beta$ and for $\beta=0$, the Euler
characteristics is equal to the dimension of the $0$th cohomology group for
generic $\beta$ and for $\beta=0$. Hence the dimension of the $0$th
cohomology group for generic $\beta$ is the same as for $\beta=0$.

\subsubsection{Proposition.}    \label{virquantum}
{\em The operator $\q^\beta: \pi_0 \arr \pi_1$ has no cokernel for generic
$\beta$. Its kernel, $\W_\beta(\sw_2)$, is a conformal vertex operator
algebra. It contains a Virasoro element $W_{-2}^\beta v_0$, and
$\W_\beta(\sw_2)$ is freely generated from $v_0$ under the action of the
Fourier components $W_n^\beta, n<-1$, of the field
$$Y(W_{-2}^\beta v_0,z) =
\sum_{n\in\Z} W_n^\beta z^{-n-2}.$$ This VOA is isomorphic to the VOA of
the Virasoro algebra with central charge $c = c(\beta) = 13 - 3\beta^2
-12\beta^{-2}$.}

\vspace{3mm}
\noindent {\em Proof.} By \propref{vir}, the operator $-\q$, which,
according to \lemref{linear}, is the limit of the operator $\beta^{-2}
\q^\beta$, when $\beta\arr 0$, has no cokernel. In other words, the
$1$st cohomology of the complex \eqref{family} vanishes, when
$\beta=0$. This complex decomposes into a direct sum of
finite-dimensional subcomplexes with respect to the $\Z-$grading. By
\corref{van}, the character of the $0$th cohomology for generic $\beta$ ( =
the kernel of the operator $\q^\beta$) is the same as for $\beta=0$:
$$\prod_{n\geq 2} (1-q^n)^{-1}.$$ This formula means that there is a vector
$W_{-2}^\beta v_0$ of degree $2$ in the kernel of $\q^\beta$. One can
deduce from the axioms of VOA and the fact that the kernel of $\q^\beta$ is
a VOA (cf. \lemref{kernel}) that this vector must be a
Virasoro element in $\pi_0$. But we can also find an explicit formula for
this vector:
\begin{equation}    \label{virel}
W_{-2}^\beta v_0 = \left(
\frac{1}{2\beta^2} b_{-1}^2 + (\frac{1}{2} - \frac{1}{\beta^2}) b_{-2}
\right) v_0.
\end{equation}
The Fourier components $W_n^\beta$ of the corresponding field generate an
action of the Virasoro algebra with central charge $c(\beta)$ on $\pi_0$
and $\pi_1$. This action commutes with the action of the differential
$\q^\beta$. The operators $W_n^\beta, n<-1$, act freely on $\pi_0$ for any
value of $\beta$, because they act freely for $\beta=0$
(cf. \propref{vir}). Therefore, by applying these operators to $v_0$ we
obtain a subspace of the kernel, which has the same character as the
kernel; hence it coincides with the kernel.

Clearly, this is the vertex operator algebra of the Virasoro algebra, which
is defined, e.g., in \cite{fz}.

\subsubsection{Corollary.} {\em The space $I_\beta(\sw_2)$ of local
integrals of motion of the quantum Liouville model is isomorphic to
the Lie algebra of residues of fields of the Virasoro vertex
operator algebra.}

\vspace{3mm}
\noindent {\em Proof.} By definition, $I_\beta(\sw_2)$ coincides with
the $1$th cohomology of the complex $\C \larr \F_0^\beta \larr
\F_1^\beta \larr \C$. We can use the spectral sequence
\eqref{doublecomplex} to compute this cohomology, in the same way as
in the proof of \propref{localint}. The Corollary then follows from
\propref{virquantum} and the fact that for any VOA $V$ the quotient of $V$
by the total derivatives and constants is isomorphic to the space of
residues of fields.

\subsubsection{General case.}

We now want to show that all classical integrals of motion of the classical
Toda field theory, associated to a finite-dimensional simple Lie algebra
$\g$, can be deformed. In order to establish that, we will construct a
quantum deformation $F^*_\beta(\g)$ of the complex $F^*(\g)$ and then use
vanishing of higher cohomologies in the classical limit $\beta=0$.

Roughly speaking, this can be achieved as follows. We will first construct
a resolution $B_*^q(\g)$ over the quantized universal enveloping algebra
(quantum group) of $\g$, $U_q(\g)$, which is a deformation of the standard
BGG resolution $B_*(\g)$. The resolution $B_*^q(\g)$ consists of Verma
modules $M_{\rho-s(\rho)}^q$ over $U_q(\g)$, and the differentials are
given by linear combinations of embeddings of Verma modules, defined by
singular vectors $P_{s',s}^q \cdot {\bold 1}_{\rho-s'(\rho)} \in
M_{\rho-s(\rho)}^q$, where $P_{s',s}^q \in U_q(\n_+)$. In the limit $q \arr
1$ this resolution coincides with the BGG resolution of $\g$.

We will define the quantum deformations of the operators $P_{s',s}(Q):
\pi_{\rho - s(\rho)} \arr \pi_{\rho - s'(\rho)}$ as
linear combinations of certain multiple integrals of products of bosonic
vertex operators $\widetilde{V}_{\al_i}(z)$, corresponding to
$P_{s',s}^q$. It turns out that the operators $\q_i^\beta$ satisfy the
$q-$deformed Serre relations, which are the relations in the quantized
universal enveloping algebra of the nilpotent subalgebra of $\g$,
$U_q(\n_+)$, with $q=\exp(\pi i \beta^2)$. This will allow us to define
differentials on the quantum complex.

Using the quantum complex, we will show that all classical integrals of
motion can be quantized.

\subsection{Quantum groups and quantum BGG resolutions.}

\subsubsection{Quantum group $U_q(\g)$.}    \label{qgroup}
Let $\g$ be a Kac-Moody Lie algebra associated to a symmetrizable Cartan
matrix $\|a_{ij}\|, i,j \in S$. The quantum group $U_q(\g)$ \cite{D,jimbo},
where $q \in \C^\times, q\neq \pm 1$, is the associative algebra over $\C$
with generators $e_i, f_i, K_i, K_i^{-1}, i \in S$, and the relations:
$$K_i K_j = K_j K_i , \quad \quad K_i K_i^{-1} = K_i^{-1} K_i = 1,$$ $$K_i
e_j K_i^{-1} = q^{(\al_i,\al_j)} e_j,
\quad K_i f_j K_i^{-1} = q^{-(\al_i,\al_j)} f_j,$$ $$[e_i,f_j] =
\delta_{i,j} \frac{K_i -
K_i^{-1}}{q^{(\al_i,\al_i)/2} - q^{-(\al_i,\al_i)/2}},$$ and the so-called
$q$-Serre relations, which can be defined as follows.

Introduce a grading on the free algebra
with generators $e_i, i=1,\ldots,l,$ with respect to the weight lattice
$P$, by putting $\deg e_i = \al_i$. If $x$ is a homogeneous element of this
algebra of weight $\gamma$, put $$\mbox{ad}_q e_i \cdot x = e_i x -
q^{(\al_i,\gamma)} x e_i.$$ Likewise, we can introduce operators
$\mbox{ad}_q f_i$ on the free algebra with generators $f_i,
i=1,\ldots,l$. Then the $q$-Serre relations read:
\begin{equation}    \label{quantumserre}
(\mbox{ad}_q e_i)^{-a_{ij}+1} \cdot e_j = 0, \quad \quad (\mbox{ad}_q
f_i)^{-a_{ij}+1} \cdot f_j = 0.
\end{equation}

If we put $$h_i = \frac{K_i - K_i^{-1}}{q^{(\al_i,\al_i)/2} -
q^{-(\al_i,\al_i)/2}},$$ then in the limit $q \arr 1$ these relations
coincide with the standard relations of $\g$ in terms of $e_i, f_i, h_i, i
\in S$.

Denote by $U_q(\n_+)$ the subalgebra of $U_q(\g)$, generated by $e_i,
i=1,\ldots,l$, with the relations \eqref{quantumserre}. This algebra is a
quantum deformation of the universal enveloping algebra of the nilpotent
subalgebra $\n_+$ of $\g$. Let $U_q(\goth{b}_-)$ be the subalgebra of
$U_q(\g)$, generated by $f_i, K_i, K_i^{-1}, i=1,\ldots,l$.

\subsubsection{Verma modules.}
Verma modules over $U_q(\g)$ are defined in the same way as Verma modules
over $\g$ (cf. \secref{verma}).

Let $\C_\la$ be the one-dimensional representation of $U_q(\goth{b}_-)$,
which is spanned by vector ${\bold 1}_\la$, such that $$f_i \cdot {\bold
1}_\la = 0, \quad \quad K_i \cdot {\bold 1}_\la = q^{(\la,\al_i)} {\bold
1}_\la, \quad \quad i=1,\ldots,l.$$

The Verma module $M_\la^q$ over $U_q(\g)$ of lowest weight $\la$ is the
module induced from the $U_q(\goth{b}_-)$-module $\C_\la$: $$M_\la^q =
U_q(\g) \otimes_{U_q(b_-)} \C_\la.$$

\subsubsection{Singular vectors.}

We want to construct a resolution $B_*^q(\g)$ over $U_q(\g)$, which would
coincide with the standard BGG resolution in the limit $q\arr 1$.
Therefore as a vector space, the $j$th group $B_j^q(\g)$ of this resolution
should be the direct sum of Verma modules $$B_j^q(\g) = \oplus_{l(s)=j}
M_{\rho - s(\rho)}^q.$$ Now we have to construct the differentials.

We want to show first that the structure of singular vectors in the modules
$M_{\rho - s(\rho)}^q$ for generic $q$ is the same as the structure of the
singular vectors for $q=1$, which is described in \secref{defres}. We will
then proceed in the same way as in the case $q=1$.

The existence of these singular vectors can be derived from the determinant
formula for the Shapovalov form on $M_\la^q$. This formula has been
established in \cite{dk} (cf. formula (1.9.3)) for finite-dimensional
simple Lie algebras. In \cite{jl} (cf. Lemma 6.4) the irreducible factors
of the formula were found for arbitrary symmetrizable Kac-Moody
algebras. This is sufficient for our purposes.

The above cited results show that the Shapovalov form on $M_\la^q$ at
weight $\eta$ vanishes, if one of the following equations is satisfied:
\begin{equation}    \label{T}
(\la-\rho,\gamma) + \frac{m}{2} (\gamma,\gamma) = 0,
\end{equation}
where $\gamma$ runs over the set of positive roots of $\g$, and $m$ runs
over the set of positive integers, such that $m\gamma < \eta$. Note that in
this formula we have signs different from those in \cite{dk,jl}, because we
work with modules of {\em lowest weight}.

Consider now the module $M_{\rho - s(\rho)}^q$. It is known that for any
$s'$, which satisfies $s\prec s'$ and $l(s') = l(s) + 1$, there exist
$\gamma$ and $m$, such that $\rho - s'(\rho) = \rho - s(\rho) + m\gamma$
and $(s(\rho),\gamma) - \frac{m}{2} (\gamma,\gamma) = 0$. Therefore the
equation \eqref{T} is satisfied and hence the determinant is equal to $0$
at weight $\rho - s'(\rho)$.

\subsubsection{Remark.} In fact, in order to prove this statement, we do
not need the exact formula for the determinant. We can proceed along the
lines of \cite{KK}, using the Casimir operators, constructed in
\cite{drinfeld} or \cite{jl} and the limit $q \arr 1$ of the
determinant formula from \cite{KK}.

\subsubsection{} On the other hand, the determinant is not equal to $0$ at
any level $\eta < \rho - s'(\rho)$ for $q=1$ \cite{KK}, and hence for
generic $q$.  Therefore in $M_{\rho - s(\rho)}^q$ there should exist a
singular vector of weight $\rho - s'(\rho)$.

This singular vector defines a map $$i_{s',s}^q: M_{\rho - s'(\rho)}^q \arr
M_{\rho - s(\rho)}^q$$ by sending the lowest weight vector ${\bold 1}_{\rho
- s'(\rho)}^q$ of $M_{\rho - s'(\rho)}^q$ to the singular vector
$P_{s',s}^q \cdot {\bold 1}_{\rho - s(\rho)}^q$ of $M_{\rho - s(\rho)}^q$
of weight $\rho - s'(\rho)$. This map is an embedding of Verma modules for
generic $q$, because it is an embedding for $q=1$.

We can map singular vectors, constructed this way, inductively to $M_0^q$.
Thus, we obtain in $M_0^q$ singular vectors of weights $\rho - s(\rho)$ for
arbitrary elements $s$ of the Weyl group.

We know that for $q=1$ there is only one singular vector of weight $\rho -
s(\rho)$ in $M_0$. But the dimension of the space of singular vectors of a
Verma module $M_\la^q$ of a certain weight for generic $q$ is greater than
or equal to that for $q=1$. Therefore there is only one singular vector of
such weight in $M_0^q$ for generic $q$.

Uniqueness implies the relation
\begin{equation}    \label{P}
P_{s'',s'_1}^q P_{s'_1,s}^q = P_{s'',s'_2}^q P_{s'_2,s}^q
\end{equation}
in $U_q(\n_+)$. Indeed, both $P_{s'',s'_1}^q P_{s'_1,s}^q {\bold 1}_{\rho -
s(\rho)}^q$ and $P_{s'',s'_2}^q P_{s'_2,s}^q {\bold 1}_{\rho - s(\rho)}^q$
are singular vectors in $M_{\rho - s(\rho)}$ of weight $\rho -
s''(\rho)$. Therefore they coincide for generic $q$.

\subsubsection{The differential.}    \label{qbgg}
Now we can construct a differential on the complex $B_*^q(\g)$. For any
pair $s, s'$ of elements of the Weyl group of $\g$, such that $s\prec s''$,
we have the embeddings of $\g$-modules $i_{s',s}^q: M_{\rho - s'(\rho)}^q
\arr M_{\rho - s(\rho)}^q$. They satisfy: $i_{s'_1,s}^q \circ i_{s'',s'_1}^q
= i_{s'_2,s}^q \circ i_{s'',s'_2}^q$, according to
\eqref{P}. We define the differential $d_j^q: B_j^q(\g) \arr B_{j-1}^q(\g)$
by the formula
\begin{equation}    \label{qdiff}
d_j^q = \sum_{l(s)=j-1,l(s')=j,s\prec s'} \ep_{s',s} \cdot i_{s',s}^q
\end{equation}
(cf. formula \eqref{diff}). By construction, this differential is nilpotent
(cf. \thmref{acyclic}).

Note that since higher cohomologies of $B_*^q(\g)$ vanish for $q=1$, they
also vanish for generic $q$, by \lemref{generic}. Therefore the $0$th
cohomology is one-dimensional for generic $q$, by \corref{van}. Thus, for
generic $q$, $B_*(\g)$ is a free resolution of the trivial representation of
$U_q(\n_+)$.

\subsubsection{Remark.}
In the same way we can $q$-deform the BGG resolution of an arbitrary
integrable representation $V_\la, \la \in P$, of a Kac-Moody algebra. Such
a resolution has also been constructed in \cite{malikov} by other methods.

Vanishing of higher cohomologies of this resolution for $q=1$ implies that
they vanish for generic $q$. But then the $0$th cohomology for generic $q$
is a module over $U_q(\g)$, which is irreducible, since $V_\la$ is
irreducible, and whose character is the same as the character of $V_\la$
for $q=1$, by \lemref{van}. This gives an alternative proof of the fact
that any integrable representation of $\g$ can be $q$-deformed, previously
proved by Lusztig \cite{lusztig} by other methods.

\subsection{Toda field theories associated to finite-dimensional simple
Lie algebras.}

\subsubsection{} Now we can use the resolution $B_*(\g)$ to define the
quantum complex $F^*_\beta(\g) = \oplus_{j\geq 0} F^j_\beta(\g)$. The $j$th
group $F^j_\beta(\g)$ of the complex is the same as the $j$th group of the
classical complex $F^*(\g)$: $$F^j_\beta(\g) = \oplus_{l(s)=j}
\pi_{\rho-s(\rho)}.$$ To define the differentials of the complex, we have
to quantize the operators $P_{s',s}(Q): \pi_{\rho-s(\rho)} \arr
\pi_{\rho-s'(\rho)}$. In order to do that we should first learn how to
compose the operators $\q_i^\beta$.

\subsubsection{}
Let $p = (p_1,\ldots,p_m)$ be a permutation of the set $(1,2,\ldots,m)$. We
define a contour of integration $C_p$ in the space $(\C^\times)^m$ with the
coordinates $z_1,\ldots,z_m$ as the product of one-dimensional contours
along each of the coordinates, going counterclockwise around the origin
starting and ending at the point $z_i = 1$, and such that the contour of
$z_{p_i}$ is contained inside the contour of $z_{p_j}$ for $i>j$.

Denote by ${\bold i} = (i_1,\ldots,i_m)$ a sequence of numbers from $1$ to
$l$, such that $i_1 \leq i_2 \leq \ldots \leq i_m$. We can apply a
permutation $p$ to this sequence to obtain another sequence ${\bold j} =
(j_1,\ldots,j_m) = (i_{p_1},\ldots,i_{p_m})$. Let us define an operator
$V_{{\bold j}}^\beta$ as the integral
$$\int_{C_p} dz_1 \ldots dz_m
\prod_{1\leq k<l\leq m} (z_k - z_l)^{\beta^2(\al_{i_k},\al_{i_l})}
\prod_{1\leq k\leq m} z_k^{\beta^2(\la,\al_{i_k})}
:\V_{\al_{i_1}}(z_1) \ldots \V_{\al_{i_m}}(z_m): =$$
\begin{equation}    \label{composition}
\sum_{n_1,\ldots,n_m \in\Z} \Gamma_{\js}^{n_1,\ldots,n_m}
:V_{\al_{i_1}}(n_1) \ldots V_{\al_{i_m}}(n_m):,
\end{equation}
where the coefficient $\Gamma_{\js}^{n_1,\ldots,n_m}$ is given by
\begin{equation}    \label{coefficient}
\Gamma_{js}^{n_1,\ldots,n_m} = \int_{C_p} dz_1 \ldots dz_m
\prod_{1\leq k<l\leq m} (z_k - z_l)^{\beta^2(\al_{i_k},\al_{i_l})}
\prod_{1\leq k\leq m} z_k^{\beta^2(\la,\al_{i_k})-n_k}.
\end{equation}

In the integrals above, we choose the branch of the power function, which
takes real values for real values of the $z_i$'s, such that $z_{j_1} >
z_{j_2} > \ldots > z_{j_m}$. Thus, $C_p$ should be viewed as an element of
the group of relative $m$-chains in $(\C^\times)^m$ modulo the diagonals,
with values in the one-dimensional local system $\xi_{\is}$, which is
defined by the multivalued function $$\prod_{1\leq k<l\leq m} (z_k -
z_l)^{-\beta^2(\al_{i_k},\al_{i_l})} \prod_{1\leq k\leq m}
z_k^{-\beta^2(\la,\al_{i_k})}.$$

The integral of the type \eqref{coefficient} over any such relative chain
is well-defined for generic values of $\beta$. Indeed, the integral
$$\int_{C} dz_1 \ldots dz_m \prod_{1\leq k<l\leq m} (z_k - z_l)^{\mu_{kl}}
\prod_{1\leq k\leq m} z_k^{\nu_k}$$ over such a chain $C$ converges in the
region $\mbox{Re} \mu_{kl} \geq 0$, and can be uniquely analytically
continued to other values of $\mu_{kl}$, which do not lie on certain
hyperplanes, cf. \cite{varchenko}, especially, Theorem (10.7.7), for
details. These hyperplanes are defined by setting some linear combinations
of $\mu_{kl}$'s, with integral coefficients, to negative integers. For
generic $\beta$ the exponents in our integral do not lie on those
hyperplanes, therefore the integral is well-defined.

Note that the integral in \eqref{coefficient} depends on $\js$ and not on
$p$. For any $N \in \Z$ the operator $$\sum_{n_1+\ldots+n_m=N}
\Gamma_{{\bold j}}^{n_1,\ldots,n_m} :V_{\al_{i_1}}(n_1) \ldots
V_{\al_{i_m}}(n_m):,$$ is a well-defined homogeneous operator acting from
$\pi_\la$ to $\pi_{\la+\gamma}$, where $\gamma = \sum_{j=1}^m
\al_{i_j}$. Therefore, the operator $V_{{\bold j}}^\beta$ is a linear
operator from $\pi_\la$ to the completion $\bar{\pi}_{\la+\gamma}$ of
$\pi_{\la+\gamma}$.

\subsubsection{}
One can interpret the operator $V_{\js}^\beta$ as a suitably
defined composition operator $\q_{j_1}^\beta \ldots \q_{j_m}^\beta$.
Indeed, let us define the bosonic vertex operator $V_{\al_i}(z)$,
acting from $\pi_\la$ to $\pi_{\la+\al_i}$ as $z^{\beta^2(\la,\al_i)}
\V_{\al_i}(z)$. One has (cf., e.g., \cite{flm}, formulas (8.4.25) and
(A.2.9))
\begin{equation}    \label{product}
V_{\al_{j_1}}(z_1) \ldots V_{\al_{j_m}}(z_m)
= \prod_{1\leq k<l\leq m} (z_k - z_l)^{\beta^2(\al_{j_k},\al_{j_l})}
:V_{\al_{j_1}}(z_1)
\ldots V_{\al_{j_m}}(z_m):,
\end{equation}
for $|z_1| > \ldots > |z_m|$. Note that unlike the Fourier components
of $V_{\al_{j_1}}(z_1) \ldots V_{\al_{j_m}}(z_m)$, the Fourier
components of $:V_{\al_{j_1}}(z_1)\ldots V_{\al_{j_m}}(z_m):$ are
well-defined linear operators, analytically depending on
$z_1,\ldots,z_m$ in the region $\C^m\backslash diag$, which do not
change under the permutations of coordinates.

We have for $\js_1 = (j^1_1,\ldots,j^1_m)$ and $\js_2 =
(j^2_1,\ldots,j^2_n)$:
\begin{equation}    \label{compose}
V_{\js_1}^\beta V_{\js_2}^\beta = V_{(\js_1,\js_2)}^\beta.
\end{equation}
where $V_{\js_2}^\beta: \pi_\la \arr \bar{\pi}_{\la+\gamma_2}$ and
$V_{\js_1}^\beta: \pi_{\la+\gamma_2} \arr
\bar{\pi}_{\la+\gamma_1+\gamma_2}$. Indeed, we can choose the contour $C_p$
in such a way that $|z_{p_i}|\geq 1$ for $i=1,\ldots,m$ and $|z_{p_i}|\leq
1$ for $i=m+1,\ldots,m+n$. Then the composition in the left hand side of
\eqref{compose} is well-defined, and concides with the right hand side.

According to \lemref{linear}, the $\beta^2$--linear term of the operator
$\q_j^\beta$ coincides with the operator $-\q_j$, therefore the leading
($\beta^{2m}$th) term of $V_{\js}^\beta$ coincides with $(-1)^m \q_{j_1}
\ldots \q_{j_m}$.

The operators $\q_i$ satisfy the Serre relations. We want to show that
the operators $V_i^\beta$ satisfy the $q$-Serre relations, where
$q=\exp (\pi i \beta^2)$.

\subsubsection{}
Consider a free algebra $A$ with generators $g_i, i=1,\ldots,l$. We
can assign to each monomial $g_{j_1} \ldots g_{j_m}$ the contour
$C_{\js}$ and hence the operator $V_{\js}^\beta$. This gives us a map
$\Delta$ from $A$ to the space of linear combinations of such
contours. Given such a linear combination $C$, we define $V_C^\beta$
as the linear combination of the corresponding operators
$V_{C_{\js}}^\beta$.

Consider the two-sided ideal $S_q$ in $A$, which is generated by the
$q$-Serre relations $(\mbox{ad} g_i)_q^{-a_{ij}+1} \cdot g_j,
i\neq j$, where $q=\exp (\pi i \beta^2)$.

\subsubsection{Lemma.}    \label{qserre}
{\em If $C$ belongs to $\Delta(S_q)$, then $V_C^\beta = 0$.}

\vspace{3mm}
\noindent {\em Proof} is given in \cite{bouwknegt}. It is based on
rewriting the integrals over the contours $C_{\js}$ as integrals over the
contours, where all variables are on the unit circle with some ordering of
their arguments.

We have to prove that $(\on{ad} \q_i)^{-a_{ij}+1} \cdot \q_j = 0$. Let
${\cal V}_{p(\is)}^\beta$ be the operator, defined by formula
\eqref{composition}, where the contour $C_p$ is replaced by contour $${\cal
C}_p = \{(z_1,\ldots,z_m) \cond |z_i| = 1, 0 < \on{arg} z_{p_1} < \ldots <
\on{arg} z_{p_m} < 2\pi \}.$$ By induction one can prove \cite{bouwknegt}
that $$(\on{ad} \q_i)^m \cdot \q_j = I_m {\cal V}_{(i,\ldots,i,j)},$$ where
$$I_m = \prod_{j=0}^{m-1} \frac{(1-q^{j (\al_i,\al_i) + 2
(\al_i,\al_j)})(1-q^{(j+1) (\al_i,\al_i)})}{1-q^{(\al_i,\al_i)}},$$ and
Lemma follows.

Thus, we obtain a well-defined map, which assigns to each element $P$
of the algebra $U_q(\n_+) \simeq A/S_q$ the operator $V_P^\beta$.

\subsubsection{Lemma.}    \label{homogeneous}
{\em Let $P \in U_q(\n_+)$ be such that $P \cdot {\bold 1}_\la$ is a
singular vector of $M_\la^q$ of weight $\la+\gamma$. Then the operator
$V_P^\beta$ is a homogeneous linear operator $\pi_\la \arr
\pi_{\la+\gamma}$.}

\vspace{3mm}
\noindent {\em Proof.} It is known that under the conditions of the
Lemma, the contour $\Delta(P)$ is a cycle in the group
$H^m((\C^\times)^m,diag;\xi_{\is})$,
cf. \cite{B1,B,bouwknegt,VS,varchenko}. Therefore, only the degree $0$
integrands (with respect to the grading $\deg z_i = \deg dz_i = 1$) of the
integral over $\Delta(P)$ give non-zero results. In other words, all
coefficients $$\int_{\Delta(P)} dz_1 \ldots dz_n
\prod_{1\leq k<l\leq m} (z_k - z_l)^{\beta^2(\al_{i_k},\al_{i_l})}
\prod_{1\leq k\leq m} z_k^{\beta^2(\la,\al_{i_k})-n_k}$$ vanish,
unless $\sum_{i=1}^m n_i = -m$. It means that in the operator
$V_P^\beta$ all homogeneous components, except one, vanish, and
Lemma follows.

\subsubsection{Remark.} The statements of \lemref{qserre} and
\lemref{homogeneous} follow from more general results on the remarkable and
not yet fully understood correspondence between quantum groups and local
systems of the type $\xi_{\is}$ on configuration spaces, established in the
works of Schechtman and Varchenko \cite{VS,varchenko,schechtman} (cf. also
\cite{felder}).

Let us also remark that in \cite{varchenko} Varchenko constructed
explicitly certain {\em absolute} chains in $(\C^\times)\backslash diag$,
which have all the nice properties of the {\em relative} chains $C_p$,
including the factorization property \eqref{compose} (cf. \S 14 of
\cite{varchenko}). One can use these chains instead of $C_p$ in the
definition of our operators.

\subsubsection{Quantum differentials.} We are ready now to define the
differentials $\delta^j_\beta: F^{j-1}_\beta(\g) \arr F^j_\beta(\g)$ of the
quantum complex $F^*_\beta(\g)$. Recall that for any pair $s, s'$ of
elements of the Weyl group there exists a singular vector $P_{s',s}^q
\cdot {\bold 1}_{\rho - s(\rho)}^q \in M_{\rho - s(\rho)}$ of weight $\rho
- s'(\rho)$, where $P_{s',s}^q \in U_q(\n_+)$. We put:
\begin{equation}    \label{qdifferential}
\delta^j_\beta = \sum_{l(s)=j-1,l(s')=j,s\prec s'} \ep_{s',s} \cdot
V_{P_{s',s}^q}^\beta,
\end{equation}
where $q = \exp (\pi i \beta^2)$. By \lemref{homogeneous}, the
differentials $\delta^j_\beta$ are well-defined homogeneous linear
operators. From the nilpotency of the differential of the quantum BGG
resolution and \lemref{qserre} we deduce that these differentials are
nilpotent.

Thus, we obtain a family of complexes $F^*_\beta(\g)$, depending on the
parameter $\beta$. As was explained above, we can rescale the
differentials by some powers of $\beta$, so that when $\beta=0$ the
complex $F^*_\beta(\g)$ becomes the classical complex $F^*(\g)$.

\subsubsection{Theorem}    \label{walgebra}
{\em For generic $\beta$ higher cohomologies of the complex $F^*_\beta(\g)$
vanish. The $0$th cohomology, $\W_\beta(\g)$, is a conformal vertex
operator algebra. There exist elements $W^{(1),\beta}_{-d_1-1} v_0$,
$\ldots, W^{(l),\beta}_{-d_l-1} v_0$ of $\W_\beta(\g)$ of degrees
$d_1+1,\ldots,d_l+1$, where the $d_i$'s are the exponents of $\g$, such
that $\W_\beta(\g)$ is freely generated (in the sense explained below) from
$v_0$ under the action of the Fourier components $W^{(i),\beta}_{n_i},
1\leq i\leq l,n_i<-d_i,$ of the corresponding fields $${\bold W}_i^\beta(z)
= Y(W^{(i),\beta}_{-d_i-1} v_0,z) = \sum_{n\in\Z} W^{(i),\beta}_n
z^{-n-d_i-1}.$$ Moreover, $W^{(1),\beta}_{-2} v_0$ is the Virasoro element,
and the other elements $W^{(i),\beta}_{-d_i-1} v_0$ can be chosen so as to
be annihilated by the corresponding Virasoro generators $W^{(1),\beta}_n,
n>0$.}

\vspace{3mm}
\noindent {\em Proof.} \lemref{generic} together with
\propref{exist} lead us to conclude that for generic $\beta$ all
higher cohomologies of the complex $F^*_\beta(\g)$ vanish.

Therefore, by \corref{van}, the $0$th cohomology, $\W_\beta(\g)$, of the
complex $F^*_\beta(\g)$ for generic $\beta$ has the same character as the
$0$th cohomology of the complex $F^*(\g)$ (cf. the proof of
\propref{exist}):
\begin{equation}    \label{character}
\prod_{1\leq i\leq l,n_i>d_i} (1-q^{n_i})^{-1}.
\end{equation}
This formula shows that there is a vector of degree $2$,
$W^{(1),\beta}_{-2} v_0$, in $\W_\beta(\g)$, which is a Virasoro
element. It is given by the following formula: $$W^{(1),\beta}_{-2} v_0 =
\left( \frac{1}{2\beta^2} \sum_{i=1}^l b^i_{-1} b^{i*}_{-1} + \rho_{-2} -
\frac{1}{\beta^2} \rho^\vee_{-2} \right) v_0.$$

The Fourier components $L_n = W^{(1),\beta}_n, n \in \Z$, of the field
corresponding to $W^{(1),\beta}_{-2} v_0$, generate an action of the
Virasoro algebra on $\pi_\la$. They commute with the action of the
differential $\delta^1_\beta$ and hence they preserve the space
$\W_\beta(\g)$, which is the kernel of $\delta^1_\beta$.

Recall that a singular vector is a vector, which is annihilated by the
positive Virasoro generators $L_n, n>0$. Verma module is a module over the
Virasoro algebra, which is freely generated from a singular vector by $L_n,
n<0$. The degree of the singular vector, from which it is generated is
called highest weight. The vacuum Verma module is the module, freely
generated by $L_n, n<-1$, from a vector, which is annihilated by $L_n,
n\geq -1$. From the structural theory of Verma modules over the Virasoro
algebra \cite{FeFu} we know that Verma modules of integral non-zero highest
weight and the vacuum Verma module are irreducible for generic central
charge. Moreover, there can be no non-trivial extensions between such
modules for generic central charge.

Therefore for generic $\beta$ each singular vector of $\pi_0$, except for
$v_0$, generates an irreducible Verma module of positive integral highest
weight under the free action of the generators $L_n, n<0$. The vector $v_0$
generates the irreducible vacuum Verma module under the free action of the
generators $L_n, n<-1$. As a module over the Virasoro algebra, $\pi_0$ is a
direct sum of the vacuum Verma module generated from $v_0$ and some Verma
modules of positive integral highest weights. Since $\W_\beta(\g)$ is a
submodule of $\pi_0$, it is also a direct sum of such modules. In
particular, we see that the character of the space of singular vectors in
$\W_\beta(\g)$ is equal to
\begin{equation}    \label{chsing}
(1-q) \prod_{2\leq i\leq l,n_i>d_i} (1-q^{n_i})^{-1} + q.
\end{equation}

{}From this fact and the character formula \eqref{character} we can derive
that there exist singular vectors $W^{(i),\beta}_{-d_i-1} v_0$ of degrees
$d_i+1$ in the $0$th cohomology, which in the limit $\beta=0$ can be chosen
as polynomial generators $W^{(i)}_{-d_i-1}$ of $\W(\g)$ from
\propref{exist}.

This can be proved by induction. Suppose, we have proved this fact for
$i<j$. Thus, we have constructed singular vectors $W^{(i),\beta}_{-d_i-1}
v_0, i=1,\ldots,j-1$, satisfying the conditions above. Let $\W(\g)'$ be the
subspace of $\W(\g)$, which consists of all polynomials in $W^{(i)}_{n_i},
n_i<-d_i, i=1,\ldots,j-1$. Consider the component $\W(\g)_{d_j+1}$ of
$\W(\g)$ of degree $d_j+1$. In this component the subspace $\W(\g)'_{d_j+1}
= \W(\g)' \cap \W(\g)_{d_j+1}$ has codimension $1$. Now consider $\beta^2$
as a formal variable and the space $W_\beta(\g)$ as a free module over the
ring $\C[[\beta^2]]$ as in \secref{quant}. Denote by $S_{d_j+1}^\beta$ the
space of singular vectors of $\W_\beta(\g)$ of degree $d_j+1$. We have a
natural projection $\W_\beta(\g) \arr
\W_\beta(\g)/\beta^2 \cdot \W_\beta(\g) \simeq \W(\g)$, the classical
limit. We will show that the image of $S_{j+1}^\beta$ in $\W(\g)_{d_j+1}$
is not contained in $\W(\g)'_{d_j+1}$.

Indeed, note that the action of the operators $L_n, n\geq -1$, on $\pi_0$
is well-defined in the limit $\beta=0$, and denote the corresponding
operators by $L_n^{(0)}$. These are derivations of $\pi_0$, which generate
a Lie subalgebra of the Virasoro algebra. The action of the operator
$L_{-1}$ does not depend on $\beta$ and coincides with the action of the
derivative $\pa$. From our inductive assumption we already know the
commutation relations between $L_n$ and $W^{(i),\beta}_{n_i},
i=1,\ldots,j-1$. They are given by formula
\eqref{primary} below. These relations give us in the limit $\beta=0$:
$$L_n^{(0)} \cdot W^{(i)}_{n_i} = (n d_i - n_i) W^{(i)}_{n+n_i},$$ where we
put $W^{(i)}_m = 0$, if $m\geq -d_i$. From these formulas it is clear that
the polynomial algebra $A^{(j)}$, generated by $W^{(i),\beta}_{n_i},
i=2,\ldots,j-1$, is preserved by the action of $L_n^{(0)}, n\geq -1$. But
the operator $L_{-1}^{(0)} = \pa$ acts freely on $\pi_0/\C$ and hence on
$A^{(j)}/\C$. If $X$ is a singular vector in $\W(\g)'$, i.e. if it is
annihilated by the operators $L_n^{(0)}, n>0$, it should lie in $A^{(j)}$.
But if $X$ is a singular vector, then $L_{-1}^{(0)} X$ can not be a
singular vector. Therefore the character of the space of singular vectors,
contained in $\W(\g)'$, is less then or equal to the character of the
quotient of $A^{(j)}$ by the total derivatives, which is equal to
\begin{equation}    \label{char}
(1-q) \prod_{2\leq i\leq j-1,n_i>d_i} (1-q^{n_i})^{-1} + q
\end{equation}
(in fact, it can be shown that it is equal to \eqref{char}).

The image of $S_{j+1}^\beta \subset \W_\beta(\g)_{d_j+1}$ in
$\W(\g)_{d_j+1}$ lies in the space of singular vectors of
$\W(\g)_{d_j+1}$. Formulas \eqref{chsing} and \eqref{char} show that the
dimension of the space $S_{j+1}^\beta$ is greater than the dimension
of the space of singular vectors of $\W(\g)'_{d_j+1}$. Hence there should
exist a singular vector in $\W_\beta(\g)_{d_j+1}$, whose image in
$\W(\g)_{d_j+1}$ is linearly independent from the subspace
$\W(\g)'_{d_j+1}$. Denote such a vector by $W^{(j),\beta}_{-d_j-1} v_0$. By
\propref{exist}, its image in $\W(\g)$ is algebraically independent from
the previously constructed $W^{(i)}_{-d_i-1}, i<j$, and hence it can be
chosen as a generator $W^{(j)}_{-d_j-1}$ of $\W(\g)$. This completes our
inductive argument.

The Fourier components $W^{(i),\beta}_{n_i}, n_i<-d_i$, of the fields
corresponding to the singular vectors $W^{(i),\beta}_{-d_i-1} v_0$ act on
$\pi_\la$ and commute with the differential $\delta^1_\beta$. Therefore
they act on $\W_\beta(\g)$. In the limit $\beta=0$ their action coincides
with the action by multiplication by the polynomials $W^{(i)}_{n_i}$.

The polynomials $W^{(i)}_{n_i}, n_i<-d_i$, were shown in \propref{exist} to
be algebraically independent. Therefore monomials $$W^{(i_1)}_{n_{i_1}}
\ldots W^{(i_m)}_{n_{i_m}} \in \W(\g),$$ where $i_1 \leq \ldots \leq i_m$
and $n_{i_j} < n_{i_{j+1}}$ for $i_j=i_{j+1}$ are linearly independent. But
these monomials are images (classical limits) of monomial elements
$$W^{(i_1,\beta)}_{n_{i_1}} \ldots W^{(i_m,\beta)}_{n_{i_m}} v_0 \in
\W_\beta(\g)$$ ordered so that $i_1 \leq \ldots \leq i_m$
and $n_{i_j} < n_{i_{j+1}}$ for $i_j=i_{j+1}$. Therefore the latters are
linearly independent in $\W_\beta(\g)$.

Hence such monomials linearly span a subspace in the $\W_\beta(\g)$, whose
character is given by \eqref{character}. But we know that this is the
character of $\W_\beta(\g)$. Hence these monomials form a basis of
$\W_\beta(\g)$. By analogy with the case of universal enveloping algebras,
where one can choose the Poincare-Birkhoff-Witt basis, one can say that the
operators $W^{(i),\beta}_{n_i}, 1\leq i\leq l,n_i<-d_i$, {\em freely
generate} $\W_\beta(\g)$ from $v_0$.

\subsubsection{Remark.} A vector in a VOA, which is a singular vector
with respect to the action of the Virasoro algebra, gives rise to a
field, which is called a {\em primary field} of conformal dimension equal
to the degree of this vector.

Suppose, $A$ is such a vector, of degree $\Delta$. Then we have the
following OPE: $$Y(T,z) Y(A,w) = \frac{\Delta Y(A,z)}{(z-w)^2} +
\frac{\pa_z Y(A,z)}{z-w} + \mbox{regular terms}.$$ This gives us the
following formula for the commutation relations between the generators
$L_n, n\in\Z$, of the Virasoro algebra, and the Fourier components, $A_m$,
of the field $Y(A,w) = \sum_{m\in\Z} A_m z^{-m-\Delta}$: $$[L_n,A_m] =
(n(\Delta-1) - m) A_{n+m}.$$ These commutation relations show that the
Fourier components of a primary field of conformal dimension $\Delta$
behave with respect to the generators $L_n = t^{-n+1} \pa_t$ as the
$(1-\Delta)-$differentials on the circle $t^{-m+\Delta-1} dt^{1-\Delta}.$

\thmref{walgebra} shows that the VOA $\W_\beta(\g)$ is ``generated'' by
$l$ fields: ${\bold W}_i^\beta(z), i=1,\ldots,l$. The first of them,
${\bold W}_1^\beta(z)$, is the Virasoro field, and the others are primary
fields of conformal dimensions equal to the exponents of $\g$ plus $1$ with
respect to this Virasoro field for generic $\beta$. This means that we
have the following commutation relations for generic $\beta$:
\begin{equation}    \label{primary}
[L_n,W^{(i),\beta}_m] = (n d_i - m) W^{(i),\beta}_{n+m}.
\end{equation}

The fields ${\bold W}_i^\beta(z), i=1,\ldots,l$, generate the VOA
$\W_\beta(\g)$ in the sense that the OPE of any two of these fields can
be expressed through normally ordered expressions of the same fields and
their derivatives. This is equivalent to the property that $\W_\beta(\g)$
is freely generated from $v_0$ under the action of the non-negative Fourier
components of the fields ${\bold W}_i^\beta(z)$.

The vertex operator algebras $\W_\beta(\g)$ were constructed by Fateev and
Zamolodchikov in the case of $\g=\sw_3$ \cite{FZ}, and by Fateev and
Lukyanov in the cases $\g=\sw_n$ \cite{Fateev} and $\goth{so}_{2n}$
\cite{Fateev1}. They found explicit expressions of ${\bold
W}_i^\beta(z)$ through the free fields and then verified that the OPE
closes. It was conjectured that such vertex operator algebras exist for
arbitrary finite-dimensional simple Lie algebras. In \thmref{walgebra} we
prove this conjecture (this was first announced in \cite{cargese}). A
similar construction for $\g = \sw_n$ was proposed in \cite{nie1}.

One can also define $\W$-algebra $\W_\beta(\g)$ through the {\em quantum
Drinfeld-Sokolov reduction} of the affine algebra $\widehat{\g}$ of level $k$
\cite{ff1,critical,cargese}. In this setting, $\W_\beta(\g)$ with
$\beta = - (k+h^\vee)^{-1/2}$ is the $0$th cohomology of the corresponding
BRST complex. We have shown in \cite{critical}, \S 4, and
\cite{cargese}, \S 3 that the complex $F^*_\beta(\g)$ appears as the first
term of a spectral sequence of this BRST complex for generic $k$.

One can also use the opposite spectral sequence of the BRST complex to
prove the existence of $\W$-algebras, cf. \cite{tjin}.

\subsubsection{Theorem.} {\em The Lie algebra
$I_\beta(\g)$ of local integrals of motion of the quantum Toda field
theory, associated to $\g$, is isomorphic to the Lie algebra of residues
of fields from the $\W$-algebra $\W_\beta(\g)$.}

\vspace{3mm}
\noindent {\em Proof.} As explained in \lemref{homogeneous}, the
differentials of the complex $F^*_\beta(\g)$ are integrals over cycles.
Therefore, the differentials of the complex $F^*_\beta(\g)$ commute with the
derivative, and we can form the double complex $$\C \larr F^*_\beta(\g)
\larr F^*_\beta(\g) \larr \C$$ with $\pm \pa$ as the vertical
differentials. By definition, the space $I_\beta(\g)$ coincides with the
$1$st cohomology of this double complex. The Theorem now follows from
\thmref{walgebra} and the analogue of the exact sequence \eqref{pa}
for the VOA $\W_\beta(\g)$. Thus, the Lie algebra $I_\beta(\g)$ is a
quantum deformation of the Poisson algebra $I_0(\g)$ in the same sense
as before: the Lie brackets in $I_\beta(\g)$ in the $\beta^2$-expansion
have no constant term, and the linear term coincides with the Poisson
bracket in $I_0(\g)$.

In the same way we can show that the space of integrals of motion in the
larger space $\f_0^\beta$ coincides with the Lie algebra of all Fourier
components of fields from $\W_\beta(\g)$.

\subsection{Affine Toda field theories.}

In this subsection we will extend the methods of the previous
subsection to the quantum affine Toda field theories.

\subsubsection{} Our task is again to construct the deformed complex
$F^*_\beta(\g)$, which becomes $F^*(\g)$ in the limit $\beta \arr 0$, for an
affine Lie algebra $\g$.

According to \secref{qbgg} the quantum BGG resolution $B_*^q(\g)$ exists
for the quantized affine algebra $U_q(\g)$. \lemref{qserre} and
\lemref{homogeneous} also hold in the affine case. Thus, we can define the
complex $F^*_\beta(\g)$ in the same way as in the case of
finite-dimensional simple Lie algebras.

As a vector space, the $j$th group of the complex $F^j_\beta(\g)$ is the
direct sum of the modules $\pi_{\rho-s(\rho)}$, where $s$ runs over the set
of elements of the Weyl group of $\g$ of length $j$. The differential
$\delta^j_\beta: F^{j-1}_\beta(\g) \arr F^j_\beta(\g)$ is given by formula
\eqref{qdifferential}. Note that this complex is $\Z-$graded with
finite-dimensional homogeneous components and the differentials are
homogeneous of degree $0$.

After proper rescaling of the differentials, we obtain a family of
complexes, defined for generic $\beta \in \C$, such that for $\beta=0$ we
obtain our classical complex $F^*(\g)$. Let us restrict ourselves with the
affine algebras, whose exponents are odd and the Coxeter number is even. It
turns out that using the result of \propref{wedgegen}, in which the
cohomologies of the complex $F^*(\g)$ were computed, and the fact that the
Euler characteristics of the cohomologies does not depend on $\beta$, we
can prove that the cohomologies of the complex $F^*_\beta(\g)$ for generic
$\beta$ are the same as for $\beta=0$.

\subsubsection{Proposition.} {\em For generic $\beta$ the cohomologies
of the complex $F^*_\beta(\g)$ are isomorphic to the exterior algebra
$\bigwedge^*(\ab^*)$ of the dual space to the principal commutative
subalgebra $\ab$ of $\n_+$.}

\vspace{3mm}
\noindent {\em Proof.} Let $F^j_\beta(\g)_m$ and $H^j_\beta(\g)_m$
be the $m$th homogeneous components of the $j$th group of the complex
$F^*_\beta(\g)$ and of its $j$th cohomology group, respectively. The Euler
characteristics of the $m$th homogeneous component of the complex
$F^*_\beta(\g)$, $$\sum_{j\geq 0} (-1)^j \dim F^j_\beta(\g)_m = \sum_{j\geq
0} (-1)^j \dim H^j_\beta(\g)_m,$$ does not depend on $\beta$.

{}From \propref{wedgegen} we know that the Euler character of the
complex $F^*(\g)$ is equal to that of $\bigwedge^*(\ab^*)$: $$\prod_{n
\equiv d_i mod h} (1-q^n),$$ where $d_1,\ldots,d_l$ are the
exponents of $\g$ and $h$ is the Coxeter number. In fact, one can compute
the Euler character of our complex by a different method. By definition, it
is equal to $$\sum_{j \geq 0} (-1)^{j} \mbox{ch} F^{j}_\beta(\g) =
\prod_{n>0} (1-q^{n})^{-l}
\sum_{s} (-1)^{l(s)} q^{(\rho-s(\rho)|\rho^{\vee})}.$$ Using the
Weyl-Kac character formula for the trivial representation of $\g$ in
the principal gradation \cite{kac}, we can reduce it to the product
formula above.

We restrict ourselves with the case when all the exponents of $\g$ are odd
and the Coxeter number is even (the general proof is technically more
complicated and it will be published separately). Then the subcomplex
$F^*(\g)_m$ has cohomologies of only even degrees, if $m$ is even, and of
only odd degrees, if $m$ is odd. It means, according to \lemref{generic},
that the same property holds for the subcomplex $F^*_\beta(\g)_m$ for
generic values of $\beta$.

But then we have: $$\sum_{j\geq 0} \dim H^{2j}_\beta(\g)_m =
\sum_{j\geq 0} \dim H^{2j}_0(\g)_m$$ for even $m$ and $$\sum_{j\geq 0}
\dim H^{2j-1}_\beta(\g)_m = \sum_{j\geq 0}
\dim H^{2j-1}_\beta(\g)_m$$ for odd $m$. If the value $\beta=0$ of the
parameter were not generic, then for generic $\beta$ we would have
$\dim H^l_\beta(\g)_m \geq \dim H^l_0(\g)_m$ for all $l$, and there
would exist such $i$ that $\dim H^i_\beta(\g)_m > \dim H^i_0(\g)_m$. But
this would contradict the equalities above. Therefore for any $m$ and
$j$ $$\dim H^j_\beta(\g)_m = \dim H^j_0(\g)_m,$$ and Proposition
follows.

\subsubsection{Theorem.} {\em All local integrals of motion of the
classical affine Toda field theory can be deformed, and so the space
$I_\beta(\g)$ of quantum integrals of motion is linearly generated by
mutually commuting elements of degrees equal to the exponents of $\g$
modulo the Coxeter number.}

\vspace{3mm}
\noindent {\em Proof.} The same as in \thmref{wedgegen}, in particular,
since the Lie bracket preserves the grading, from the fact that
they all have odd degrees follows that they commute with each other.

\subsection{Concluding remarks.}

\subsubsection{The duality $\beta \arr -(r^\vee)^{\frac{1}{2}}/\beta$ and
the limit $\beta \arr \infty$.}

There is a remarkable duality in $\W$-algebras
\cite{critical,thesis,duality}. Let $\g$ be a simple Lie algebra and $\g^L$
be the Langlands dual Lie algebra, whose Cartan matrix is the transpose of
the Cartan matrix of $\g$. Let $r^\vee$ be the maximal number of edges,
connecting two vertices of the Dynkin diagram of $\g$.

For generic values of $\beta$ the vertex operator algebra $\W_\beta(\g)$ is
isomorphic to the vertex operator algebra $\W_{\beta^L}(\g^L)$, where
$\beta^L = - (r^\vee)^{\frac{1}{2}}/\beta$.

Accordingly, under these conditions $I_\beta(\g) \simeq I_{\beta^L}(\g^L)$.

Clearly, $\g \simeq \g^L$, unless $\g$ is of types $B_n$ or $C_n$, in which
case they are dual to each other. The duality means then that there is only
one family of $\W$-algebras associated to the Lie algebras $B_n$ and
$C_n$.

The proof of this duality \cite{critical,thesis} is based on the explicit
computation in the rank one case, which follows from the proof of
\propref{virquantum}. Indeed, it is clear that the Virasoro element in
$\W_\beta(\sw_2)$ given by formula \eqref{virel} is invariant under the
transformation $\beta \arr -2/\beta$. Therefore for generic $\beta$ we have
the isomorphism $\W_\beta(\sw_2) \simeq \W_{-2/\beta}(\sw_2)$.

General case can be reduced to the case of $\sw_2$. Let $\pi_0^{(i)}$ be
the subspace in $\pi_0$, which is generated from $v_0$ by the operators
$b_n^{j*}, n<0, j\neq i$. These operators commute with $b_m^i, m\in\Z,$ and
hence with $\q_i^\beta = \int V_{\beta \al_i}(z) dz$.  Therefore the kernel
of the operator $\q_i^\beta$ on $\pi_0$ coincides with the tensor product
of $\pi_0^{(i)}$ and the kernel of the operator $\q_i^\beta$ on the
subspace of $\pi_0$, generated from $v_0$ by the operators $b_n^i,
n<0$. But the latter is isomorphic to $\W_{\beta \| \al_i
\|}(\sw_2)$, and hence does not change, if we replace $\beta$ by $-2/(\beta
\| \al_i \|^2)$.

Thus we see that the kernel of the operator $\q_i^\beta = \int V_{\beta
\al_i}(z) dz$ coincides with the kernel of the operator $\int
V_{-\al_i^\vee/\beta}(z) dz$, where $\al_i^\vee =
2\al_i/(\al_i,\al_i)$. It is known \cite{Kac} that the scalar product
$(\al_i^\vee,\al_j^\vee)$ in $\h^* \subset \g$ equals $r^\vee$ times the
scalar product $(\al_i^L,\al_j^L)$ of the simple roots $\al_i^L \in \h^L
\subset \g^L$ of the Langlands dual Lie algebra $\g^L$. We can
therefore identify the Heisenberg algebras $\widehat{\h}$ and
$\widehat{\h}^L$ by identifying $\al_i^\vee$ with $(r^\vee)^{\frac{1}{2}}
\al_i^L$. But then the operator $\int V_{-\al_i^\vee/\beta}(z) dz$ becomes
the operator $\q_i^{\beta^L}$, where $\beta^L
=-(r^\vee)^{\frac{1}{2}}/\beta$. Therefore for generic values of $\beta$
the kernel of the operator $\q_i^\beta$ coincides with the kernel of the
operator $\q_i^{\beta^L}$ of the Langlands dual Lie algebra. Thus,
$\W_\beta(\g) \simeq \W_{\beta^L}(\g^L)$.

This duality has a remarkable limit when $\beta \arr \infty$. Of course, in
this case the operator $\q_i^\beta$ is not well-defined and has to be
regularized.

It suffices describe this regularization in the rank one case. To this end,
consider the Heisenberg algebra with generators $b_n, n\in\Z$ with the
commutation relations $$[b_n,b_m] = n
\delta_{n,-m}.$$ We will introduce bases in the Fock spaces $\pi_0$ and
$\pi_\beta$ in such a way that the matrix elements of the operator $\q^\beta =
\int V_\beta(z) dz: \pi_0 \arr \pi_\beta$ are well-defined when $\beta \arr
\infty$.

As the basis elements in $\pi_0$ we will take monomials in $b'_n =
b_n/\beta, i<0$.  Denote $W_N = \int V_\beta z^{N-1} dz, N\in\Z$. The basis
in $\pi_\beta$ consists of elements
\begin{equation}    \label{monomial}
b'_{n_1} \ldots b'_{n_m} w_N, \quad \quad n_1\leq n_2\leq \ldots
\leq n_m < -1, \quad N \leq 0,
\end{equation}
where for finite $\beta$ we put $w_N = W_N \cdot v_0$. Here $v_0$ is the
vacuum vector of $\pi_0$. In particular, $w_0$ is the vacuum vector $v_1$
of $\pi_\beta$.

In order to find out how $\q^\beta$ acts on a monomial basis element of
$\pi_0$, we can use the commutation relations $$[W_N,b'_n] = - W_{N+n},$$
which follow from formula \eqref{vertexcom}. So, when we apply $\q^\beta =
W_1$ to a monomial in $b'_n$'s, we obtain a linear combination of terms of
the form \eqref{monomial}, but with $n_m\leq -1$. However, we can
re-express elements $P \cdot (b'_{-1})^k w_N$, where $P$ is a polynomial in
$b'_n, n<-1$, in terms of elements of the form \eqref{monomial}, using the
identity
\begin{equation}    \label{identity}
\sum_{n+M=N,n<0,M\leq 0} b'_n w_M = - \frac{1}{\beta^2} N w_N, \quad N<0.
\end{equation}
This identity can be obtained by applying to $v_0$ the negative Fourier
components of the formula $$:b(z) V_\beta(z): = \frac{1}{\beta}
\frac{\pa}{\pa z} V_\beta(z).$$

But then see that the matrix elements of the operator $\q^\beta$ in this
new basis are polynomials in $\beta^{-2}$, and therefore they define a
certain linear operator, when $\beta^{-2} = 0$. We will denote this
operator by $\q^\infty$.

We can show that the kernel of the operator $\q^\infty$ coincides with the
$\beta \arr \infty$ limit of the kernel of the operator $\q^\beta$ for
generic $\beta$ \cite{thesis}. We can then check that this kernel is
isomorphic to the $\beta \arr 0$ limit of the kernel of the operator
$\q^\beta$ for generic $\beta$, which is described in \propref{vir}.

By extending this result to higher rank case in the same way as for generic
$\beta$, we can prove that $\W_\infty(\g) \simeq \W_0(\g^L)$, where
$\W_\infty(\g)$ denotes the intersection of kernels of the operators
$\q_i^\infty$ associated to $\g$ \cite{critical,thesis,duality}.

The quotient $I_\infty(\g)$ of $\W_\infty(\g)$ by total derivatives and
constants has a Poisson bracket, which is equal to the $\beta^{-2}$--linear
term in the commutator in $I_\beta(\g)$. This isomorphism implies that
$I_\infty(\g) \simeq I_0(\g^L)$.

In \cite{critical,thesis} we showed that $I_\infty(\g)$ is isomorphic to
the center $Z(\widehat{\g})$ of a certain completion of the universal
enveloping algebra of the affine algebra $\widehat{\g}$ at the critical
level. Thus, we see that the center $Z(\widehat{\g})$ is isomorphic to the
classical $\W-$algebra of the Langlands dual Lie algebra $\g^L$.

In the same way we can prove the isomorphism $I_\infty(\g) \simeq
I_0(\g^L)$, where $\g$ is an affine Kac-Moody algebra and $\g^L$ is the
affine algebra, whose Cartan matrix is obtained by transposing the Cartan
matrix of $\g$.

\subsubsection{Explicit formulas.} It is an interesting problem to find
explicit formulas for the quantum integrals of motion of the affine Toda
field theories.

Explicit formulas for the classical ones are known in many cases. For
instance, there are many effective methods to compute explicitly the KdV
hamiltonians, which are the local integrals of motion of the sine-Gordon
model. However, it seems none of those methods can be used to produce the
quantum integrals, cf. e.g. \cite{km}. So far, only partial results have
been obtained in this direction.

First of all, a few quantum integrals of motion of the sine-Gordon theory of
low degrees are known for any value of the central charge of the Virasoro
algebra, cf. e.g. \cite{SY}. When the central charge is equal to
$1-3(2n-1)^2/(2n+1)$ (the $(2,2n+1)$ minimal model of the Virasoro
algebra), it is known that the quantum integral of motion of degree $2n-1$
can be obtained as the residue of the field, corresponding to the
singular vector of degree $2n$ in the vacuum Verma module of the Virasoro
algebra \cite{fkm,EY1,dm1}. An explicit formula is known for this singular
vector, and this allows one to write down the corresponding integral of
motion for this value of the deformation parameter. A similar phenomenon
has been observed in other theories \cite{dm1}.

Finally, the quantum integrals of motion are known for the central charge
$c=-2$, which corresponds to $\beta=2$. The reason for that is that in this
case the operators $\q_1$ and $\q_0$ have a simple realization in terms of
the Clifford algebra with the generators $\psi_i, \psi^*_i, i \in \Z$ and
the anti-commutation relations $$[\psi_i,\psi^*_j]_+ = \delta_{i,-j}.$$
Indeed, let $\bigwedge^* = \oplus_{n\in\Z} \bigwedge^n$ be the Fock
representation of this algebra with the vacuum vector $v$, satisfying
$$\psi_i v = 0, i\geq 0, \quad \psi^*_i v = 0, i>0.$$

This representation is $\Z-$graded in accordance with the convention $\deg
\psi_i = 1, \deg \psi^*_i = -1$. One can introduce an action of this
Clifford algebra on the space $\oplus_{n\in\Z} \pi_n$ using the vertex
operators by the formulas $$\psi(z) = \sum_{m\in\Z} \psi_m z^{-m-1} =
V_1(z), \quad \quad \psi^*(z) =
\sum_{m\in\Z} \psi^*_m z^{-m} = V_{-1}(z).$$ Since $\beta$ is an integer,
all Fourier components of these vertex operators are well-defined on any of
the modules $\pi_n$. This boson-fermion correspondence allows us to
identify our complex $F^*_2(\widehat{\sw}_2) \simeq \oplus_{n\in\Z} \pi_{2n}$
with the even part of $\bigwedge^*$. The operators $\q_1$ and $\q_0$ then
become
\begin{equation}    \label{ferm}
\q_1 = \int V_2(z) dz = \int \psi(z) \pa_z \psi(z) dz, \quad \quad \q_0 =
\int V_{-2}(z) dz = \int \psi^*(z) \pa_z \psi^*(z) dz.
\end{equation}

It is not difficult to prove directly that they satisfy the Serre relations
with $q=1$. Recall that for generic values of $\beta$ only the compositions
of the operators $\q_1^\beta$ and $\q_0^\beta$, corresponding to the
singular vectors in the Verma modules over the quantum group, are
well-defined as linear operators acting between the spaces $\pi_n$. The
operators \eqref{ferm} are always well-defined and they generate an action
of the nilpotent Lie subalgebra $\n_+$ of $\widehat{\sw}_2$ on
$\bigwedge^*$. In fact, this action can be extended to an action of the
whole Lie algebra $\widehat{\sw}_2$ \cite{leclair}.

It is possible to write down explicit formulas for the integrals of motion
for $\beta=2$ in terms of the fermions $\psi(z),
\psi^*(z)$ \cite{dm2}: $${\cal H}_{2n+1} = \int \psi(z) \pa_z^{2n-1}
\psi^*(z) dz.$$ These formulas can be converted into nice formulas in terms
of the generators of the Virasoro algebra, which first appeared in
\cite{SY}. Other features of the case $\beta=2$ have been studied in
\cite{leclair}.

Similar, but more complicated is the case when $\beta = N$, a positive
integer. Then the operators $\q_1$ and $\q_0$ can be written in terms of
$\psi(z)$ and $\psi^*(z)$ as follows: $$\q_1 = \int
\psi(z) \pa_z \psi(z) \ldots \pa_z^{N-1} \psi(z) dz,$$  $$\q_0 = \int
\psi^*(z) \pa_z \psi^*(z) \ldots \pa_z^{N-1} \psi^*(z) dz.$$ In these
cases explicit formulas for the integrals of motion are still lacking.
Finding such formulas for infinitely many integers $N$ would lead to an
independent proof of the existence of quantum integrals of motion.

\subsubsection{Special values of $\beta$ for finite-dimensional $\g$.} So
far, we have only been interested in the generic values of the deformation
parameter $\beta$. In this subsection we will discuss briefly what happens
for special values of $\beta$, that is the values, for which the kernel of
the operator $\sum_i \Q_i^\beta$ on $\F_0^\beta$ becomes larger.

Let us first look at the case of finite-dimensional $\g$. In the simplest
case of $\g=\sw_2$ our complex is $\pi_0 \larr \pi_1$, and we have proved
that for generic values of $\beta$ the first cohomology of this complex is
trivial. In fact, this statement can be proved directly, using the
description of the structure of the modules $\pi_0$ and $\pi_1$ over the
Virasoro algebra from \cite{FeFu}.  According to this description, the
module $\pi_1$ is irreducible for generic $\beta$, while the module $\pi_0$
contains an irreducible submodule, such that the quotient by this submodule
is isomorphic to $\pi_1$. If $\beta^2$ is a positive rational number,
however, the modules $\pi_0$ and $\pi_1$ may become highly reducible, and
that leads to the appearance of the first cohomology and the enlargement of
the $0$th cohomology.

The most interesting situation occurs when $\beta^2 = 2p/q$, where $p,q>1$
are two relatively prime integers. The corresponding central charge $c = 1
- 6(p-q)^2/pq$ is the central charge of the $(p,q)$ minimal model
\cite{bpz}. In that case the composition structure of the modules $\pi_0$
and $\pi_1$ becomes very complicated \cite{FeFu}, and our complex has very
large cohomology groups. It turns out, however, that one can extend this
complex to an infinite two-sided complex, whose cohomologies are
concentrated in one dimension and are isomorphic to the irreducible
representation of the Virasoro algebra of highest weight $0$. This
representation is the quotient of the vacuum Verma module of the Virasoro
algebra by its submodule, generated by a unique singular vector, which it
contains. Note that this irreducible representation is at the same time the
VOA of the corresponding minimal model.

Such a complex was constructed by Felder \cite{F}. It has one Fock space
$\pi_\la$ with an appropriate $\la$ in each group, and the only cohomology
occurs in the $0$th group of the complex.

The situation with the $\W-$algebras, associated to general
finite-dimensional Lie algebras, is apparently very similar. The vacuum
Verma module $\W_\beta(\g)$, freely generated from the vacuum vector by the
operators $W^{(i),\beta}_{n_i}, n_i<-d_i$ (cf. \thmref{walgebra}), which is
irreducible for generic values of $\beta$, may contain singular vectors, if
$\beta^2$ is a positive rational number. The quotient $L_\beta(\g)$ of
$\W_\beta(\g)$ by the submodule, generated by these singular vectors, is
irreducible. One should be able to construct two-sided complexes, which
consist of the Fock spaces $\pi_\la$, labeled by elements of the {\em
affine} Weyl group of $\g$, in which the $0$th cohomology would be
isomorphic to $L_\beta(\g)$ and all other cohomologies would be trivial.

Such complexes have been conjectured in \cite{fkw}, Conjecture 3.5.2. These
complexes should appear \cite{fkw} as the result of the quantum
Drinfeld-Sokolov reduction of similar complexes (two-sided BGG resolutions)
over the corresponding affine Kac-Moody algebra $\widehat{\g}$
\cite{FeFr,bf,B}. This has been proved for $\g=\sw_2$ in \cite{ff2,bo}.

These complexes are closely connected with certain complexes \cite{B} of
modules over the quotient of the quantum group $U_q(\g)$ with $q$ a root of
unity, $q = \exp (2\pi i p/q)$, by a big central subalgebra, cf. \cite{dk}.

There is also another interesting value of $\beta$, namely, $\beta=1$ for a
{\em simply-laced} Lie algebra $\g$ \cite{F1,BBSS,B,FKRW}. In this case
$q=-1$, and slightly redefined operators $\Q_i^\beta$ generate the
nilpotent Lie algebra $\n_+$. This nilpotent subalgebra lies in the
constant subalgebra of the whole affine algebra $\widehat{\g}$ acting on
the direct sum of the Fock modules $\pi_\la$ where the summation is over
the root lattice $Q$ of $\g$, by vertex operators \cite{FK}. The
$\W$--algebra $\W_1(\g)$ can then be interpreted as the space of invariants
of the constant subalgebra of $\widehat{\g}$ in $\pi = \oplus_{\la \in Q}
\pi_\la$. This implies that $I_1(\g)$ (for which the central charge is the
rank of $\g$) is the commutant of $\g$ in $\f^1_0$. A version of $I_1(\g)$
was defined for the first time by I.~Frenkel in \cite{F1}.

It was proved in \cite{FKRW}, Theorem 4.2 (cf. also \cite{Bo,bs}), that
$\W_1(\g)$ has the same character as $\W_\beta(\g)$ for generic $\beta$,
i.e. that the intersection of kernels of the operators $\Q_i^\beta$ does
not increase at the point $\beta=1$. Let us show that the higher
cohomologies of the complex $F^*_1(\g)$ vanish. This has been conjectured
(and proved for $\g=\sw_2$) in \cite{F1}.

Consider the complex $\widetilde{F}^*(\g)$, in which the $j$th group
consists of $\# \{ w| l(w)=j \}$ copies of $\pi$, and the differentials are
given by the same formulas as the differentials of the complex
$F^*_1(\g)$. The cohomology of the complex $\widetilde{F}^*(\g)$ it is the
cohomology of the Lie algebra $\n_+$ with coefficients in $\pi$. The
complex $F^*_1(\g)$ is a subcomplex of $\widetilde{F}^*(\g)$. It is easy to
show that its cohomology is the subspace of the cohomology of
$\widetilde{F}^*(\g)$ of weight $0$ with respect to the Cartan subalgebra
of $\g$ acting on $\widetilde{F}^*(\g)$ and commuting with the
differentials. But $\pi$ is a direct sum of finite-dimensional
$\g$--modules. By Borel-Weil-Bott-Kostant theorem, weight $0$ cohomology
classes in $\widetilde{F}^*(\g)$ can occur only in dimension $0$ and those
are the invariants of $\g$ in $\pi$. Therefore the higher cohomologies of
$F_1^*(\g)$ vanish. This implies that the character of $\W_1(\g)$ coincides
with the character of $\W_\beta(\g)$ for generic $\beta$, and thus we
obtain an alternative proof of Theorem 4.2 of \cite{FKRW}.

It was shown in \cite{FKRW} that $I_1(\sw_N)$ is the quotient of the local
completion of the universal enveloping algebra of the Lie algebra
$\W_\infty$ with central charge $N-1$.

\subsubsection{Special values of $\beta$ for affine $\g$.}
Now let us turn to the space $I_\beta(\g)$ of the integrals of motion of
the affine Toda field theory associated to an affine algebra $\g$. The space
of these integrals was defined as the intersection of kernels of the
operators $\Q_i^\beta, i=0,\ldots,l,$ on $\F_0^\beta$. For generic values
of $\beta$ the intersection of kernels of the operators $\Q_i^\beta$
with $i=1,\ldots,l$ coincides with the $\W-$algebra $I_\beta(\bar{\g})$, and
so the space $I_\beta(\g)$ can be defined as the kernel of the operator
$\Q_0^\beta$ on $I_\beta(\bar{\g})$.

Recall that the $\W-$algebra $I_\beta(\bar{\g})$ is the quotient of the
vacuum Verma module $\W_\beta(\bar{\g})$ by the total derivatives and
constants. If $\g$ is untwisted, the operator $\q_0^\beta$ can be
interpreted as the residue $\int \Phi_{1,1,Adj}(z) dz$ of a certain primary
field $\Phi_{1,1,Adj}(z)$, acting from $\W_\beta(\bar{\g})$ to another
module $M_\beta(\bar{\g})$ over the $\W-$algebra \cite{Zam,fl,EY,HM}, so
that the operator $\Q_0^\beta$ is the corresponding operator on the
quotients by the total derivatives and constants. Therefore, for generic
$\beta$ the space $I_\beta(\g)$ consists of the elements $P^- \in
\W_\beta(\bar{\g})$, for which $\q_0^\beta \cdot P^-$ is a total derivative
in $M_\beta(\bar{\g})$:
\begin{equation}    \label{perturb}
\int \Phi_{1,1,Adj}(z) dz \cdot P^- = \pa P^+.
\end{equation}
This equation shows that the pair $(P^-,P^+)$ can be interpreted as a
conservation law (compare with \remref{conslaws}) in the deformation of the
corresponding conformal field theory obtained by adding $\la \int
\Phi_{1,1,Adj}(z) dz$, where $\la$ is a parameter of deformation, to the
action \cite{Zam}.

When $\beta^2$ is a positive rational number, the module
$\W_\beta(\bar{\g})$ may become reducible. Because of that, the
cohomologies of the complex $F^*(\g)$ increase. In such a case it is
appropriate to redefine integrals of motion as elements $P^-$ of the {\em
irreducible} module $L_\beta(\bar{\g})$, which satisfy the equation
\eqref{perturb} \cite{Zam}. This may result in dropping out of some of the
``generic'' integrals of motion. At the same time some new ones may
appear.

For instance, for $\g=\widehat{\sw}_2, \beta^2 = 4/(2n+1)$ (the $(2,2n+1)$
model) the density of the integral of motion of the quantum sine-Gordon
theory of degree $2n-1$ coincides with the field, corresponding to the
singular vector of degree $2n$. Since we take the quotient by the
submodule, generated by this vector, this integral of motion drops out
(cf. the previous section). It has been argued that the integrals of motion
of degrees, which are divisible by $2n-1$, also drop out in this case
\cite{fkm,kns,EY1}.

Another example of dropping out of integrals of motion is (in our
terminology) the Toda field theory associated to the twisted algebra
$A_2^{(2)}$ for the value $\beta^2 = 3/2$. The integrals of motion of this
Toda theory for generic values of $\beta$ have all positive integral
degrees, which are not divisible by $2$ and $3$. They are elements of the
Virasoro algebra $I_\beta(\sw_2)$, because for $\g = A_2^{(2)}$, $\bar{\g}
= \sw_2$. They have the property \eqref{perturb} with the field
$\Phi_{(1,3)}(z) =
\Phi_{1,1,Adj}(z)$ replaced by $\Phi_{(1,2)}(z)$ \cite{Zam}. The value
$\beta^2 = 3/2$ corresponds to the Ising model $(3,4)$ with central charge
$c=1/2$. It was found in \cite{Zam,fz} that the integral of motion of
degree $5$ drops out for this special value of parameter. It was
conjectured that the degrees of the integrals of motion which should occur
are relatively prime with $30$, so that they are the exponents of
$E_8^{(1)}$ modulo the Coxeter number.

On the other hand, in the same theory with $\beta^2 = 8/5$ (the Ising
tri-critical point, the $(4,5)$ minimal model) the appearance of an integral
of motion of degree $9$ was observed \cite{fz,cm} and it was conjectured
that there should also be integrals of motion of degrees $9n$, where $n$ is
an arbitrary positive odd integer.

In these examples, the dropping out or appearance of new integrals of
motion is caused by the existence of a larger vertex operator algebra of
symmetries of the model. For instance, it is known that the $(2,2n+1)$
minimal model of the Virasoro algebra coincides with $(2n-1,2n+1)$ minimal
model of the $\W$-algebra $\W(\sw_{2n-1})$ \cite{kns}, and that the Ising
model has a hidden symmetry of $\W(E_8)$ \cite{bg}. Therefore one should
expect that the degrees of integrals of motion in such a model should
satisfy ``exclusion rules'' of the larger symmetry algebra as well. It is
interesting whether there are other reasons for dropping out or appearance
of new integrals of motion.

It seems plausible that for the special values of $\beta$ one can construct
a two-sided complex, consisting of the modules $\pi_\la$, whose first
cohomology would give the space of integrals of motion, corresponding to
the irreducible representation $L_\beta(\bar{\g})$. We have constructed a
candidate for such a complex for the $(2,2n+1)$ model. The computation of
the Euler character of this complex suggests that its first cohomology is
indeed generated by elements of all odd degrees, which are not divisible by
$2n-1$. We will discuss this complex elsewhere.

\subsubsection{Spectrum of the integrals of motion.} Our integrals of motion,
both classical and quantum, act on the spaces $\pi_\la$. They are not
diagonalizable, since they are all of negative degrees (in particular, the
first of them is the operator of derivative $\pa$). However, one can define
a transformation on the space $\f_0^\beta$ of Fourier components of
fields \cite{nahm}, which maps the set of integrals of motion to a set of
mutually commuting elements of degree $0$ (for example, the first integral
of motion, ${\cal H}_1=\pa=L_{-1}$ maps to $L_0-c/24$), cf. \cite{nie4}.

It would be very interesting to find the spectrum of these operators on the
modules $\pi_\la$.

\vspace{15mm}
\noindent {\bf Acknowledgments.} E.F. would like to thank the organizers of
the C.I.M.E. Summer School, M. Francaviglia and S. Greco, for giving him
the opportunity to present these lectures, and for creating a very
stimulating atmosphere at the School.

The main part of this work was done during our visits to the Research
Institute for Mathematical Sciences and Yukawa Institute of Kyoto
University in 1991-1993. We express our deep gratitude to these
institutions, and especially to T. Inami, M. Kashiwara, and T. Miwa, for
financial support and generous hospitality.

We thank J. Bernstein, V. Kac, D. Kazhdan, F. Smirnov, and A. Varchenko for
their interest in this work and valuable comments.

E.F. was supported by a Junior Fellowship from the Society of Fellows of
Harvard University and in part by NSF grant DMS-9205303.

\end{document}